\documentclass[twocolumn]{aastex631}

\usepackage[T1]{fontenc}
\DeclareRobustCommand{\VAN}[3]{#2}
\let\VANthebibliography\thebibliography
\def\thebibliography{\DeclareRobustCommand{\VAN}[3]{##3}\VANthebibliography}

\usepackage{graphicx}	
\usepackage{amsmath}	
\usepackage{amssymb}	
\usepackage{acronym} 

\usepackage{xspace}
\usepackage{multirow}

\newcommand{\Zsun}{\ensuremath{\,\rm{Z}_{\odot}}\xspace}

\newcommand{\alphaCE}{\ensuremath{\alpha_{\rm{CE}}}\xspace}

\newcommand*\diff{\mathop{}\!\mathrm{d}}
\defcitealias{Boesky_2023}{Boesky et al. (2024)} 

\acrodef{DCO}{double compact object}
\acrodef{BNS}{binary neutron star}
\acrodef{BBH}{binary black hole}
\acrodef{BHNS}{black hole--neutron star}
\acrodef{NS}{neutron star}
\acrodef{BH}{black hole}
\acrodef{GW}{gravitational wave}
\acrodef{SFR}{star formation rate}
\acrodef{SFRD}{star formation rate density}
\acrodef{CE}{common-envelope}

\begin{document}

\title{The Binary Black Hole Merger Rate Deviates From the Cosmic Star Formation Rate: A Tug of War Between Metallicity and Delay Times}

\author[0009-0005-9830-9966]{Adam P. Boesky}
\affiliation{Center for Astrophysics \textbar{} Harvard \& Smithsonian,
60 Garden St., Cambridge, MA 02138, USA}

\author[0000-0002-4421-4962]{Floor S. Broekgaarden}
\affiliation{Center for Astrophysics \textbar{} Harvard \& Smithsonian,
60 Garden St., Cambridge, MA 02138, USA}
\affiliation{AstroAI at the Center for Astrophysics \textbar{} Harvard \& Smithsonian,
60 Garden St., Cambridge, MA 02138, USA}
\affiliation{Simons Society of Fellows, Simons Foundation, New York, NY 10010, USA}
\affiliation{Department of Astronomy and Columbia Astrophysics Laboratory, Columbia University, 550 W 120th St, New York, NY 10027, USA}
\affiliation{William H. Miller III Department of Physics and Astronomy, Johns Hopkins University, Baltimore, Maryland 21218, USA}

\author[0000-0002-9392-9681]{Edo Berger}
\affiliation{Center for Astrophysics \textbar{} Harvard \& Smithsonian,
60 Garden St., Cambridge, MA 02138, USA}
\affiliation{The NSF AI Institute for Artificial Intelligence and Fundamental Interactions}

\begin{abstract}
Gravitational-wave detectors are now making it possible to investigate how the merger rate of binary black holes (BBHs) evolves with redshift.
In this study, we examine whether the BBH merger rate of isolated binaries deviates from a scaled star formation rate density (SFRD)---a frequently used model in state-of-the-art research.
To address this question, we conduct population synthesis simulations using COMPAS with a grid of stellar evolution models, calculate their cosmological merger rates, and compare them to a scaled SFRD.
We find that our simulated rates deviate by factors up to $3.5\times$ at $z\sim0$ and $5\times$ at $z\sim 9$ due to two main phenomena: (i) The formation efficiency of BBHs is an order of magnitude higher at low metallicities than at solar metallicity; and (ii) BBHs experience a wide range of delays (from a few Myr to many Gyr) between formation and merger.
Deviations are similar when comparing to a \textit{delayed} SFRD, and even larger (up to $\sim 10\times$) when comparing to SFRD-based models scaled to the local merger rate.
Interestingly, our simulations find that the BBH delay time distribution is redshift-dependent, increasing the complexity of the redshift distribution of mergers.
We find similar results for simulated merger rates of BHNSs and BNSs.
We conclude that the rate of BBH, BHNS, and BNS mergers from the isolated channel can significantly deviate from a scaled SFRD, and that future measurements of the merger rate will provide insights into the formation pathways of gravitational-wave sources.
\end{abstract}
\section{Introduction} \label{sec:intro}
The rapidly increasing sample of \ac{GW} events detected by the Advanced LIGO and Virgo interferometers offers a new opportunity to explore the formation and properties of \acp{BH} and \acp{NS} as a function of redshift. 
The most recent \ac{GW} catalogs (GWTC-3, OGC-4) and independent \ac{GW} data analyses already contain about $100$ \ac{BBH} mergers out to redshifts $z\sim 1.5$ \citep{Abbott:2021GWTC1, GWTC2, Abbott:2021GWTC3, Abbott:2021-GWTC-2-1, Nitz:2023-4-OGC, Venumadhav:2019, Venumadhav:2020, 2019PhRvD.100b3007Z, Olsen:2022, mehta2023new, Wadekar:2023}, 
and next-generation \ac{GW} detectors, such as Cosmic Explorer and Einstein Telescope, are poised to detect stellar-mass black hole mergers beyond $z\gtrsim 10$ \citep[e.g.,][]{EinsteinTelescope:2010Punturo, EinsteinTelescope:2012Sathyaprakash, CosmicExplorer:2019reitze, EinsteinTelescope:2020Maggiore, CosmicExplorer:2021evans, CosmicExplorer:2023,  Singh_2022, Gupta:2023}.
The rate and properties of \ac{BBH}, \ac{BHNS}, and \ac{BNS} mergers as a function of redshift can provide invaluable insights into the physical processes underlying \ac{BH} and \ac{NS} formation, the massive (binary) stars that lead to their formation, their host galaxies, and the different formation channels at play \citep[e.g.,][]{Fishbach:2018, RodriguezLoeb:2018, Vitale:2019, Bavera_2021, Ng_2021, Chu:2022, chruslinska_2022, Mapelli:2022, van_Son_2022, Ray:2023, Santoliquido:2023, Vijaykumar:2023}.
A challenge, however, is that information about the formation pathway and progenitor system is not directly imprinted in the GW observations. 
Inferring such properties, or making predictions for future \ac{GW} detectors, therefore often requires making assumptions about the underlying merger population.

A common assumption in the literature is that the BBH (and \ac{BHNS}/
\ac{BNS}) merger rate, $\mathcal{R}_{\rm{merge}}(z)$, can be described by scaled versions of the cosmic \ac{SFRD}, matched to the local observed merger rate \citep[e.g.,][]{Iacovelli:2022, Gupta:2023, Lehoucq:2023}. 
This assumption is based on the notion that \ac{GW} sources are formed from massive stars, whose formation rate is described by the \ac{SFRD}, which rapidly form binary compact object systems ($\sim 10$ Myr; e.g., \citealt{KippenhahnWeigert:1990}) that merge after some time.
There are many astrophysical processes in the evolution of binary stars, however, that can drastically alter this paradigm, leading to a merger rate that might \textit{not} follow a scaled \ac{SFRD}.  
As will be the focal point of this paper, there are two key processes within the isolated binary evolution formation channel that can cause significant deviations from the \ac{SFRD}.

First, the formation efficiency of compact object binaries is metallicity-dependent. The evolutionary outcome of massive (binary) stars can rely strongly on birth metallicity, as metallicity drives mass loss through stellar winds which impacts, for example, radial extension of stars and the remnant mass of compact objects \citep[e.g.,][]{Vink_2000, Vink_2001, Langer:2012}. Recent work has shown that the formation yield of \acp{BBH} is strongly dependent on their birth metallicity (\citealt{Chruslinska_2018, Giacobbo_2018, Broekgaarden_2022}), resulting in a \ac{BBH} merger rate that does not follow the \ac{SFRD}, but an \ac{SFRD} convolved with this metallicity dependence. 

Second, BBH mergers may occur with a significant delay relative to the formation epoch, and the distribution of delay times can itself be metallicity-dependent \citep[e.g.,][]{Belczynski:2016, Giacobbo_2018}. Although massive binary stars form \ac{BBH} systems (or BHNS and BNS systems) in just $\sim 10$ Myr, the merger delay times can be as long as many Gyr \citep[e.g.][]{Peters_1964, Neijssel_2019}, resulting in a merger rate that may follow a \textit{delayed} \ac{SFRD}. 
Indeed, some \ac{GW} studies model $\mathcal{R}_{\rm{merge}}(z)$ as a scaled \ac{SFRD} combined with a delay time distribution, which is often assumed to follow a simple power law, $\diff n/ \diff t \propto t^{-1}$, with a minimum delay time of $\sim 10-100\,\rm{Myr}$ \citep[e.g.,][]{Regimbau:2012, Belgacem:2019, Borhanian:2022arXiv220211048B, Colombo:2022, Iacovelli:2022, Naidu:2022, Lehoucq:2023}. 

Here, we investigate how the two aforementioned effects impact $\mathcal{R}_{\rm{merge}}(z)$. We employ a large set of population synthesis models for the isolated binary evolution formation pathway to simulate BBH (as well as \ac{BHNS} and \ac{BNS}) mergers and investigate the resulting merger rates as a function of redshift for a wide range of assumptions about stellar and binary evolution.   The paper is structured as follows.  
We describe our models and methods in Section~\ref{sec:methods}. We investigate the effects of metallicity-dependent formation yield and delay times on $\mathcal{R}_{\rm{merge}}(z)$ in Section~\ref{sec:shift-to-the-right} and Section~\ref{sec:t_delay}, respectively. 
In Section~\ref{sec:Comparing-to-a-delayed-SFR}, we assess the accuracy of models from the literature by comparing our simulated rates to \ac{SFRD}-based models scaled to the inferred local BBH merger rate and convolved with different delay time distributions. 
We discuss our results for \ac{BHNS} and \ac{BNS} mergers in Section~\ref{The-shape-of-the-BHNS-and-BNS-merger-rate}. 
We end with a discussion in Section~\ref{sec:discussion} and a summary of our key findings in Section~\ref{sec:conclusion}.

\section{Methods} 
\label{sec:methods}
\subsection{Population Synthesis Simulated Merger Rates}\label{sec:methods-pop-synth-merger-rate}
%
We use the simulation output and methodology from
 \citet{Boesky_2023}, 
which are briefly summarized here. \citetalias{Boesky_2023} presents a population of simulated compact object mergers formed through the isolated binary evolution pathway. 
The simulations are generated using COMPAS\footnote{Compact Object Mergers: Population Astrophysics and Statistics, \url{https://compas.science}}\citep{COMPAS_2022}, a rapid population synthesis code.
COMPAS simulates the evolution of massive stars based on analytical fitting formulae for single and binary star evolution by \citet{Hurley_2000, Hurley_2002}, derived from stellar evolution tracks presented in \citet{Pols_1998}, as well as earlier work by \citet{Eggleton_1989} and \citet{Tout_1996}. 

\citetalias{Boesky_2023} studies two two-parameter-varied grids of simulations. Here, we only focus on the grid that varies the \ac{CE} efficiency, $\alphaCE$, and mass transfer efficiency, $\beta$, parameters. 
The \ac{CE} phase results from dynamically unstable mass transfer in which the companion star is engulfed in the donor's envelope and tightens the binary through drag \citep[see][and references therein]{ivanova:2020book}. 
COMPAS employs the `$\alphaCE - \lambda$' formalism \citep{Webbink_1984, deKool_1990} to parameterize the \ac{CE} phase. The $\alphaCE$ parameter is of particular interest when investigating the merger rate because it is capable of significantly reducing binary separation, and thus potentially altering the expected $\diff n/ \diff t \propto t^{-1}$ delay time distribution \citep[e.g.][]{Belczynski:2018, VignaGomez_2018}. 
Our grid uses four \ac{CE} efficiency values, $\alphaCE =$ $0.1$, $0.5$, $2$, and $10$, which is a range representative of prior studies \citep[e.g.,][]{VignaGomez_2018, Bavera_2021, Dorozsmai:2024, Santoliquido_2021, Broekgaarden_2022}. 
In tandem with $\alphaCE$, we also vary the accretion efficiency $\beta = \Delta M_{\rm acc}/M_{\rm donor}$, where $\Delta M_{\rm donor}$ and $\Delta M_{\rm acc}$ are the changes in the mass of the donor and accretor stars, respectively. 
We use $\beta =$ $0.25$, $0.5$, and $0.75$ to reflect the parameter's theoretical range of $[0,1]$ cf. earlier work \citep{Broekgaarden_2022, vanSon:2022, Dorozsmai:2024}.
Henceforth, we refer to the model with $\alphaCE = 2$ and $\beta = 0.5$ as the ``fiducial'' model.

Throughout our study, we adopt a popular model of the \ac{SFRD} from \citet{Madau_Dickinson_2014} (Equation 15):
\begin{equation} \label{eq:SFR}
    \psi(z) = a \frac{(1+z)^{b}}{1+[(1+z)/c]^d}\,\,\, \rm{M}_\odot\,\rm{yr}^{-1}\,\rm{Mpc}^{-3},
\end{equation}
with $a = 0.01$, $b = 2.6$, $c = 3.2$, and $d = 6.2$ \citep{Madau_Fragos_2017}. 
The true SFRD is uncertain, and exploring the impact of assumed \ac{SFRD} model is left for future research.
We expect the qualitative results in this work, however, to be robust under different choices for the \ac{SFRD}.

To calculate the merger rate of a simulated binary population of BBH, BHNS, or BNS mergers,
we convolve a metallicity-specific SFRD, $\mathcal{S}(Z_i,z)$, with the formation rate of the population:

\begin{equation}
\begin{split}
    &\mathcal{R}_{\rm{merge}}(t_m, M_1, M_1) \equiv  \frac{\textrm{d}^4N_{\textrm{merge}}}{\textrm{d}t_m\textrm{d}V_c\textrm{d}M_1\textrm{d}M_2} (t_\textrm{m}, M_1, M_2)                       
    \\ &= \int \textrm{d}Z_i \int_0^{t_m}\textrm{d}t_{\textrm{delay}}\mathcal{S}(Z_i, z(t_\textrm{form} = t_m  - t_\textrm{delay})) \times \\ &\frac{\textrm{d}^4N_{\textrm{form}}}{\textrm{d}M_\textrm{SFR}\textrm{d}t_\textrm{delay}\textrm{d}M_1\textrm{d}M_2}(Z_i, t_\textrm{delay}, M_1, M_2),
\end{split}
\label{eq:merger_rate}
\end{equation}
where $t_m$ is the time of the merger in the comoving frame, $t_\textrm{delay}$ is the time between the formation and merger of a binary, and $M_1$ and $M_2$ are the binary component masses. 
To obtain the metallicity-dependent star formation rate $\mathcal{S}(Z_i, z(t_\textrm{form}))$, we convolve the commonly-used galaxy stellar mass function from \citet{Panter_2004} with the mass-metallicity relation from \citet{Ma_2016}, and multiply the result with the SFRD in Equation~\ref{eq:SFR} \citep[cf.][]{Neijssel_2019, Broekgaarden_2022}. 
We use \texttt{astropy} to transform $t_m$ to a redshift using the WMAP9 cosmology\footnote{We find that the rates are not significantly impacted by the choice of cosmology \citep[cf.][]{Neijssel_2019}.}. 
For more details on the merger rate and its calculation, see the methodology in \citet{COMPAS_2022}.

\subsection{SFRD-Based Toy Model Merger Rates}\label{sec:method-toy-model-merger-rates}
To investigate whether popular, simplistic models of the merger rate are representative of population synthesis results, we compare our simulations to three models from the literature. Throughout this study, we will refer to these as ``toy models''. The first, and simplest, toy model is a scaled \ac{SFRD}. The second is a scaled SFRD convolved with a constant delay time of $t_{\rm{delay}} = 20\ \rm{Myr}$. The third is a SFRD convolved with a $\diff n / \diff t \propto 1/t$ delay time distribution that has a minimum delay time of $\sim 20\,\rm{Myr}$ \citep[similar to studies including][]{Regimbau:2012, Borhanian:2022arXiv220211048B, Colombo:2022, Iacovelli:2022, Lehoucq:2023}. 

For all toy models, we use the SFRD described in Equation~\ref{eq:SFR} to match the SFRD assumed for calculating the merger rate. 
Throughout this paper, we compare toy models to our population synthesis simulated merger rates (Equation~\ref{eq:merger_rate}) by normalizing both such that their areas integrate to unity, and then dividing the simulated $\mathcal{R}_{\rm{merge}}(z)$ by the toy model.
We first discuss results for the scaled \ac{SFRD} 
toy model, and then we describe results for the two delayed \ac{SFRD} toy models in Section~\ref{sec:Comparing-to-a-delayed-SFR} and beyond.

\section{Results: deviation from the SFRD} \label{sec:results}

\begin{figure*}
  \includegraphics[width=\linewidth]{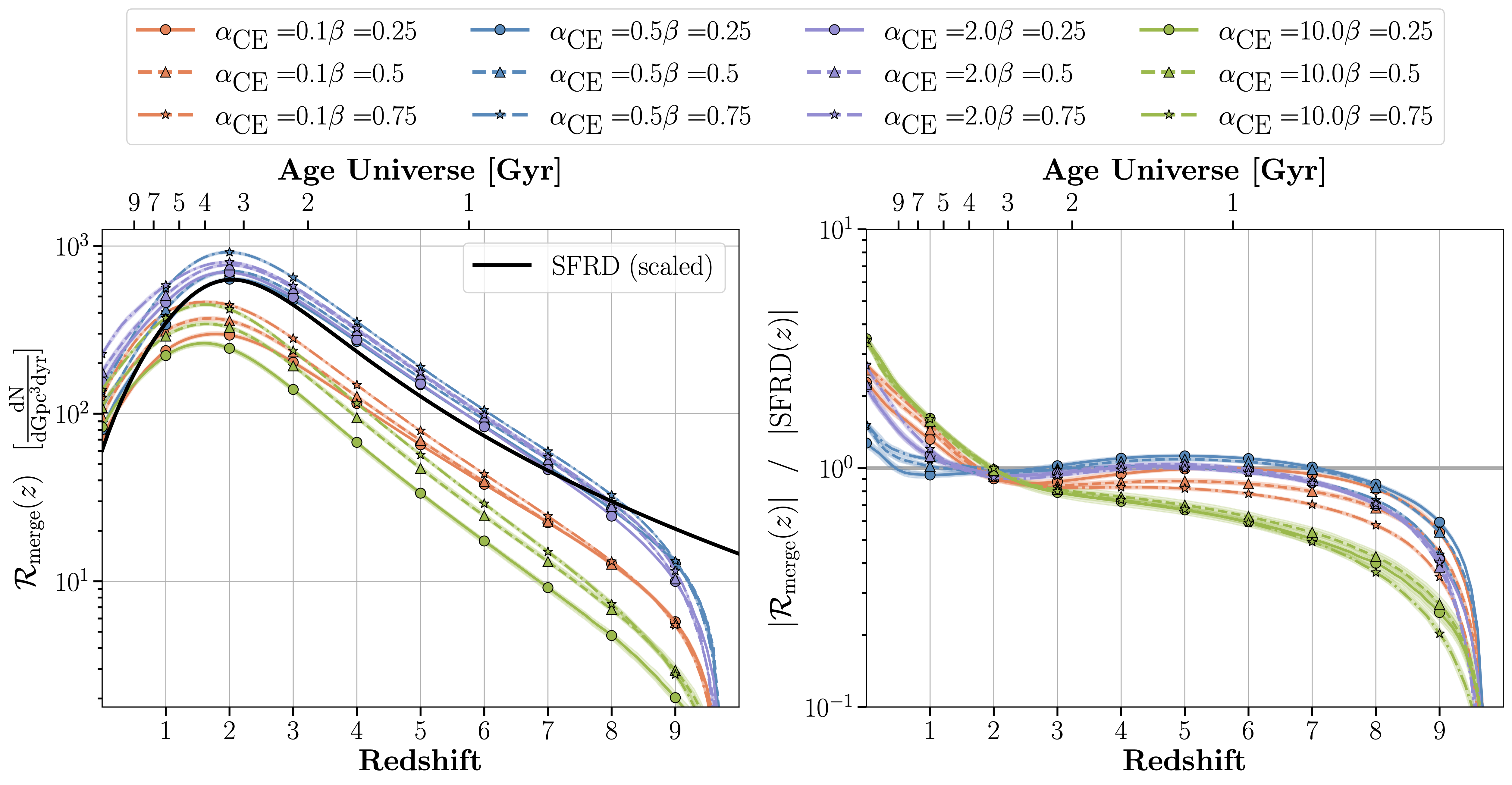}
  \caption{{\it Left:} The \ac{BBH} {\it merger} rate as a function of redshift for the $\alphaCE$ and $\beta$ model variations explored in this paper compared to the \ac{SFRD} scaled arbitrarily (black line). {\it Right:} 
  The BBH merger rate divided by the \ac{SFRD}, where we normalize the areas of both to 1. The horizontal grey line indicates a merger rate that follows the SFRD. 
  In both panels we include the $1\sigma$ and $2\sigma$ confidence intervals calculated by bootstrapping the simulation results to show the sampling uncertainty (note that uncertainties are typically the same size as the width of the lines).
  The sharp decline in the merger rate at $z\sim 9$ is due to star formation starting at $z=10$ in our calculations and the fact that mergers are delayed.}
  \label{fig:merger_rate_A}
\end{figure*}

In Figure~\ref{fig:merger_rate_A}, we show the BBH merger rates produced with our suite of models, and compare them to an arbitrarily scaled SFRD.
In the left panel, we find that our simulated models produce different merger rate distributions as a function of redshift, and also lead to different merger rate magnitudes. 
Models with $\alphaCE = 0.5, 2.0$ produce merger rates of factors $\sim 3-10$ higher than models with $\alphaCE=0.1, 10.0$. 
This is a result of $\alphaCE=0.1$ causing binaries to merge before stars are able to become \acp{BBH}, while $\alphaCE=10$ does not tighten the binaries enough to merge within a Hubble time\footnote{We elaborate on this ``sweet-spot'' of $\alphaCE$ in Section~\ref{sec:t_delay}.}. 

Most importantly for the scope of this study, we find that the redshift evolution of our simulated merger rates deviate from the scaled SFRD toy model.
These deviations from the SFRD are highlighted in the right panel of Figure~\ref{fig:merger_rate_A}, in which we normalize the merger rates and divide them by the normalized SFRD\footnote{We note that this normalization is chosen to investigate the \textit{relative} differences between our simulations and the SFRD toy model, but leads in practice to minimal differences around the merger rate peak $z\sim2$. 
See Section~\ref{sec:Comparing-to-a-delayed-SFR} for results that scale by the local merger rate instead.}.
We find that the normalized BBH merger rates tend to be higher (up to a factor of $3.5$) compared to the SFRD at low redshift $z \lesssim 1.5$, as is visible by all lines exceeding unity. 
The merger rates approach the scaled SFRD around redshift $2$, and then fall below the SFRD above redshifts $\gtrsim 2$, with the exact $z$ at which this occurs depending on the model. 
Some models only fall below the SFRD line at high redshifts $z \gtrsim 7$, such as those with $\alphaCE = 0.5$.
We find deviations of our simulated merger rates from the SFRD as high as factors $5\times$ (for the $\alphaCE = 10$ models around $z \sim 9$).
The first three columns in Table~\ref{tab:merger_rate_SFR_deviations} in the appendix provide statistics on the relative ratios between the (normalized) simulated BBH merger rates and SFRD from $z=0$ to $z=9$.

Despite model-to-model variations, the general trend among our simulated merger rates in Figure~\ref{fig:merger_rate_A} indicates shared underlying physical processes that cause $\mathcal{R}_{\rm{merge}}(z)$ to deviate from the \ac{SFRD}. 
As stated in Section~\ref{sec:intro}, the physical sources of merger rate-SFRD deviations is the effect of metallicity and delay times, which we explore comprehensively in the following two sections.

\subsection{The Effect of Metallicity on the Merger Rate}
\label{sec:shift-to-the-right}
\begin{figure*}
    \centering
    \includegraphics[width=\linewidth]{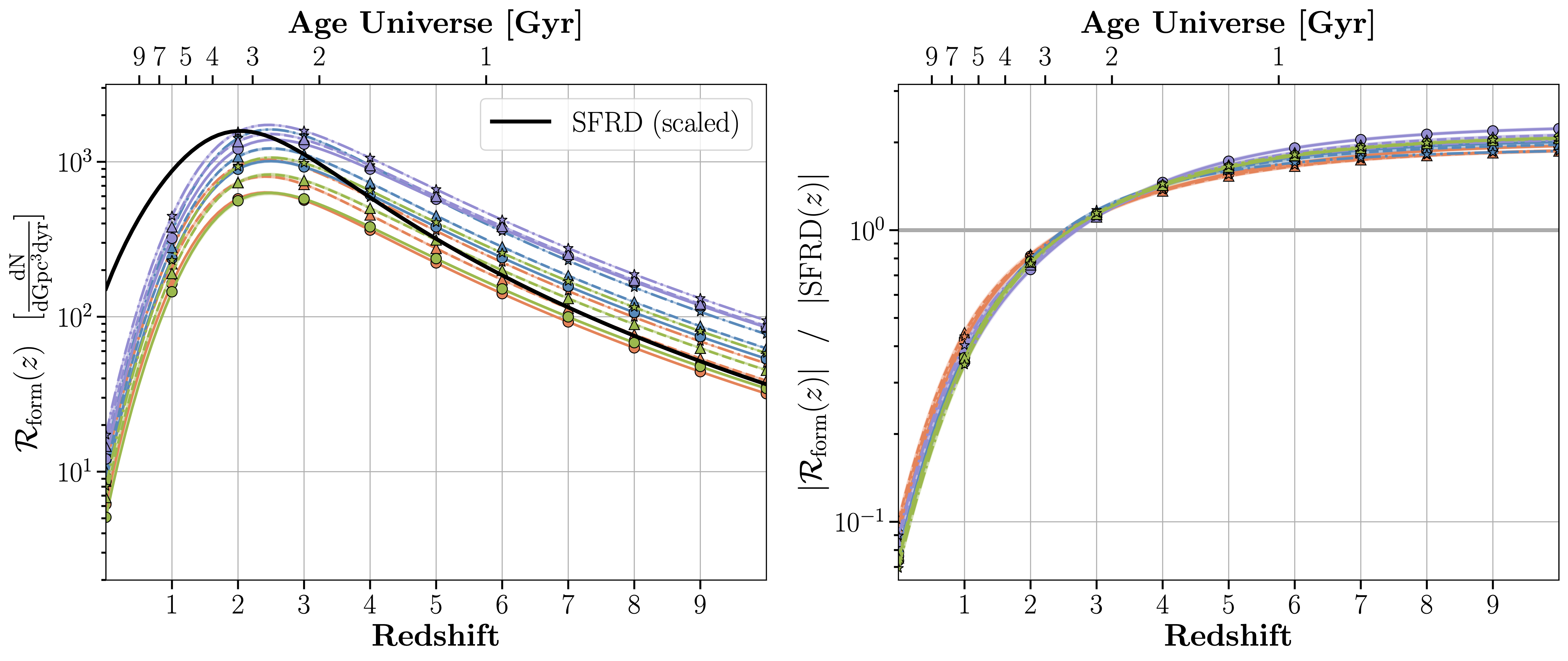}
    \caption{Same as Figure~\ref{fig:merger_rate_A}, but for the \ac{BBH} {\it formation} rate (formation is defined as the moment when the second BH forms). The peak of the BBH formation rate is clearly shifted to higher redshifts than the SFRD due to a higher BBH formation efficiency at lower metallicities. 
    }
    \label{fig:formation_rate_A}
\end{figure*}

First, we focus on the effect of the metallicity-dependent formation efficiency on $\mathcal{R}_{\rm{merge}}(z)$ for \acp{BBH}.  
In the left panel of Figure~\ref{fig:formation_rate_A}, we plot the \ac{BBH} \textit{formation} rate as a function of redshift, $\mathcal{R}_\textrm{form}(z)$ (Equation 1 in \citealt{Broekgaarden_2021}), for all our models
and in the right panel of Figure~\ref{fig:formation_rate_A}, we plot the ratio of \ac{BBH} formation rate relative to the scaled SFRD (the toy model). 
If the formation yield of \acp{BBH} were independent of metallicity, the formation rate of \acp{BBH} as a function of redshift would exactly follow the \ac{SFRD} since the simulations assume all other initial conditions (e.g., initial mass, initial separation) to be independent of initial metallicity, and the time required for massive stars to form \acp{BBH} is negligible on this scale. We find the following results.

\begin{figure*}
    \centering
    \includegraphics[width=1\linewidth]{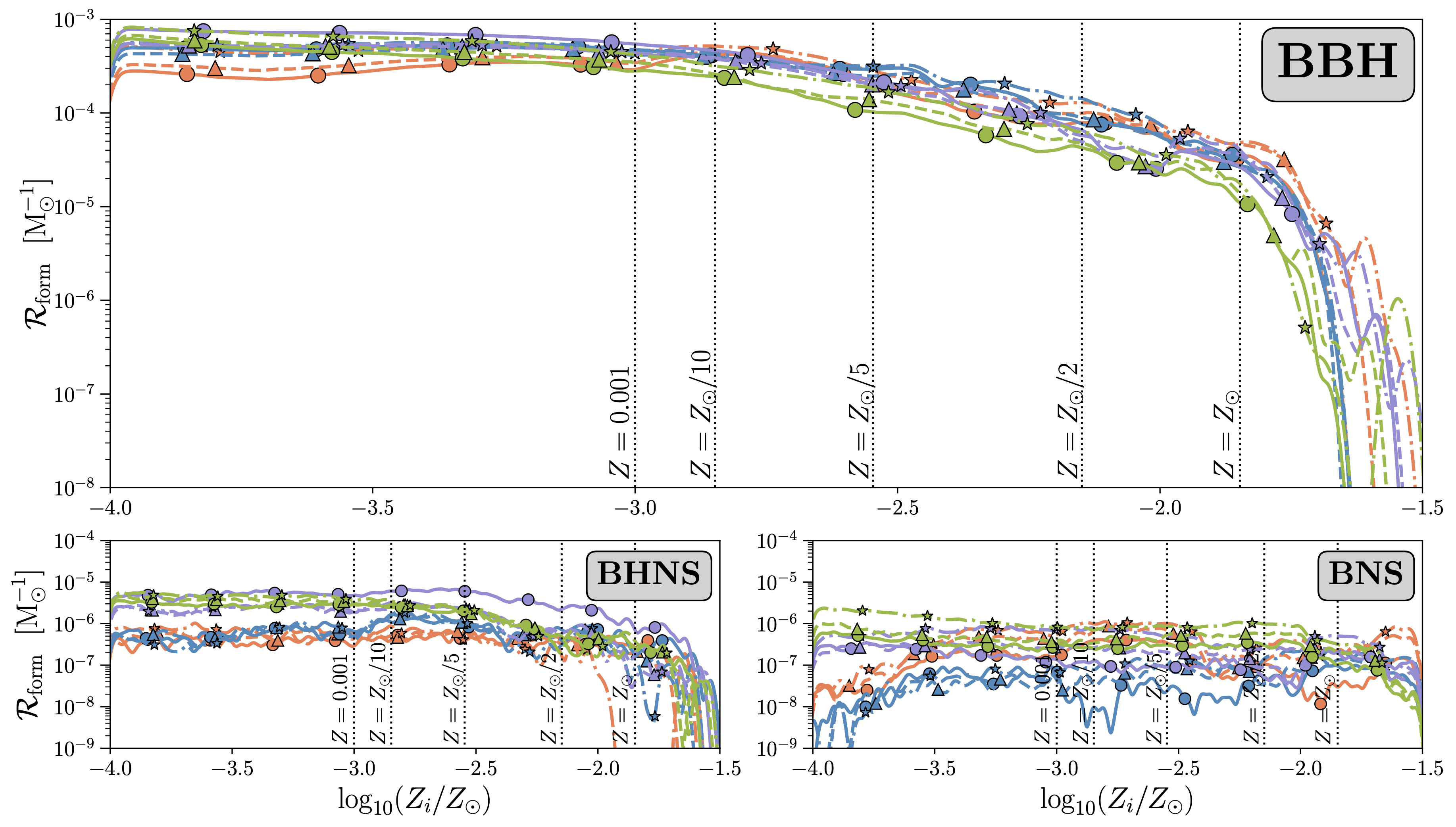}
    \caption{The formation yield of \acp{BBH}, \acp{BHNS}, and \acp{BNS} that will merge within a Hubble time, per solar mass of star formation, $\diff \mathcal{R}_\textrm{form} / \diff M_\textrm{SFRD}$, as a function of birth metallicity, $Z_i$, for all models.  We choose initial metallicities for each binary by sampling randomly from a log-uniform metallicity distribution from $Z_i = 0.0001$ to $Z_i = 0.03$. We then obtain $\mathcal{R}_\textrm{form}(Z_i)$ using a kernel density estimation over the initial metallicities of the simulated binaries. The decline in BBH formation efficiency at $Z_i \gtrsim Z_{\odot}/5$ is clearly visible.}
    \label{fig:formation_efficiency_per_metallicity_A}
\end{figure*}

First, we observe that, similarly to $\mathcal{R}_{\rm{merge}}(z)$ in Figure~\ref{fig:merger_rate_A}, models with $\alphaCE = 0.5, 2.0$ tend have higher $\mathcal{R}_{\rm{form}}(z)$ than the other two $\alphaCE$ values by a factor of up to $\sim 5$. 
Notably, there is very little variation in the redshift evolution of $\mathcal{R}_{\rm{form}}(z)$ between models, and they all deviate from the scaled SFRD.
Specifically, the peak of the \ac{BBH} formation rate from all models is at a higher redshift, $z\approx 2.5$, than the peak of the \ac{SFRD}, $z\approx 2$. 
As a result, we find that the BBH formation rate in the simulated $\mathcal{R}_{\rm{form}}(z)$ is suppressed at $z \lesssim z_{\rm{peak}}$ compared to the scaled SFRD with deviations of up to an order of magnitude at $z\sim 0$. The simulated BBH rate is relatively boosted for $z \gtrsim z_{\rm{peak}}$ by a factor of $\sim 2 \times$ compared to the SFRD.
All in all, our results find a metallicity-dependence in the formation rates of population synthesis simulations which favors high $z$ BBH formation, thus favoring the rate of \textit{merging} \acp{BBH} at higher redshifts compared to lower redshifts.

To further probe the metallicity dependence of \ac{BBH} formation, we show the simulated formation rates of \acp{BBH} as a function of initial metallicity, $Z_i$, in the top panel of Figure~\ref{fig:formation_efficiency_per_metallicity_A}. We find that as $Z_i$ increases, the \ac{BBH} formation yield falls off more and more rapidly, as much as an order of magnitude by $Z_\odot/2$, and three orders of magnitude by $2Z_\odot$. This rapid decline is consistent with findings for \ac{BBH} mergers formed from isolated binary evolution in recent literature (e.g., \citealt{Giacobbo_2018, Klencki_2018, Giacobbo_2018_2, Chruslinska_2018, Neijssel_2019, Santoliquido_2021, Broekgaarden_2022}, and \citealt{Dorozsmai:2022}).
Since the mean metallicity increases with cosmic time, higher \ac{BBH} formation efficiency at low progenitor metallicity results in a higher \ac{BBH} formation rate relative to the \ac{SFRD} at higher redshifts ($z\gtrsim 2.5$) (as is visible in Figure~\ref{fig:formation_rate_A}). 
The redshift at which we observe notable increases in formation rate is around where the proportion of formed stars has a considerable change in birth metallicity based on the Madau-Dickinson prescription (see Appendix \ref{sec:appendix-metallicity-dependent-star-formation-history} for details on the metallicity-dependent SFRD).
Low model-to-model variation in the relationship between metallicity and formation yield leads to the tight spread (within factor few) of BBH formation rates among models that we observe in Figure \ref{fig:formation_rate_A} \citep[cf.][]{Chruslinska_2018, Santoliquido_2021, Broekgaarden_2022}. %
Boosted BBH formation efficiency at low birth metallicity is predominantly caused by less wind loss through line-driven stellar winds at low metallicity, which leads to tighter binary systems (e.g., because more mass needs to be expelled during a CE event) and fewer systems that disrupt during supernova. We discuss this in more detail in Appendix~\ref{sec:impact_of_metallicity}. %

\subsection{The Effect of Delay Times on the Merger Rate}
\label{sec:t_delay}

The delay times ($t_{\rm{delay}}$) between the formation and merger of \acp{BBH}, can range from a few Myr to greater than the Hubble time \citep{Mennekens_2016, Belczynski_2016, Lipunov_2017, Stevenson_2017,Eldridge_2018} and lead the BBH merger rate to peak at a lower redshift than the formation rate. 
It is generally thought that the delay time distribution roughly follows $\diff n/ \diff t \propto t^{-1}$ with a minimum delay time, $t_{\rm{min}}$, typically bound between $10$ Myr -- $500$ Myr \citep[e.g.,][]{Regimbau:2012, Borhanian:2022arXiv220211048B, Colombo:2022, Iacovelli:2022, Naidu:2022}. 

\begin{figure*}
    \centering
    \includegraphics[width=\linewidth]{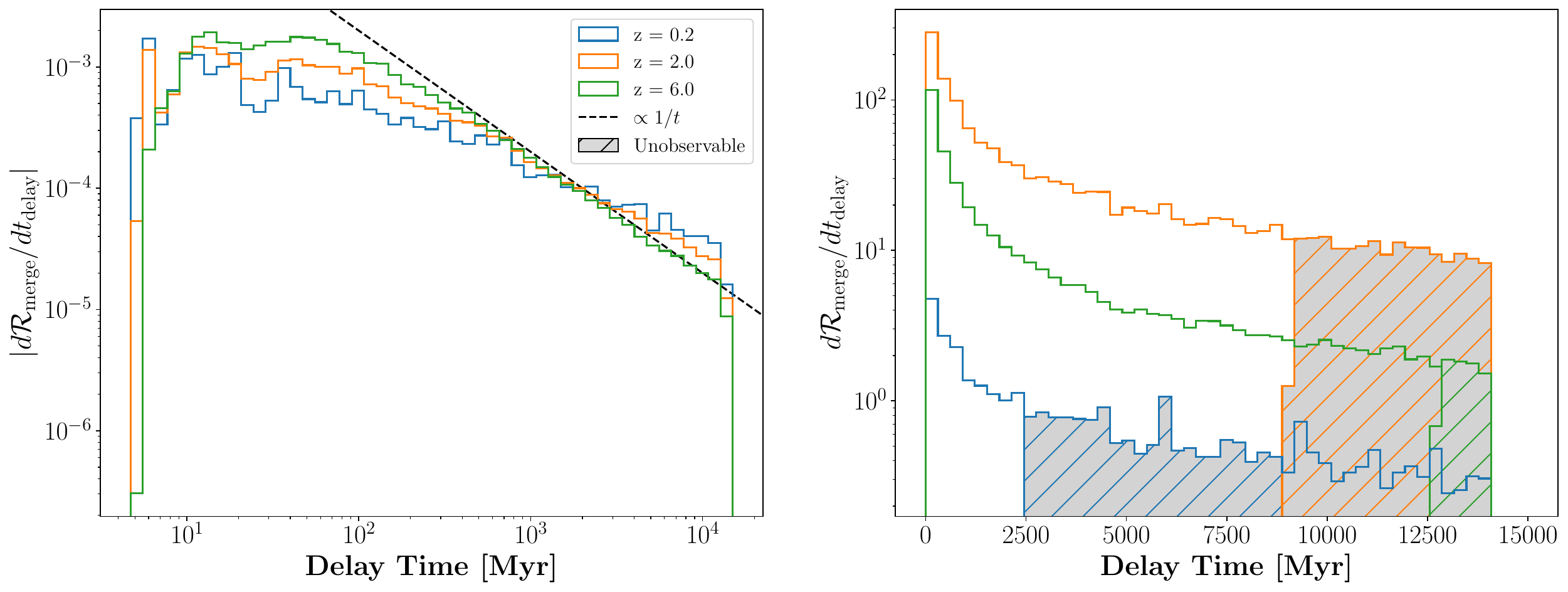}
    \caption{The BBH delay time distributions from our fiducial model for binaries formed at redshift $0.2$, $2$, and $6$. ({\it Left}) The distributions normalized to an area of 1. The bin widths are log-uniformly and uniformly spaced in the left and right panels, respectively. ({\it Right}) The gray hatched regions are the portion of each distribution that would not merge by $z=0$. The lowest $t_{\rm{delay}}$ bins (those with $t_{\rm{delay}} \lesssim 10 \ \rm{Myr}$) suffer from sampling uncertainty.}
\label{fig:delay_time_distribution_at_redshifts}
\end{figure*}

To understand the effect of delay times, we first investigate whether BBH delay times are dependent on formation redshift. 
In the left panel of Figure~\ref{fig:delay_time_distribution_at_redshifts}, we show the normalized distributions of delay times for binaries formed at $z=0.2,\hspace{0.05cm} 2,\hspace{0.05cm} 6$ for our fiducial model. 
We do indeed find redshift dependence, as the distributions for $z=0.2$ and $z=2$ have flatter slopes than $1/t$---specifically, the best fits are $t^{-0.60},\ t^{-0.77}$, and $t^{-0.95}$ for $z=0.2,\ 2.0$, and $6.0$, respectively, in the region $t_{\rm{delay}} > 100\,\rm{Myr}$.
We also find that the delay time distributions at all three redshifts flatten out for delay times $\lesssim 500\,\rm{Myr}$ (compared to $t_{\rm{delay}} > 500\,\rm{Myr}$).

In the right panel of Figure~\ref{fig:delay_time_distribution_at_redshifts} we show the distributions of delay times from our fiducial model, differentiating observable and unobservable mergers, where the latter are those formed at a given redshift with delay times longer than the time between its formation and Hubble time.
We find that the majority of binaries formed at $z=0.2$ merge after $z=0$. Furthermore, while there are around an order of magnitude more BBH mergers at $z=2.0$ than to $z=6.0$, the proportion of \textit{observable} mergers at $z=2$ is less than that from $z=6.0$ due to differences in the range of delay times that lead to mergers we can detect.
\begin{figure}
    \centering
    \includegraphics[width=1\columnwidth]{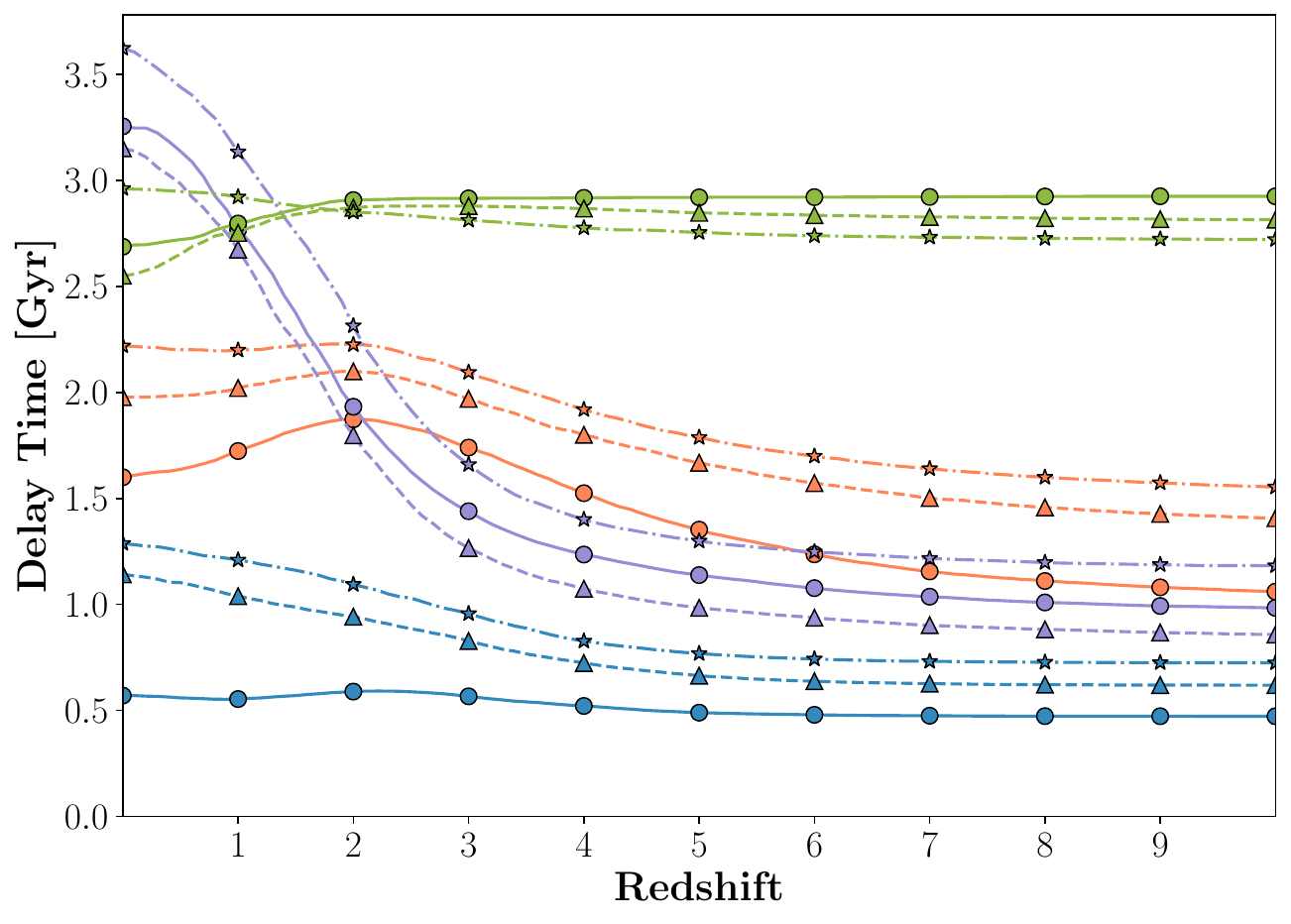}
    \caption{The median delay times of BBH mergers for each model in grid A as a function of redshift. The median only includes BBHs that merge in Hubble time for each redshift.}
    \label{fig:median_delay_time_cosmic_history}
\end{figure}

In Figure \ref{fig:median_delay_time_cosmic_history}, we show how the median delay time for BBHs merging in Hubble time evolves with redshift for all models. We find two main results.
First, the median delay time evolves differently as a function of redshift for different models.
Among our simulations there are four sets of models with similar median $t_\textrm{delay}$ behavior throughout redshift, which can be differentiated by $\alphaCE$ value. 
This suggests that the \ac{CE} mass transfer phase and the \ac{CE} efficiency parameter used in the simulations are important in determining a binary's delay time \citep[cf.][]{Chu:2022} and that the effect of $\alphaCE$ on the delay time distribution is redshift-dependent. 
We find that the median delay time in Figure  \ref{fig:median_delay_time_cosmic_history} does not increase monotonically with the $\alphaCE$ value. 
This is due to the relationship between $\alphaCE$ and the post-CE separation: smaller $\alphaCE$ values require the binary's orbit to shrink more for successful envelope ejection, which in turn leads to significantly shorter delay times \citep[][]{Peters_1964}. 
There is, however, a ``sweet-spot'' for the $\alphaCE$ value: values that are too low prevent the \ac{CE} from being successfully ejected, and the system instead undergoes a stellar merger, leading to the non-monotonic behavior (cf. \citealt{Kruckow:2018, Bavera_2021, Broekgaarden_2022}).
Second, the delay time medians are not constant throughout cosmic history, as can be seen by the decrease in the median BBH delay time as a function of redshift for nearly all models in Figure~\ref{fig:median_delay_time_cosmic_history}. 
This is because at higher redshifts, stars form with lower metallicities so experience weaker stellar winds and less radial expansion. 
These conditions typically lead binary systems to retain more mass, experience less orbital widening, and undergo the same evolutionary pathway in a tighter orbit.
Combined, this causes binaries to form \ac{BBH} systems with shorter orbits at lower metallicity, which reduces the typical simulated BBH delay time at higher redshifts.
The models with $\alphaCE = 2.0$ have the most distinct median delay time $z$-evolution; this unique redshift dependence is caused by a drastic increase in the proportion of mergers formed through the classic \ac{CE} channel at higher redshift that leads to shorter delay times compared to the ``only stable mass transfer channel'' for these simulations \citep[in agreement with, e.g., ][]{Olejak:2022supernova, van_Son_2022}. 
See Appendix~\ref{sec:appendix:BBH-formation-channels-as-a-function-of-redshift} for more details on the formation channel breakdown of our simulations throughout cosmic history.

\subsection{Comparison to Delayed SFRD Models and Models Scaled to the Local Merger Rate} \label{sec:Comparing-to-a-delayed-SFR}
\begin{figure*}
    \centering
    \includegraphics[width=\linewidth]{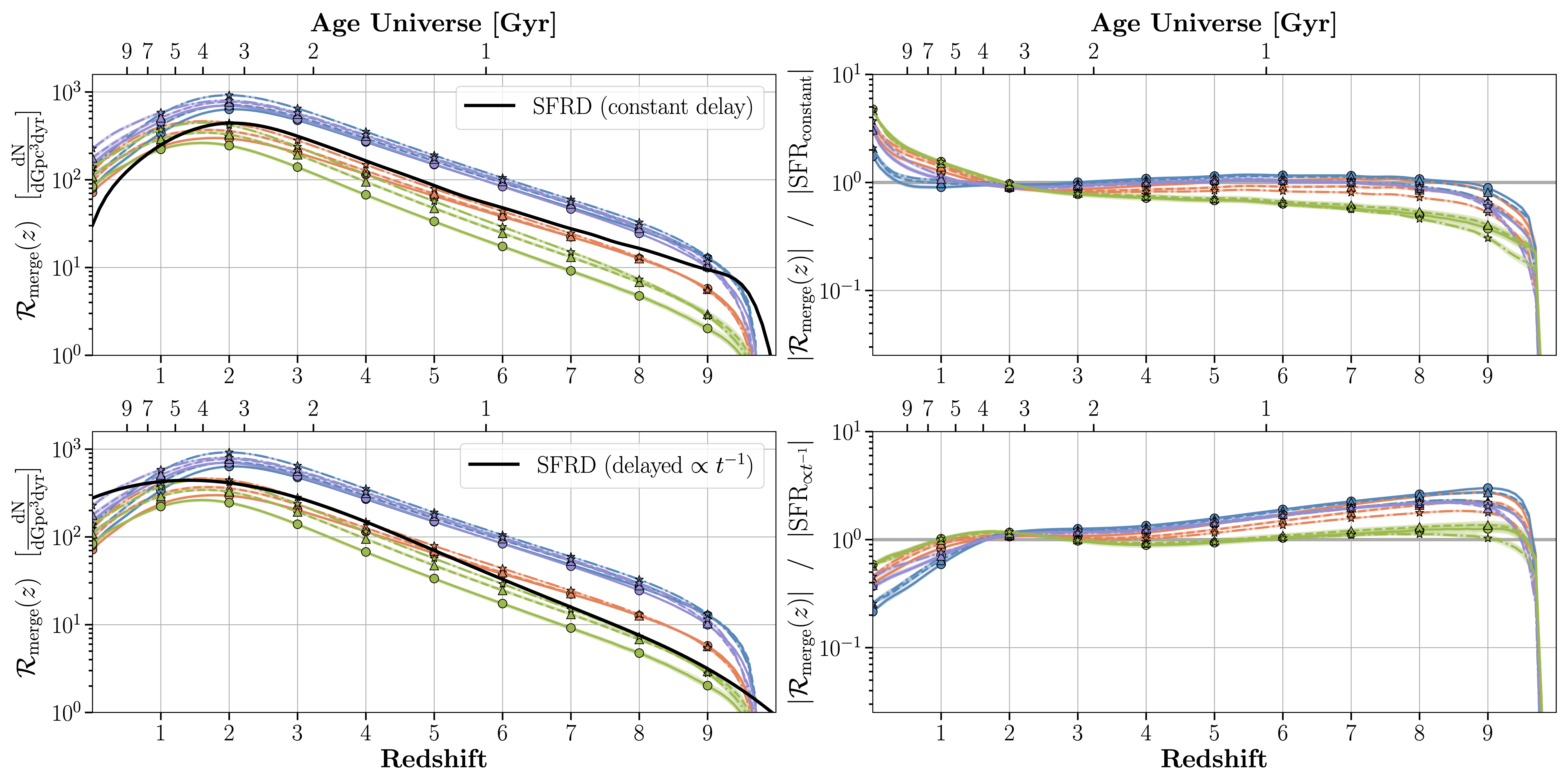}
    \caption{BBH merger rate as a function of redshift compared to toy models assuming a scaled \acp{SFRD} with a constant delay time of $20 \ \rm{ Myr}$ (top row) and  a delay time distribution of $\diff n / \diff t \propto 1/t$ with a minimum delay time of $20 \ \rm{ Myr}$ (bottom row). Labels and colors are the same as in Figure~\ref{fig:merger_rate_A}. }
    \label{fig:merger_rate_A_delayed}
\end{figure*}

So far, we have presented the deviations of population synthesis simulated BBH merger rates from a scaled SFRD, but we now consider whether our simulated merger rates deviate from a \textit{delayed} SFRD.
This is motivated by---in addition to support for and use of delays in models from the literature---our findings in Section~\ref{sec:t_delay} that the delay time distribution of BBH mergers can be complex and variable between models. 
We use two different delayed SFRD models: (i) 
a scaled SFRD convolved with a delay time distribution $\diff n / \diff t \propto \delta(t - c)$, where $c = 20\,\rm{Myr}$, and  
(ii) a scaled SFRD convolved with a delay time distribution  $\diff n / \diff t \propto 1/t$ with a minimum delay time of $20\,\rm{Myr}$. See Section~\ref{sec:method-toy-model-merger-rates} for more details and references on delayed toy models.

\begin{figure*}
    \centering
    \includegraphics[width=\linewidth]{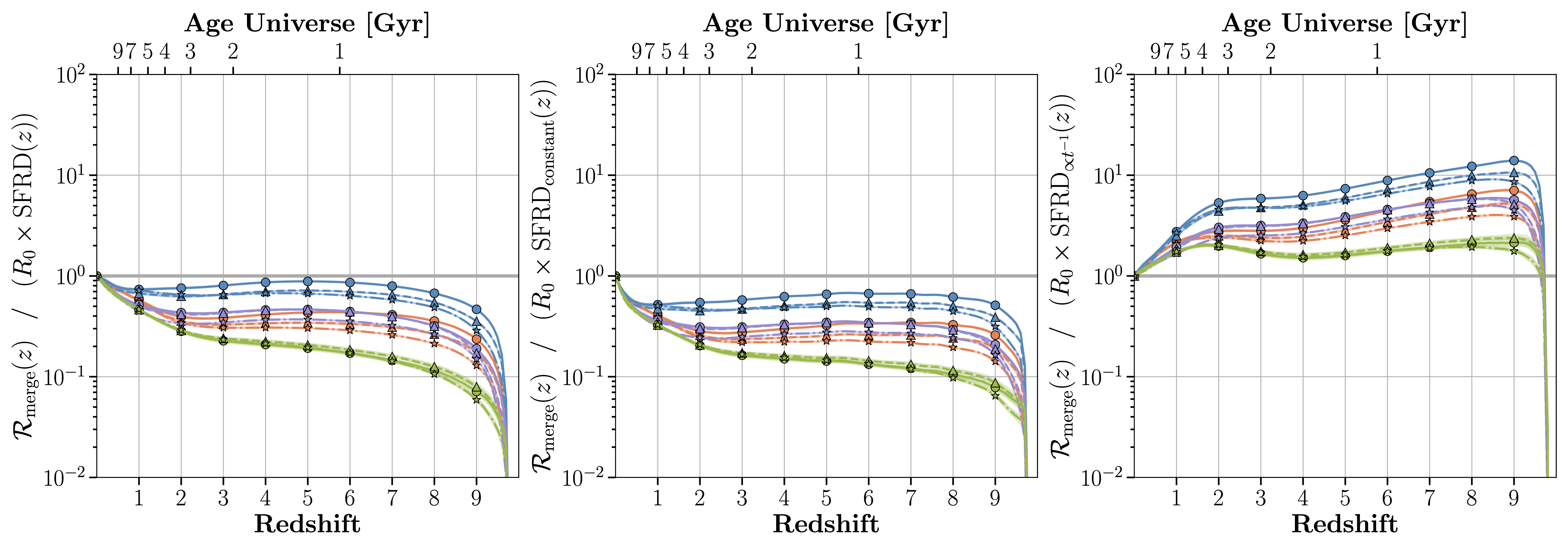}
    \caption{BBH merger rate as a function of redshift compared to toy model merger rates assuming a scaled SFRD (left), a scaled SFRD with constant delay times (middle), and a scaled  SFRD with a $\propto t^{-1}$ delay time distribution (right). All simulations are normalized to the same intrinsic merger rate $R_{\rm{merge}, 0}$. We assumed minimum delay times for the middle and right panel of $20\,\rm{Myr}$. Labels and colors are the same as in Figure ~\ref{fig:merger_rate_A}. }
    \label{fig:delayed_R0_norm_merger_rates}
\end{figure*}

The top panels in Figure~\ref{fig:merger_rate_A_delayed} show the deviations of our simulated $\mathcal{R}_{\rm{merge}}(z)$ from the SFRD model with constant $t_{\rm{delay}}$.
We find that at high redshifts ($z\gtrsim 5$), the population synthesis merger rates deviate by factors up to $\sim 3\times$ from the constant delay toy model---this is slightly less, $2\%$ versus $4\%$ at $z=6$ and $1\%$ versus $13\%$ at $z=7$, than deviations from the SFRD model with no delays (see Table~\ref{tab:merger_rate_SFR_deviations}).
On the other hand, at low redshifts, particularly for $z\sim 0$, deviations are up to a factor $5\times$, which is larger than deviations without delay times (which has deviations up to $3\times$ at $z\sim 0$). 

In the bottom panels of Figure~\ref{fig:merger_rate_A_delayed} we show the deviations of our simulated BBH merger rates from a SFRD with a $1/t$ delay time distribution. 
This toy model peaks at a lower redshift of $z \sim 1.5$ compared to $z \sim 2$ for the non-delayed or constant delayed SFRD merger rate models.
We find that our population synthesis model merger rates deviate with factors up to $5\times$ at low redshifts $z \lesssim 2$ from the $1/t$ delayed SFRD merger rate model and up to factors $3\times$ at high redshift $z\sim9$ (see Table~\ref{tab:merger_rate_SFR_deviations}).
Finally, it is important to note that there is a considerable difference between the effect of convolving the \ac{SFRD} with a $1/t$ versus constant delayed distribution; therefore, it is crucial that the high $t_{\rm{delay}}$ tail is considered by studies in the future.


Lastly, in Figure~\ref{fig:delayed_R0_norm_merger_rates} we show the deviations of our simulated merger rates from the toy models where we compare by scaling to the local merger rate $\mathcal{R}_0 \equiv
\mathcal{R}_{\rm{merge}}(0)$---a scaling often chosen for SFRD-based merger rate models in the literature \citep[e.g.,][]{Belgacem:2019, Borhanian:2022arXiv220211048B, Colombo:2022, Iacovelli:2022, Lehoucq:2023}.
Interestingly, we find that the population synthesis simulated merger rates deviate by factors up to $20 \times$ compared to each of the \ac{SFRD} models, with the most significant deviations occurring in the range of redshifts $z \gtrsim 3$. 
These deviations are \textit{significantly} larger compared to those experienced when we scaled the rates by area. Hence, using phenomenological models for the BBH merger rate that scale SFRDs by the local merger rate can lead to the largest deviations from simulated population synthesis merger rates, especially for $z\gtrsim3$.

\subsection{The BHNS and BNS Merger Rate}
\label{The-shape-of-the-BHNS-and-BNS-merger-rate}
\begin{figure*}
    \centering
    \includegraphics[width=\linewidth]{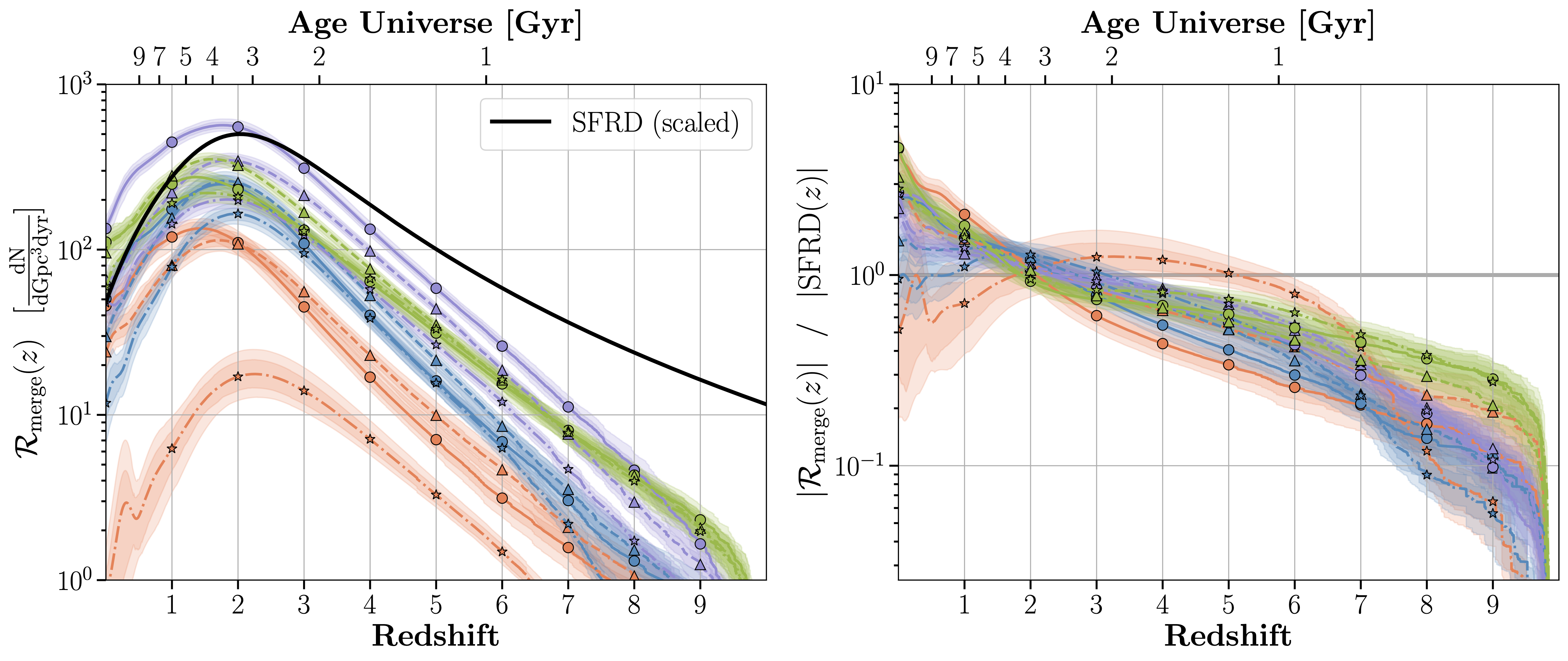}
    \caption{Same as Figure~\ref{fig:merger_rate_A} for \acp{BHNS} mergers.}
    \label{fig:merger_rate_A_BHNS}
\end{figure*}
\begin{figure*}
    \centering
    \includegraphics[width=\linewidth]{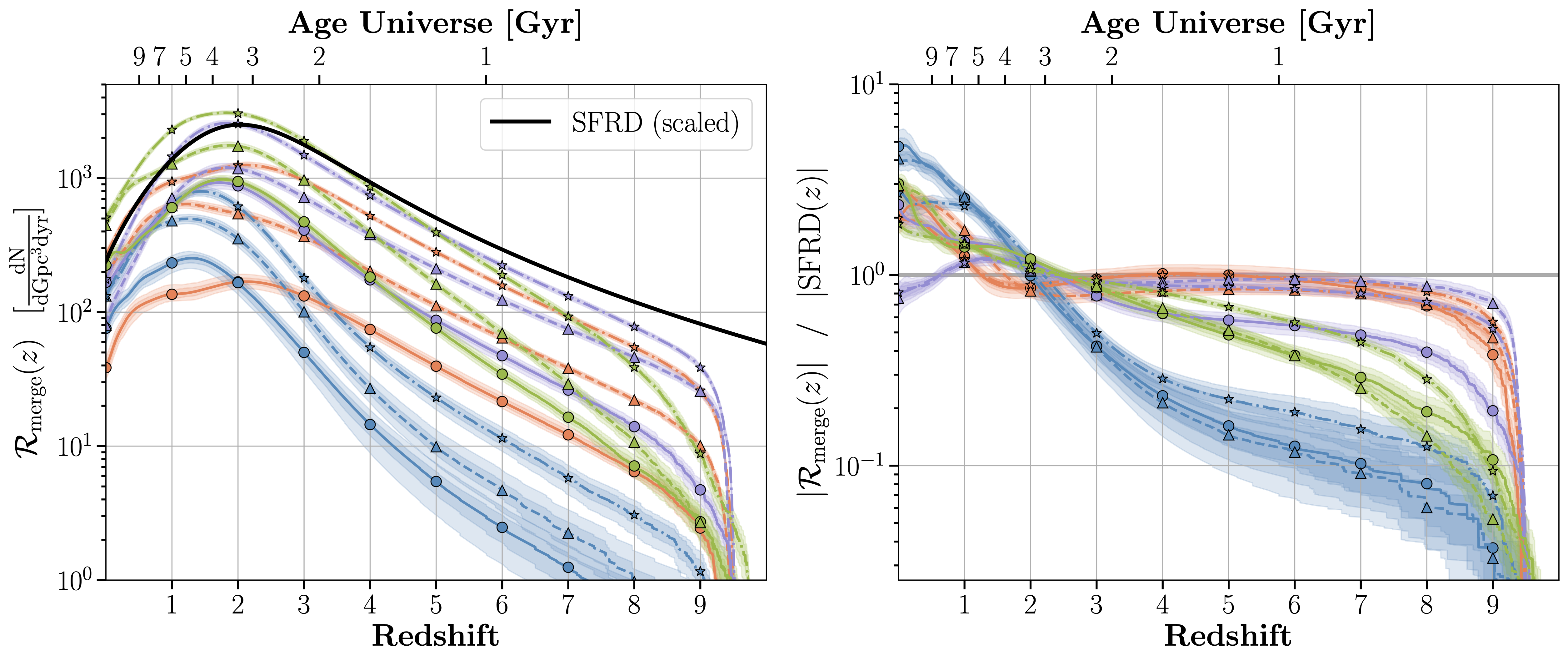}
    \caption{Same as Figure~\ref{fig:merger_rate_A} for \acp{BNS} mergers.}
    \label{fig:merger_rate_A_BNS}
\end{figure*}
%


In Figure~\ref{fig:merger_rate_A_BHNS} and Figure~\ref{fig:merger_rate_A_BNS}, we investigate the merger rate for \acp{BHNS} and \acp{BNS}, respectively. 
For both the \ac{BHNS} and \ac{BNS} merger rates, deviations from the \ac{SFRD} are generally similar to those of the \ac{BBH} merger rate---most of our simulations have a proportionally higher rate of mergers than stars formed at low redshift ($z \lesssim 2$), and a lower rate at high redshift ($z \gtrsim 4$), which are caused by our choice of scaling and the fact that the merger rates peak at lower redshift compared to the SFRD. 
There are, however, considerably more model-to-model variations for the \ac{BNS} and \ac{BHNS} merger rates than the \ac{BBH}. 
This stems from the fact that the formation efficiency-metallicity relationship has a markedly greater influence on BHNS and BNS mergers---differing by as much as two to three orders of magnitude---compared to BBH mergers, which vary within an order of magnitude, as can be found in Figure~\ref{fig:formation_efficiency_per_metallicity_A}. These findings are consistent with similar outcomes reported in the literature \citep{Klencki_2018, Chruslinska_2018, Giacobbo_2018, Neijssel_2019, RomanGarza:2020, Broekgaarden_2022}.
In our simulations, BHNS and BNS mergers are more commonly formed by binaries that experience a \ac{CE} phase than BBHs, leading BHNS and BNS simulations to be particularly sensitive to the value of $\alphaCE$. 

Interestingly, we find that many of the BHNS and BNS merger rates in Figure~\ref{fig:merger_rate_A_BHNS} and Figure~\ref{fig:merger_rate_A_BNS} demonstrate significantly different behavior from the BBH rates and SFRD-based toy models.
For example, most of the BHNS merger rates have a significantly steeper drop off after the merger rate peak in Figure~\ref{fig:merger_rate_A_BHNS}. In addition, some of the BNS models in Figure~\ref{fig:merger_rate_A_BNS} show evidence of an additional break in the merger rate slope before the peak, such as the model with $\alphaCE=0.1$ and $\beta = 0.25$ around $z=0.5$. This phenomenon is due to the complex convolution of the SFRD, formation efficiency of BHNSs/BNSs, and delay time distributions, and it lends weight to considering merger rates that deviate from a SFRD-like rate in future studies \citep[cf.][]{Callister:2023, Payne:2023}.

\section{Discussion} \label{sec:discussion}

\subsection{Other Formation Channels}\label{sec:other-formation-channels-discussion}

This study has investigated the redshift evolution of the merger rate of BBHs (and BNSs and BHNSs) formed by the isolated binary evolution channel.  
Formation pathways beyond the isolated binary evolution channel, however, can contribute and impact the  BBH merger rate as a function of redshift beyond the effects discussed in this paper \citep[e.g.,][]{Zevin:2021, Franciolini:2022, MandelBroekgaarden:2022, QiuCheng:2023}\footnote{The evidence for multiple formation channels from observations is still under debate. 
Recent studies \citep[e.g.][]{Fishbach:2022, Godfrey:2023} find evidence for substantial contributions from the isolated binary evolution channel of BBHs, but large uncertainties remain \citep[e.g.][]{Franciolini:2022, Tong:2022}.}. Studying their effects in detail is beyond the scope of this paper, but we discuss the dominant redshift behavior of these formation channels below. 

BBHs formed from population III stars (i.e., extremely metal-poor stars) are thought to merger at a significantly different rate than the isolated binary evolution channel: they begin merging at redshifts as early as $z\sim 20$ and peak around $z\sim 10$ (\citealt{Kinugawa:2020, Belczynski:2017, Hijikawa:2021, Liu:2021, Santoliquido:2023}).
The redshift evolution of population III mergers is still debated and depends strongly on assumed binary formation and interaction mechanisms that can bring black holes close together \citep[e.g.][]{Kinugawa:2014, Kinugawa:2020, Belczynski:2017, Liu:2021}

The merger rate for BBHs formed in dense star clusters is also still under debate and is heavily influenced by simulations assumptions and host cluster properties \citep[see][and references therein]{Choksi:2019}. Some studies find that the globular cluster BBH merger rate gradually increases at small redshifts until a peak around $z \sim 2$--$3$ and a sharp drop-off beyond $z \gtrsim 4$ (see \citealt{RodriguezLoeb:2018} and \citealt{Choksi:2019}).
\citet{Kritos:2022} and \citet{Mapelli:2022} found something similar, but with a longer high-redshift tail, as they assume higher cluster formation rates at high $z$.
Other studies instead find a monotonically increasing function of redshift starting from $z \sim 0$ out to their assumed starting epoch for single burst of globular cluster formation around $z \sim 3$--$4$, leading to a merger rate that peaks around $z \sim 3$--$4$ and then drops immediately to zero \citep{Rodriguez:2016, FragioneKocsis:2018}.
The sharp decrease in merger rate beyond $z\gtrsim 4$ seen in most simulations is because clusters have not had sufficient time to undergo core-collapse, a process during which \acp{BH} sink to the center of clusters and undergo the many dynamical interactions that dominate BBH merger production.
Finally, \citet{Ye_2024} recently found that the redshift distribution of the \ac{BBH} merger rate from globular clusters is correlated with primary mass, and that the peak of the merger rate is at higher $z$ for binaries with primary mass $> 30\ \rm{M}_\odot$, and even higher $z$ for primary mass $> 40\ \rm{M}_\odot$.

The merger rate of BBHs formed from chemically homogeneous evolution is still under debate and could either lead to a peak earlier or later than isolated binary evolution, mostly depending on the binary interaction and initial conditions assumed in simulation (for more details see \citealt[]{deMinkMandel:2016, MandeldeMink:2016, duBuisson:2020, Riley:2021}).

BBHs formed from primordial origins can merge as early as $z \sim 1000$ and are thought to follow a redshift rate that increases monotonically with increasing redshift as $\mathcal{R}_{\rm{merge}} \propto {t(z)}^{-34/37} $  \citep[e.g.,][]{Raidal:2019, DeLuca:2020, DeLuca:2021, Franciolini:2022}.

It will be important for future studies to further explore differences between the BBH merger rates  different formation pathways \citep[cf.][]{Martinez:2020, Zevin:2021, Bavera:2022, Mapelli:2022}.

\subsection{Observing the Merger Rate}
It is challenging to infer the merger rate as a function of redshift from observations. The \textit{local} BHNS, BNS and BBH merger rates have been constrained using GWs, pulsars binary neutron stars (and the lack of black hole-pulsar binaries), as well as short gamma-ray bursts (host galaxies), and r-process enrichment arguments---nevertheless, large uncertainties in these methods remain \citep[see][and references therein]{MandelBroekgaarden:2022}. 
Observations of the rate as a function of redshift remain scarce.
The authors of the latest GWTC-3 catalog population inference study found that the rate in the range $0 \lesssim z \lesssim 1.5$ increases proportionally to $(1+z)^{\kappa}$ with $\kappa = 2.9_{-1.8}^{+1.7}$ \citep{Ligo_Virgo_2021}.
This value of $\kappa$ is similar to the slope of the \ac{SFRD}, indicating that the \ac{BBH} redshift rate follows the \ac{SFRD} at low redshift (which was also confirmed by \citealt{edelman:2023, Nitz:2023-4-OGC}), though large uncertainties remain due to the limited sample of \ac{GW} observations and lack of models considered (e.g., the analysis from \citealt{Ligo_Virgo_2021} assumed that the BBH mass distribution does not change as a function of redshift). Other studies using different models to analyze the GW data found evidence that the slope of the BBH merger rate might deviate from that of the SFRD \citep{Callister:2023, Payne:2023}. 
We leave direct comparison to observations for future studies, and refer the reader to the discussion in \citet{Ray:2023} for more on this topic.

Besides \acp{BBH}, the merger rates of BHNS and BNS systems as a function of redshift will also be constrained by future observing runs and next-generation detectors. The merger rate of \acp{BNS} (and potentially \acp{BHNS}) can also be constrained from electromagnetic observations of gamma-ray bursts \citep[see e.g.,][and references therein]{Berger:2014, Fong:2015, Nugent:2022, Zevin_2022}---although, in practice, this will require a large sample of sGRB with associated host-galaxies as well as a better understanding of sGRB jet physics.

In coming years, new observing runs, including O4 and O5, are poised to increase the number of BBH merger detections to $\sim 500$ out to $z\sim 2$, which will help constrain the BBH merger rate as a function of redshift \citet{Aasi:2013prospects}\footnote{See \url{https://observing.docs.ligo.org/plan/index.html} for the most up-to-date information.}.
Moreover, next-generation \ac{GW} detectors, such as Cosmic Explorer and Einstein Telescope, are expected to detect stellar-mass black hole mergers out to (and beyond) redshifts $z\gtrsim 10$ and measure the merger rate with percent level precision \citep[e.g.,][]{EinsteinTelescope:2010Punturo, EinsteinTelescope:2012Sathyaprakash, CosmicExplorer:2019reitze, EinsteinTelescope:2020Maggiore, CosmicExplorer:2021evans, CosmicExplorer:2023,   Singh_2022, Gupta:2023}.


\subsection{Caveats}
The behavior of the redshift-dependent merger rate of BHNS, BNS, and BBH binaries is impacted by effects beyond those studied in the scope of the paper. We mention the most important ones here. 

First, as mentioned in Section~\ref{sec:other-formation-channels-discussion}, formation channels beyond the isolated formation channel can impact the merger rate by contributing additional binaries. Each of these channels comes with their own unique uncertainties, increasing the complexity of understanding the overall merger rate behavior \citep[e.g.,][and references therein]{Dominik:2013, Zevin:2021, Franciolini:2022, MandelBroekgaarden:2022, ArcaSedda:2023agnReview}.  

Second, not only the merger rate, but also the properties (i.e., masses, black hole spins, mass ratio) of the compact objects are redshift-dependent as a result of both their birth environments depending on redshift (e.g., metallicity), as well as the varying range of delay times can merge by a given $z$ \citep[e.g.,][]{Dominik:2013, Kinugawa:2020, Mapelli:2022, McKernan:2022, van_Son_2022}. Future work should further investigate the correlation between the merger rate and properties of \acp{NS}/\acp{BH} in source binaries as a function of redshift to further aid in understanding the underlying physics and formation channels of mergers \citep[e.g.][]{Qin:2018, Fishbach:2021, GWTC-3_population_inference, Belczynski:2022massredshift, Bavera:2022spinredshift, Biscoveanu:2022, van_Son_2022, Callister:2023}.

Third, the underlying metallicity-dependent star formation history is uncertain, especially at higher redshift $z \gtrsim 4$ \citep[see][and references therein]{Chruslinska:2019obsSFRD, chruslinska_2022}. Although we expect the qualitative results in this study to be the same for different $\mathcal{S}(Z_i,z)$ models, the quantitative results (i.e. with what factor the merger rate deviates from an SFRD's redshift evolution) will likely depend on the choice of the metallicity-dependent star formation rate assumed. This is because $\mathcal{S}(Z_i,z)$ describes the number of systems formed at a given metallicity as a function of redshift, and will therefore govern when and how much the drop in BBH formation efficiency (Figure~\ref{fig:formation_efficiency_per_metallicity_A}) will impact the merger rate distribution.

Fourth, there are many uncertainties in the physics underlying single and binary stellar evolution that can impact our study beyond the parameters considered here. A complete study is beyond the scope of this paper (and any single study), so we refer the interested reader to studies including \citet{Mapelli:2021review, Belczynski:2022, Santoliquido_2021,  Broekgaarden_2022, MandelBroekgaarden:2022, Spera:2022} and references therein. The most important effects that can impact this study will be uncertainties relating to the formation efficiency behavior (Figure~\ref{fig:formation_efficiency_per_metallicity_A}) and/or that drastically alter the delay time distributions of BBH, BHNS, and BNS mergers. In particular, models with weak stellar wind loss have found a more gradual decrease in the BBH merger efficiency at high metallicity \citep[e.g.][]{Broekgaarden_2022}. For delay time distributions, the angular momentum transport and mass transfer physics (and effect from radius expansion, e.g. \citet{Laplace:2020, Romagnolo:2023}) are important uncertainties that impact the separation at which binaries form and thus the merger time. These should be the focus of future work \citep[e.g.][]{Dorozsmai:2022, Agrawal:2023}.

\section{Conclusions} \label{sec:conclusion}

In this paper we studied how the BBH merger rate expected from isolated binary evolution deviates from the cosmic star formation rate density, focusing on two key effects: metallicity-dependent formation rates and delay times. 
To achieve this, we conducted a grid of $4 \times 3$ simulations using the population synthesis code COMPAS, varying the \ac{CE} efficiency and mass transfer efficiency parameters. 
We compared our simulated BBH rates to different \ac{SFRD}-based models from the literature, and performed similar analysis with BHNS and BNS mergers from our simulations.
Below, we summarize our main findings.

\begin{enumerate}

    \item Simulated BBH merger rates can deviate significantly (factors up to $3.5{-}5 \times$) from a merger rate model described by a scaled SFRD (Figure~\ref{fig:merger_rate_A} and  Table~\ref{tab:merger_rate_SFR_deviations}). 
    
    \item These deviations are caused by simulations experiencing (i) more efficient formation of BBHs at low metallicity, leading to a shift of the BBH formation rate peak to $z\sim 2.5$ as opposed to $z \sim 2$ for the SFRD  (Figure~\ref{fig:formation_rate_A} and Figure~\ref{fig:formation_efficiency_per_metallicity_A}) and (ii) a broad distribution of delay times that create time gaps from formation to merger, boosting the merger rate at lower redshifts (Figure~\ref{fig:delay_time_distribution_at_redshifts}). 

    \item Many of our simulations display a redshift-dependence in the BBH delay time distribution, favoring shorter delay times at higher redshifts (Figure~\ref{fig:median_delay_time_cosmic_history}). 
    This is caused by a complex interplay between metallicity and formation channels causing more orbital shrinking for binaries formed at low metallicity and high redshift. 
    We find that the common envelope efficiency parameter has a strong impact on this redshift evolution.

    \item Our simulated BBH merger rates deviate from SFRD models convolved with delay times (Figure~\ref{fig:merger_rate_A_delayed} and Table~\ref{tab:merger_rate_SFR_deviations}). 
    This is because \textit{delayed} SFRD models for the BBH rate fail to include the metallicity dependence of BBH formation found in our simulations which causes a decrease in BBH formation at low $z$. 

    \item We find even larger deviations (up to factors $\sim 10 \times$) when comparing the simulated BBH merger rates to SFRD models scaled to the local merger rate, with the largest deviations at higher redshifts $z \gtrsim 3$ (Figure~\ref{fig:delayed_R0_norm_merger_rates}). This means studies that use SFRD-based that are matched to the local GW merger rate might under- or overestimate the underlying BBH merger rate by an order of magnitude at these higher redshifts. The high-$z$ regime will be of particular relevance in coming years when the frontier of GW observation is pushed to the early years of our Universe.

    \item Similarly to BBHs, we find that our simulated BHNS and BNS merger rates deviate from a scaled SFRD rate (Figures~\ref{fig:merger_rate_A_BHNS}, ~\ref{fig:merger_rate_A_BNS}). 
    We do find, however, that the simulated BHNS and BNS merger rates are more impacted by our our model's parameter prescriptions than BBHs.

    \item Some of the simulated BHNS and BNS merger rates deviate from a simple SFRD-like redshift evolution parameterized by two slopes and a peak, and may actually include breaks before the peak (e.g. our $\alphaCE = 0.1$ models in Figure~\ref{fig:merger_rate_A_BNS}).
\end{enumerate}

Overall, we find that the simulated BBH, BHNS, and BNS merger rates of the isolated binary evolution channel can significantly deviate from a scaled cosmic star formation rate.  This motivates the use of non-SFRD-based merger rate models for future studies and exploration of the merger population.

\begin{acknowledgments}
APB acknowledges support from the Harvard PRISE and HCRP fellowships.
FSB acknowledges support for this work through the NASA FINESST scholarship 80NSSC22K1601 and from the Simons Foundation as part of the Simons Foundation Society of Fellows under award number 1141468. FSB also thanks the Steward Astronomy Department at the University of Arizona and the CCA at the Simons Foundation for providing a work place during this project. 
\end{acknowledgments}

%

\vspace{5mm}


\software{
This paper made use of simulations from \citetalias{Boesky_2023} which are publicly available at \href{https://gwlandscape.org.au/compas/}{GW Landscape}. To create this dataset, we used the COMPAS rapid binary population synthesis code version {2.31.04}, which is available for free at \url{http://github.com/TeamCOMPAS/COMPAS} \citep{COMPAS_2022}. 
The authors used {\sc{STROOPWAFEL}} from \citep{Broekgaarden_2019}, publicly available at \url{https://github.com/FloorBroekgaarden/STROOPWAFEL}.
The authors' primary programming language was \textsc{Python} from the Python Software Foundation available at \url{http://www.python.org} \citep{CS-R9526}. In addition, the following Python packages were used: \textsc{Matplotlib} \citep{2007CSE.....9...90H},  \textsc{NumPy} \citep{2020NumPy-Array}, \textsc{SciPy} \citep{2020SciPy-NMeth}, \textsc{IPython$/$Jupyter} \citep{2007CSE.....9c..21P, kluyver2016jupyter}, 
\textsc{Astropy} \citep{2018AJ....156..123A}  and   \href{https://docs.h5py.org/en/stable/}{\textsc{hdf5}} \citep{collette_python_hdf5_2014}. 
}




\bibliography{sample631}{}

\providecommand{\noopsort}[1]{}
\begin{thebibliography}{}
\expandafter\ifx\csname natexlab\endcsname\relax\def\natexlab#1{#1}\fi
\providecommand{\url}[1]{\href{#1}{#1}}
\providecommand{\dodoi}[1]{doi:~\href{http://doi.org/#1}{\nolinkurl{#1}}}
\providecommand{\doeprint}[1]{\href{http://ascl.net/#1}{\nolinkurl{http://ascl.net/#1}}}
\providecommand{\doarXiv}[1]{\href{https://arxiv.org/abs/#1}{\nolinkurl{https://arxiv.org/abs/#1}}}

\bibitem[{Abbott {et~al.}(2018)Abbott, Abbott, Abbott,
  {et~al.}}]{Aasi:2013prospects}
Abbott, B., Abbott, R., Abbott, T., {et~al.} 2018, Living Rev Relativ, 21, 3,
  \dodoi{10.1007/s41114-018-0012-9}

\bibitem[{{Abbott} {et~al.}(2019)}]{Abbott:2021GWTC1}
{Abbott}, B.~P., {et~al.} 2019, Physical Review X, 9, 031040,
  \dodoi{10.1103/PhysRevX.9.031040}

\bibitem[{{Abbott} {et~al.}(2021){Abbott}, {Abbott}, {Abraham}, {Acernese},
  {et~al.}}]{GWTC2}
{Abbott}, R., {Abbott}, T.~D., {Abraham}, S., {Acernese}, F., {et~al.} 2021,
  Physical Review X, 11, 021053, \dodoi{10.1103/PhysRevX.11.021053}

\bibitem[{{Abbott} {et~al.}(2024){Abbott}, {Abbott}, {Abraham}, {Acernese},
  {et~al.}}]{Abbott:2021-GWTC-2-1}
---. 2024, Phys. Rev. D, 109, 022001, \dodoi{10.1103/PhysRevD.109.022001}

\bibitem[{Abbott {et~al.}(2021)Abbott, Abbott, Abraham, Acernese, Ackley,
  Adams, Adams, Adhikari, Adya, Affeldt, Agathos, Agatsuma, Aggarwal, Aguiar,
  Aiello, Ain, Ajith, Allen, Allocca, Altin, Amato, Anand, Ananyeva, Anderson,
  Anderson, Angelova, Ansoldi, Antelis, Antier, Appert, Arai, Araya, Areeda,
  Arène, Arnaud, Aronson, Arun, Asali, Ascenzi, Ashton, Aston, Astone, Aubin,
  Aufmuth, AultONeal, Austin, Avendano, Babak, Badaracco, Bader, Bae, Baer,
  Bagnasco, Baird, Ball, Ballardin, Ballmer, Bals, Balsamo, Baltus, Banagiri,
  Bankar, Bankar, Barayoga, Barbieri, Barish, Barker, Barneo, Barnum, Barone,
  Barr, Barsotti, Barsuglia, Barta, Bartlett, Bartos, Bassiri, Basti, Bawaj,
  Bayley, Bazzan, Becher, Bécsy, Bedakihale, Bejger, Belahcene, Beniwal,
  Benjamin, Bennett, Bentley, Bergamin, Berger, Bergmann, Bernuzzi, Berry,
  Bersanetti, Bertolini, Betzwieser, Bhandare, Bhandari, Bhattacharjee, Bidler,
  Bilenko, Billingsley, Birney, Birnholtz, Biscans, Bischi, Biscoveanu, Bisht,
  Bitossi, Bizouard, Blackburn, Blackman, Blair, Blair, Blair, Blanch, Bobba,
  Bode, Boer, Boetzel, Bogaert, Boldrini, Bondu, Bonilla, Bonnand, Booker,
  Boom, Bork, Boschi, Bose, Bossilkov, Boudart, Bouffanais, Bozzi, Bradaschia,
  Brady, Bramley, Branchesi, Brau, Breschi, Briant, Briggs, Brighenti, Brillet,
  Brinkmann, Brockill, Brooks, Brooks, Brown, Brunett, Bruno, Bruntz, Buikema,
  Bulik, Bulten, Buonanno, Buscicchio, Buskulic, Byer, Cabero, Cadonati,
  Caesar, Cagnoli, Cahillane, Bustillo, Callaghan, Callister, Calloni, Camp,
  Canepa, Cannon, Cao, Cao, Carapella, Carbognani, Carney, Carpinelli, Carullo,
  Carver, Diaz, Casentini, Caudill, Cavaglià, Cavalier, Cavalieri, Cella,
  Cerdá-Durán, Cesarini, Chaibi, Chakravarti, Chan, Chan, Chandra, Chanial,
  Chao, Charlton, Chase, Chassande-Mottin, Chatterjee, Chattopadhyay,
  Chaturvedi, Chatziioannou, Chen, Chen, Chen, Chen, Cheng, Cheong, Chia,
  Chiadini, Chierici, Chincarini, Chiummo, Cho, Cho, Cho, Choate, Christensen,
  Chu, Chua, Chung, Chung, Ciani, Ciecielag, Cieślar, Cifaldi, Ciobanu,
  Ciolfi, Cipriano, Cirone, Clara, Clark, Clark, Clarke, Clearwater, Clesse,
  Cleva, Coccia, Cohadon, Cohen, Colleoni, Collette, Collins, Colpi,
  Constancio, Conti, Cooper, Corban, Corbitt, Cordero-Carrión, Corezzi,
  Corley, Cornish, Corre, Corsi, Cortese, Costa, Cotesta, Coughlin, Coughlin,
  Coulon, Countryman, Couvares, Covas, Coward, Cowart, Coyne, Coyne, Creighton,
  Creighton, Croquette, Crowder, Cudell, Cullen, Cumming, Cummings, Cunningham,
  Cuoco, Curylo, Canton, Dálya, Dana, DaneshgaranBajastani, D’Angelo,
  Danilishin, D’Antonio, Danzmann, Darsow-Fromm, Dasgupta, Datrier, Dattilo,
  Dave, Davier, Davies, Davis, Daw, Dean, DeBra, Deenadayalan, Degallaix,
  Laurentis, Deléglise, Favero, Lillo, Lillo, Pozzo, DeMarchi, Matteis,
  D’Emilio, Demos, Denker, Dent, Depasse, Pietri, Rosa, Rossi, DeSalvo,
  de~Varona, Dhurandhar, Díaz, Diaz-Ortiz, Didio, Dietrich, Fiore, DiFronzo,
  Giorgio, Giovanni, Giovanni, Girolamo, Lieto, Ding, Pace, Palma, Renzo,
  Divakarla, Dmitriev, Doctor, D’Onofrio, Donovan, Dooley, Doravari,
  Dorrington, Downes, Drago, Driggers, Du, Ducoin, Dupej, Durante, D’Urso,
  Duverne, Dwyer, Easter, Eddolls, Edelman, Edo, Edy, Effler, Eichholz,
  Eikenberry, Eisenmann, Eisenstein, Ejlli, Errico, Essick, Estellés, Estevez,
  Etienne, Etzel, Evans, Evans, Ewing, Fafone, Fair, Fairhurst, Fan, Farah,
  Farinon, Farr, Farr, Fauchon-Jones, Favata, Fays, Fazio, Feicht, Fejer, Feng,
  Fenyvesi, Ferguson, Fernandez-Galiana, Ferrante, Ferreira, Fidecaro, Figura,
  Fiori, Fiorucci, Fishbach, Fisher, Fishner, Fittipaldi, Fitz-Axen, Fiumara,
  Flaminio, Floden, Flynn, Fong, Font, Forsyth, Fournier, Frasca, Frasconi,
  Frei, Freise, Frey, Frey, Fritschel, Frolov, Fronzé, Fulda, Fyffe, Gabbard,
  Gadre, Gaebel, Gair, Gais, Galaudage, Gamba, Ganapathy, Ganguly, Gaonkar,
  Garaventa, García-Quirós, Garufi, Gateley, Gaudio, Gayathri, Gemme, Gennai,
  George, George, Gergely, Ghonge, Ghosh, Ghosh, Ghosh, Giacomazzo, Giacoppo,
  Giaime, Giardina, Gibson, Gier, Gill, Giri, Glanzer, Gleckl, Godwin, Goetz,
  Goetz, Gohlke, Goncharov, González, Gopakumar, Gossan, Gosselin, Gouaty,
  Grace, Grado, Granata, Granata, Grant, Gras, Grassia, Gray, Gray, Greco,
  Green, Green, Gretarsson, Griggs, Grignani, Grimaldi, Grimes, Grimm, Grote,
  Grunewald, Gruning, Guerrero, Guidi, Guimaraes, Guixé, Gulati, Guo, Gupta,
  Gupta, Gupta, Gustafson, Gustafson, Guzman, Haegel, Halim, Hall, Hamilton,
  Hammond, Haney, Hanke, Hanks, Hanna, Hannuksela, Hannuksela, Hansen, Hansen,
  Hanson, Harder, Hardwick, Haris, Harms, Harry, Harry, Hartwig, Hasskew,
  Haster, Haughian, Hayes, Healy, Heidmann, Heintze, Heinze, Heinzel, Heitmann,
  Hellman, Hello, Helmling-Cornell, Hemming, Hendry, Heng, Hennes, Hennig,
  Hennig, Vivanco, Heurs, Hild, Hill, Hines, Hochheim, Hofgard, Hofman,
  Hohmann, Holgado, Holland, Hollows, Holmes, Holt, Holz, Hopkins, Horst,
  Hough, Howell, Hoy, Hoyland, Huang, Hübner, Huddart, Huerta, Hughey, Hui,
  Husa, Huttner, Hutzler, Huxford, Huynh-Dinh, Idzkowski, Iess, Imperato,
  Inchauspe, Ingram, Intini, Isi, Iyer, JaberianHamedan, Jacqmin, Jadhav,
  Jadhav, James, Jani, Janssens, Janthalur, Jaranowski, Jariwala, Jaume,
  Jenkins, Jeunon, Jiang, Johns, Jones, Jones, Jones, Jones, Jones, Jonker, Ju,
  Junker, Kalaghatgi, Kalogera, Kamai, Kandhasamy, Kang, Kanner, Kapadia,
  Kapasi, Karathanasis, Karki, Kashyap, Kasprzack, Kastaun, Katsanevas,
  Katsavounidis, Katzman, Kawabe, Kéfélian, Keitel, Key, Khadka, Khalili,
  Khan, Khan, Khazanov, Khetan, Khursheed, Kijbunchoo, Kim, Kim, Kim, Kim, Kim,
  Kim, Kimball, King, Kinley-Hanlon, Kirchhoff, Kissel, Kleybolte, Klimenko,
  Knowles, Knyazev, Koch, Koehlenbeck, Koekoek, Koley, Kolstein, Komori,
  Kondrashov, Kontos, Koper, Korobko, Korth, Kovalam, Kozak, Krämer, Kringel,
  Krishnendu, Królak, Kuehn, Kumar, Kumar, Kumar, Kumar, Kuns, Kwang, Lackey,
  Laghi, Lalande, Lam, Lamberts, Landry, Lane, Lang, Lange, Lantz, Lanza, Rosa,
  Lartaux-Vollard, Lasky, Laxen, Lazzarini, Lazzaro, Leaci, Leavey, Lecoeuche,
  Lee, Lee, Lee, Lee, Lehmann, Leon, Leroy, Letendre, Levin, Li, Li, Li, Li,
  Li, Linde, Linker, Linley, Littenberg, Liu, Liu, Llorens-Monteagudo, Lo,
  Lockwood, London, Longo, Lorenzini, Loriette, Lormand, Losurdo, Lough,
  Lousto, Lovelace, Lück, Lumaca, Lundgren, Ma, Macas, MacInnis, Macleod,
  MacMillan, Macquet, Hernandez, Magaña-Sandoval, Magazzù, Magee, Majorana,
  Maksimovic, Maliakal, Malik, Man, Mandic, Mangano, Mansell, Manske,
  Mantovani, Mapelli, Marchesoni, Marion, Márka, Márka, Markakis, Markosyan,
  Markowitz, Maros, Marquina, Marsat, Martelli, Martin, Martin, Martinez,
  Martinez, Martynov, Masalehdan, Mason, Massera, Masserot, Massinger,
  Masso-Reid, Mastrogiovanni, Matas, Mateu-Lucena, Matichard, Matiushechkina,
  Mavalvala, Maynard, McCann, McCarthy, McClelland, McCormick, McCuller,
  McGuire, McIsaac, McIver, McManus, McRae, McWilliams, Meacher, Meadors,
  Mehmet, Mehta, Melatos, Melchor, Mendell, Menendez-Vazquez, Mercer, Mereni,
  Merfeld, Merilh, Merritt, Merzougui, Meshkov, Messenger, Messick, Metzdorff,
  Meyers, Meylahn, Mhaske, Miani, Miao, Michaloliakos, Michel, Middleton,
  Milano, Miller, Miller, Millhouse, Mills, Milotti, Milovich-Goff, Minazzoli,
  Minenkov, Mir, Mishkin, Mishra, Mistry, Mitra, Mitrofanov, Mitselmakher,
  Mittleman, Mo, Mogushi, Mohapatra, Mohite, Molina, Molina-Ruiz, Mondin,
  Montani, Moore, Moraru, Morawski, Moreno, Morisaki, Mours, Mow-Lowry, Mozzon,
  Muciaccia, Mukherjee, Mukherjee, Mukherjee, Mukherjee, Mukund, Mullavey,
  Munch, Muñiz, Murray, Nadji, Nagar, Nardecchia, Naticchioni, Nayak, Neil,
  Neilson, Nelemans, Nelson, Nery, Neunzert, Ng, Ng, Nguyen, Nguyen, Nguyen,
  Nichols, Nissanke, Nocera, Noh, North, Nothard, Nuttall, Oberling, O’Brien,
  O’Dell, Oganesyan, Ogin, Oh, Oh, Ohme, Ohta, Okada, Olivetto, Oppermann,
  Oram, O’Reilly, Ormiston, Ormsby, Ortega, O’Shaughnessy, Ossokine,
  Osthelder, Ottaway, Overmier, Owen, Pace, Pagano, Page, Pagliaroli, Pai, Pai,
  Palamos, Palashov, Palomba, Pan, Panda, Pang, Pankow, Pannarale, Pant,
  Paoletti, Paoli, Paolone, Parker, Pascucci, Pasqualetti, Passaquieti,
  Passuello, Patel, Patricelli, Payne, Pechsiri, Pedraza, Pegoraro, Pele, Penn,
  Perego, Perez, Périgois, Perreca, Perriès, Petermann, Petterson, Pfeiffer,
  Pham, Phukon, Piccinni, Pichot, Piendibene, Piergiovanni, Pierini, Pierro,
  Pillant, Pilo, Pinard, Pinto, Piotrzkowski, Pirello, Pitkin, Placidi,
  Plastino, Pluchar, Poggiani, Polini, Pong, Ponrathnam, Popolizio, Porter,
  Poverman, Powell, Pracchia, Prajapati, Prasai, Prasanna, Pratten, Prestegard,
  Principe, Prodi, Prokhorov, Prosposito, Puecher, Punturo, Puosi, Puppo,
  Pürrer, Qi, Quetschke, Quinonez, Quitzow-James, Raab, Raaijmakers, Radkins,
  Radulesco, Raffai, Rafferty, Rail, Raja, Rajan, Rajbhandari, Rakhmanov,
  Ramirez, Ramirez, Ramos-Buades, Rana, Rao, Rapagnani, Rapol, Ratto, Raymond,
  Razzano, Read, Regimbau, Rei, Reid, Reitze, Rettegno, Ricci, Richardson,
  Richardson, Richardson, Ricker, Riemenschneider, Riles, Rizzo, Robertson,
  Robinet, Rocchi, Rocha, Rodriguez, Rodriguez-Soto, Rolland, Rollins, Roma,
  Romanelli, Romano, Romel, Romero, Romero-Shaw, Romie, Ronchini, Rose, Rose,
  Rose, Rosell, Rosińska, Rosofsky, Ross, Rowan, Rowlinson, Roy, Roy, Ruggi,
  Ryan, Sachdev, Sadecki, Sakellariadou, Salafia, Salconi, Saleem, Samajdar,
  Sanchez, Sanchez, Sanchez, Sanchis-Gual, Sanders, Santiago, Santos,
  Saravanan, Sarin, Sassolas, Sathyaprakash, Sauter, Savage, Savant, Sawant,
  Sayah, Schaetzl, Schale, Scheel, Scheuer, Schindler-Tyka, Schmidt, Schnabel,
  Schofield, Schönbeck, Schreiber, Schulte, Schutz, Schwarm, Schwartz, Scott,
  Scott, Seglar-Arroyo, Seidel, Sellers, Sengupta, Sennett, Sentenac, Sequino,
  Sergeev, Setyawati, Shaffer, Shahriar, Sharifi, Sharma, Sharma, Shawhan,
  Shen, Shikauchi, Shink, Shoemaker, Shoemaker, Shukla, ShyamSundar,
  Sieniawska, Sigg, Singer, Singh, Singh, Singha, Singhal, Sintes, Sipala,
  Skliris, Slagmolen, Slaven-Blair, Smetana, Smith, Smith, Somala, Son, Soni,
  Sorazu, Sordini, Sorrentino, Sorrentino, Soulard, Souradeep, Sowell, Spencer,
  Spera, Srivastava, Srivastava, Staats, Stachie, Steer, Steinke, Steinlechner,
  Steinlechner, Steinmeyer, Stevenson, Stolle-McAllister, Stops, Stover,
  Strain, Stratta, Strunk, Sturani, Stuver, Südbeck, Sudhagar, Sudhir, Suh,
  Summerscales, Sun, Sun, Sunil, Sur, Suresh, Sutton, Swinkels, Szczepańczyk,
  Tacca, Tait, Talbot, Tanasijczuk, Tanner, Tao, Tapia, Martin, Tasson, Taylor,
  Tenorio, Terkowski, Thirugnanasambandam, Thomas, Thomas, Thomas, Thompson,
  Thondapu, Thorne, Thrane, Tiwari, Tiwari, Tiwari, Toland, Tolley, Tonelli,
  Tornasi, Torres-Forné, Torrie, e~Melo, Töyrä, Tran, Trapananti, Travasso,
  Traylor, Tringali, Tripathee, Trovato, Trudeau, Tsai, Tsang, Tse, Tso,
  Tsukada, Tsuna, Tsutsui, Turconi, Ubhi, Udall, Ueno, Ugolini, Unnikrishnan,
  Urban, Usman, Utina, Vahlbruch, Vajente, Vajpeyi, Valdes, Valentini, Valsan,
  van Bakel, van Beuzekom, van~den Brand, Broeck, Vander-Hyde, van~der Schaaf,
  van Heijningen, Vardaro, Vargas, Varma, Vass, Vasúth, Vecchio, Vedovato,
  Veitch, Veitch, Venkateswara, Venneberg, Venugopalan, Verkindt, Verma, Veske,
  Vetrano, Viceré, Viets, Villa-Ortega, Vinet, Vitale, Vo, Vocca, Vorvick,
  Vyatchanin, Wade, Wade, Wade, Walet, Walker, Wallace, Wallace, Walsh, Wang,
  Wang, Wang, Wang, Ward, Warner, Was, Washington, Watchi, Weaver, Wei,
  Weinert, Weinstein, Weiss, Wellmann, Wen, Weßels, Westhouse, Wette, Whelan,
  White, White, Whiting, Whittle, Wilken, Williams, Williams, Williamson,
  Willis, Willke, Wilson, Wimmer, Winkler, Wipf, Woan, Woehler, Wofford, Wong,
  Wrangel, Wright, Wu, Wysocki, Xiao, Yamamoto, Yang, Yang, Yang, Yap, Yeeles,
  Yoon, Yu, Yu, Yuen, Zadrożny, Zanolin, Zelenova, Zendri, Zevin, Zhang,
  Zhang, Zhang, Zhang, Zhao, Zhao, Zhou, Zhou, Zhu, Zimmerman, Zucker, Zweizig,
  Collaboration, \& the Virgo~Collaboration}]{GWTC-3_population_inference}
Abbott, R., Abbott, T.~D., Abraham, S., {et~al.} 2021, The Astrophysical
  Journal Letters, 913, L7, \dodoi{10.3847/2041-8213/abe949}

\bibitem[{{Abbott} {et~al.}(2023){Abbott}, {Abbott}, {Acernese}, {Ackley},
  {Adams}, {Adhikari}, {Zucker}, {Zweizig}, {et~al.}}]{Abbott:2021GWTC3}
{Abbott}, R., {Abbott}, T.~D., {Acernese}, F., {et~al.} 2023, Physical Review
  X, 13, 041039, \dodoi{10.1103/PhysRevX.13.041039}

\bibitem[{Abbott {et~al.}(2023)Abbott, Abbott, Acernese, Ackley, Adams,
  Adhikari, Adhikari, Adya, Affeldt, Agarwal, Agathos, Agatsuma, Aggarwal,
  Aguiar, Aiello, Ain, Ajith, Akutsu, de~Alarc\'on, Akcay, Albanesi, Allocca,
  Altin, Amato, Anand, Anand, Ananyeva, Anderson, Anderson, Ando, Andrade,
  Andres, Andri\ifmmode~\acute{c}\else \'{c}\fi{}, Angelova, Ansoldi, Antelis,
  Antier, Antonini, Appert, Arai, Arai, Arai, Araki, Araya, Araya, Areeda,
  Ar\`ene, Aritomi, Arnaud, Arogeti, Aronson, Arun, Asada, Asali, Ashton, Aso,
  Assiduo, Aston, Astone, Aubin, Austin, Babak, Badaracco, Bader, Badger, Bae,
  Bae, Baer, Bagnasco, Bai, Baiotti, Baird, Bajpai, Ball, Ballardin, Ballmer,
  Balsamo, Baltus, Banagiri, Bankar, Barayoga, Barbieri, Barish, Barker,
  Barneo, Barone, Barr, Barsotti, Barsuglia, Barta, Bartlett, Barton, Bartos,
  Bassiri, Basti, Bawaj, Bayley, Baylor, Bazzan, B\'ecsy, Bedakihale, Bejger,
  Belahcene, Benedetto, Beniwal, Bennett, Bentley, BenYaala, Bergamin, Berger,
  Bernuzzi, Berry, Bersanetti, Bertolini, Betzwieser, Beveridge, Bhandare,
  Bhardwaj, Bhattacharjee, Bhaumik, Bilenko, Billingsley, Bini, Birney,
  Birnholtz, Biscans, Bischi, Biscoveanu, Bisht, Biswas, Bitossi, Bizouard,
  Blackburn, Blair, Blair, Blair, Bobba, Bode, Boer, Bogaert, Boldrini,
  Bonavena, Bondu, Bonilla, Bonnand, Booker, Boom, Bork, Boschi, Bose, Bose,
  Bossilkov, Boudart, Bouffanais, Bozzi, Bradaschia, Brady, Bramley, Branch,
  Branchesi, Brandt, Brau, Breschi, Briant, Briggs, Brillet, Brinkmann,
  Brockill, Brooks, Brooks, Brown, Brunett, Bruno, Bruntz, Bryant, Bulik,
  Bulten, Buonanno, Buscicchio, Buskulic, Buy, Byer, Cadonati, Cagnoli,
  Cahillane, Bustillo, Callaghan, Callister, Calloni, Cameron, Camp, Canepa,
  Canevarolo, Cannavacciuolo, Cannon, Cao, Cao, Capocasa, Capote, Carapella,
  Carbognani, Carlin, Carney, Carpinelli, Carrillo, Carullo, Carver, Diaz,
  Casentini, Castaldi, Caudill, Cavagli\`a, Cavalier, Cavalieri, Ceasar, Cella,
  Cerd\'a-Dur\'an, Cesarini, Chaibi, Chakravarti, Subrahmanya, Champion, Chan,
  Chan, Chan, Chan, Chan, Chandra, Chanial, Chao, Chapman-Bird, Charlton,
  Chase, Chassande-Mottin, Chatterjee, Chatterjee, Chatterjee, Chaturvedi,
  Chaty, Chatziioannou, Chen, Chen, Chen, Chen, Chen, Chen, Chen, Chen, Cheng,
  Cheong, Cheung, Chia, Chiadini, Chiang, Chiarini, Chierici, Chincarini,
  Chiofalo, Chiummo, Cho, Cho, Choudhary, Choudhary, Christensen, Chu, Chu,
  Chu, Chua, Chung, Ciani, Ciecielag, Cie\ifmmode~\acute{s}\else \'{s}\fi{}lar,
  Cifaldi, Ciobanu, Ciolfi, Cipriano, Cirone, Clara, Clark, Clark, Clarke,
  Clearwater, Clesse, Cleva, Coccia, Codazzo, Cohadon, Cohen, Cohen, Colleoni,
  Collette, Colombo, Colpi, Compton, Constancio, Conti, Cooper, Corban,
  Corbitt, Cordero-Carri\'on, Corezzi, Corley, Cornish, Corre, Corsi, Cortese,
  Costa, Cotesta, Coughlin, Coulon, Countryman, Cousins, Couvares, Coward,
  Cowart, Coyne, Coyne, Creighton, Creighton, Criswell, Croquette, Crowder,
  Cudell, Cullen, Cumming, Cummings, Cunningham, Cuoco, Cury\l{}o, Dabadie,
  Canton, Dall'Osso, D\'alya, Dana, DaneshgaranBajastani, D'Angelo, Danila,
  Danilishin, D'Antonio, Danzmann, Darsow-Fromm, Dasgupta, Datrier, Datta,
  Dattilo, Dave, Davier, Davies, Davis, Davis, Daw, Dean, DeBra, Deenadayalan,
  Degallaix, De~Laurentis, Del\'eglise, Del~Favero, De~Lillo, De~Lillo,
  Del~Pozzo, DeMarchi, De~Matteis, D'Emilio, Demos, Dent, Depasse, De~Pietri,
  De~Rosa, De~Rossi, DeSalvo, De~Simone, Dhurandhar, D\'{\i}az, Diaz-Ortiz,
  Didio, Dietrich, Di~Fiore, Di~Fronzo, Di~Giorgio, Di~Giovanni, Di~Giovanni,
  Di~Girolamo, Di~Lieto, Ding, Di~Pace, Di~Palma, Di~Renzo, Divakarla,
  Dmitriev, Doctor, D'Onofrio, Donovan, Dooley, Doravari, Dorrington, Drago,
  Driggers, Drori, Ducoin, Dupej, Durante, D'Urso, Duverne, Dwyer, Eassa,
  Easter, Ebersold, Eckhardt, Eddolls, Edelman, Edo, Edy, Effler, Eguchi,
  Eichholz, Eikenberry, Eisenmann, Eisenstein, Ejlli, Engelby, Enomoto, Errico,
  Essick, Estell\'es, Estevez, Etienne, Etzel, Evans, Evans, Ewing, Fafone,
  Fair, Fairhurst, Farah, Farinon, Farr, Farr, Farrow, Fauchon-Jones, Favaro,
  Favata, Fays, Fazio, Feicht, Fejer, Fenyvesi, Ferguson, Fernandez-Galiana,
  Ferrante, Ferreira, Fidecaro, Figura, Fiori, Fishbach, Fisher, Fittipaldi,
  Fiumara, Flaminio, Floden, Fong, Font, Fornal, Forsyth, Franke, Frasca,
  Frasconi, Frederick, Freed, Frei, Freise, Frey, Fritschel, Frolov, Fronz\'e,
  Fujii, Fujikawa, Fukunaga, Fukushima, Fulda, Fyffe, Gabbard, Gadre, Gair,
  Gais, Galaudage, Gamba, Ganapathy, Ganguly, Gao, Gaonkar, Garaventa,
  Garc\'{\i}a, Garc\'{\i}a-N\'u\~nez, Garc\'{\i}a-Quir\'os, Garufi, Gateley,
  Gaudio, Gayathri, Ge, Gemme, Gennai, George, George, Gerberding, Gergely,
  Gewecke, Ghonge, Ghosh, Ghosh, Ghosh, Ghosh, Giacomazzo, Giacoppo, Giaime,
  Giardina, Gibson, Gier, Giesler, Giri, Gissi, Glanzer, Gleckl, Godwin,
  Golomb, Goetz, Goetz, Gohlke, Goncharov, Gonz\'alez, Gopakumar, Gosselin,
  Gouaty, Gould, Grace, Grado, Granata, Granata, Grant, Gras, Grassia, Gray,
  Gray, Greco, Green, Green, Gretarsson, Gretarsson, Griffith, Griffiths,
  Griggs, Grignani, Grimaldi, Grimm, Grote, Grunewald, Gruning, Guerra, Guidi,
  Guimaraes, Guix\'e, Gulati, Guo, Guo, Gupta, Gupta, Gupta, Gustafson,
  Gustafson, Guzman, Ha, Haegel, Hagiwara, Haino, Halim, Hall, Hamilton,
  Hammond, Han, Haney, Hanks, Hanna, Hannam, Hannuksela, Hansen, Hansen,
  Hanson, Harder, Hardwick, Haris, Harms, Harry, Harry, Hartwig, Hasegawa,
  Haskell, Hasskew, Haster, Hattori, Haughian, Hayakawa, Hayama, Hayes, Healy,
  Heidmann, Heidt, Heintze, Heinze, Heinzel, Heitmann, Hellman, Hello,
  Helmling-Cornell, Hemming, Hendry, Heng, Hennes, Hennig, Hennig, Hernandez,
  Vivanco, Heurs, Hild, Hill, Himemoto, Hines, Hiranuma, Hirata, Hirose,
  Hochheim, Hofman, Hohmann, Holcomb, Holland, Hollows, Holmes, Holt, Holz,
  Hong, Hopkins, Hough, Hourihane, Howell, Hoy, Hoyland, Hreibi, Hsieh, Hsu,
  Huang, Huang, Huang, Huang, Huang, Huang, H\"ubner, Huddart, Hughey, Hui,
  Hui, Husa, Huttner, Huxford, Huynh-Dinh, Ide, Idzkowski, Iess, Ikenoue, Imam,
  Inayoshi, Ingram, Inoue, Ioka, Isi, Isleif, Ito, Itoh, Iyer, Izumi,
  JaberianHamedan, Jacqmin, Jadhav, Jadhav, James, Jan, Jani, Janquart,
  Janssens, Janthalur, Jaranowski, Jariwala, Jaume, Jenkins, Jenner, Jeon,
  Jeunon, Jia, Jin, Johns, Jones, Jones, Jones, Jones, Jones, Jonker, Ju, Jung,
  Jung, Junker, Juste, Kaihotsu, Kajita, Kakizaki, Kalaghatgi, Kalogera, Kamai,
  Kamiizumi, Kanda, Kandhasamy, Kang, Kanner, Kao, Kapadia, Kapasi, Karat,
  Karathanasis, Karki, Kashyap, Kasprzack, Kastaun, Katsanevas, Katsavounidis,
  Katzman, Kaur, Kawabe, Kawaguchi, Kawai, Kawasaki, K\'ef\'elian, Keitel, Key,
  Khadka, Khalili, Khan, Khazanov, Khetan, Khursheed, Kijbunchoo, Kim, Kim,
  Kim, Kim, Kim, Kim, Kimball, Kimura, Kinley-Hanlon, Kirchhoff, Kissel, Kita,
  Kitazawa, Kleybolte, Klimenko, Knee, Knowles, Knyazev, Koch, Koekoek, Kojima,
  Kokeyama, Koley, Kolitsidou, Kolstein, Komori, Kondrashov, Kong, Kontos,
  Koper, Korobko, Kotake, Kovalam, Kozak, Kozakai, Kozu, Kringel, Krishnendu,
  Kr\'olak, Kuehn, Kuei, Kuijer, Kulkarni, Kumar, Kumar, Kumar, Kumar, Kume,
  Kuns, Kuo, Kuo, Kuromiya, Kuroyanagi, Kusayanagi, Kuwahara, Kwak, Lagabbe,
  Laghi, Lalande, Lam, Lamberts, Landry, Landry, Lane, Lang, Lange, Lantz,
  La~Rosa, Lartaux-Vollard, Lasky, Laxen, Lazzarini, Lazzaro, Leaci, Leavey,
  Lecoeuche, Lee, Lee, Lee, Lee, Lee, Lee, Lehmann, Lema\^{\i}tre, Leonardi,
  Leroy, Letendre, Levesque, Levin, Leviton, Leyde, Li, Li, Li, Li, Li, Li,
  Lin, Lin, Lin, Lin, Lin, Linde, Linker, Linley, Littenberg, Liu, Liu, Liu,
  Liu, Llamas, Llorens-Monteagudo, Lo, Lockwood, Loh, London, Longo, Lopez,
  Portilla, Lorenzini, Loriette, Lormand, Losurdo, Lott, Lough, Lousto,
  Lovelace, Lucaccioni, L\"uck, Lumaca, Lundgren, Luo, Lynam, Macas, MacInnis,
  Macleod, MacMillan, Macquet, Hernandez, Magazz\`u, Magee, Maggiore, Magnozzi,
  Mahesh, Majorana, Makarem, Maksimovic, Maliakal, Malik, Man, Mandic, Mangano,
  Mango, Mansell, Manske, Mantovani, Mapelli, Marchesoni, Marchio, Marion,
  Mark, M\'arka, M\'arka, Markakis, Markosyan, Markowitz, Maros, Marquina,
  Marsat, Martelli, Martin, Martin, Martinez, Martinez, Martinez, Martinovic,
  Martynov, Marx, Masalehdan, Mason, Massera, Masserot, Massinger, Masso-Reid,
  Mastrogiovanni, Matas, Mateu-Lucena, Matichard, Matiushechkina, Mavalvala,
  McCann, McCarthy, McClelland, McClincy, McCormick, McCuller, McGhee, McGuire,
  McIsaac, McIver, McRae, McWilliams, Meacher, Mehmet, Mehta, Meijer, Melatos,
  Melchor, Mendell, Menendez-Vazquez, Menoni, Mercer, Mereni, Merfeld, Merilh,
  Merritt, Merzougui, Meshkov, Messenger, Messick, Meyers, Meylahn, Mhaske,
  Miani, Miao, Michaloliakos, Michel, Michimura, Middleton, Milano, Miller,
  Miller, Miller, Miller, Millhouse, Mills, Milotti, Minazzoli, Minenkov, Mio,
  Mir, Miravet-Ten\'es, Mishra, Mishra, Mistry, Mitra, Mitrofanov,
  Mitselmakher, Mittleman, Miyakawa, Miyamoto, Miyazaki, Miyo, Miyoki, Mo,
  Modafferi, Moguel, Mogushi, Mohapatra, Mohite, Molina, Molina-Ruiz, Mondin,
  Montani, Moore, Moraru, Morawski, More, Moreno, Moreno, Mori, Morisaki,
  Moriwaki, Morr\'as, Mours, Mow-Lowry, Mozzon, Muciaccia, Mukherjee,
  Mukherjee, Mukherjee, Mukherjee, Mukherjee, Mukund, Mullavey, Munch, Mu\~niz,
  Murray, Musenich, Muusse, Nadji, Nagano, Nagano, Nagar, Nakamura, Nakano,
  Nakano, Nakashima, Nakayama, Napolano, Nardecchia, Narikawa, Naticchioni,
  Nayak, Nayak, Negishi, Neil, Neilson, Nelemans, Nelson, Nery, Neubauer,
  Neunzert, Ng, Ng, Nguyen, Nguyen, Nguyen, Quynh, Ni, Nichols, Nishizawa,
  Nissanke, Nitoglia, Nocera, Norman, North, Nozaki, Siles, Nuttall, Oberling,
  O'Brien, Obuchi, O'Dell, Oelker, Ogaki, Oganesyan, Oh, Oh, Oh, Ohashi,
  Ohishi, Ohkawa, Ohme, Ohta, Okada, Okutani, Okutomi, Olivetto, Oohara, Ooi,
  Oram, O'Reilly, Ormiston, Ormsby, Ortega, O'Shaughnessy, O'Shea, Oshino,
  Ossokine, Osthelder, Otabe, Ottaway, Overmier, Pace, Pagano, Page,
  Pagliaroli, Pai, Pai, Palamos, Palashov, Palomba, Pan, Pan, Panda, Pang,
  Pang, Pankow, Pannarale, Pant, Panther, Paoletti, Paoli, Paolone, Parisi,
  Park, Park, Parker, Pascucci, Pasqualetti, Passaquieti, Passuello, Patel,
  Pathak, Patricelli, Patron, Paul, Payne, Pedraza, Pegoraro, Pele, Arellano,
  Penn, Perego, Pereira, Pereira, Perez, P\'erigois, Perkins, Perreca,
  Perri\`es, Petermann, Petterson, Pfeiffer, Pham, Phukon, Piccinni, Pichot,
  Piendibene, Piergiovanni, Pierini, Pierro, Pillant, Pillas, Pilo, Pinard,
  Pinto, Pinto, Piotrzkowski, Piotrzkowski, Pirello, Pitkin, Placidi, Planas,
  Plastino, Pluchar, Poggiani, Polini, Pong, Ponrathnam, Popolizio, Porter,
  Poulton, Powell, Pracchia, Pradier, Prajapati, Prasai, Prasanna, Pratten,
  Principe, Prodi, Prokhorov, Prosposito, Prudenzi, Puecher, Punturo, Puosi,
  Puppo, P\"urrer, Qi, Quetschke, Quitzow-James, Raab, Raaijmakers, Radkins,
  Radulesco, Raffai, Rail, Raja, Rajan, Ramirez, Ramirez, Ramos-Buades, Rana,
  Rapagnani, Rapol, Ray, Raymond, Raza, Razzano, Read, Rees, Regimbau, Rei,
  Reid, Reid, Reitze, Relton, Renzini, Rettegno, Reza, Rezac, Ricci, Richards,
  Richardson, Richardson, Riemenschneider, Riles, Rinaldi, Rink, Rizzo,
  Robertson, Robie, Robinet, Rocchi, Rodriguez, Rolland, Rollins, Romanelli,
  Romano, Romel, Romero-Rodr\'{\i}guez, Romero-Shaw, Romie, Ronchini, Rosa,
  Rose, Rosi\ifmmode~\acute{n}\else \'{n}\fi{}ska, Ross, Rowan, Rowlinson, Roy,
  Roy, Roy, Rozza, Ruggi, Ryan, Sachdev, Sadecki, Sadiq, Sago, Saito, Saito,
  Sakai, Sakai, Sakellariadou, Sakuno, Salafia, Salconi, Saleem, Salemi,
  Samajdar, Sanchez, Sanchez, Sanchez, Sanchis-Gual, Sanders, Sanuy, Saravanan,
  Sarin, Sassolas, Satari, Sathyaprakash, Sato, Sato, Sauter, Savage, Sawada,
  Sawant, Sawant, Sayah, Schaetzl, Scheel, Scheuer, Schiworski, Schmidt,
  Schmidt, Schnabel, Schneewind, Schofield, Sch\"onbeck, Schulte, Schutz,
  Schwartz, Scott, Scott, Seglar-Arroyo, Sekiguchi, Sekiguchi, Sellers,
  Sengupta, Sentenac, Seo, Sequino, Sergeev, Setyawati, Shaffer, Shahriar,
  Shams, Shao, Sharma, Sharma, Shawhan, Shcheblanov, Shibagaki, Shikauchi,
  Shimizu, Shimoda, Shimode, Shinkai, Shishido, Shoda, Shoemaker, Shoemaker,
  ShyamSundar, Sieniawska, Sigg, Singer, Singh, Singh, Singha, Sintes, Sipala,
  Skliris, Slagmolen, Slaven-Blair, Smetana, Smith, Smith, Soldateschi, Somala,
  Somiya, Son, Soni, Soni, Sordini, Sorrentino, Sorrentino, Sotani, Soulard,
  Souradeep, Sowell, Spagnuolo, Spencer, Spera, Srinivasan, Srivastava,
  Srivastava, Staats, Stachie, Steer, Steinhoff, Steinlechner, Steinlechner,
  Stevenson, Stops, Stover, Strain, Strang, Stratta, Strunk, Sturani, Stuver,
  Sudhagar, Sudhir, Sugimoto, Suh, Sullivan, Summerscales, Sun, Sun, Sunil,
  Sur, Suresh, Sutton, Suzuki, Suzuki, Swinkels, Szczepa\ifmmode~\acute{n}\else
  \'{n}\fi{}czyk, Szewczyk, Tacca, Tagoshi, Tait, Takahashi, Takahashi,
  Takamori, Takano, Takeda, Takeda, Talbot, Talbot, Tanaka, Tanaka, Tanaka,
  Tanaka, Tanaka, Tanasijczuk, Tanioka, Tanner, Tao, Tao, Mart\'{\i}n, Taranto,
  Tasson, Telada, Tenorio, Terhune, Terkowski, Thirugnanasambandam, Thomas,
  Thomas, Thomas, Thompson, Thondapu, Thorne, Thrane, Tiwari, Tiwari, Tiwari,
  Toivonen, Toland, Tolley, Tomaru, Tomigami, Tomura, Tonelli, Torres-Forn\'e,
  Torrie, e~Melo, T\"oyr\"a, Trapananti, Travasso, Traylor, Trevor, Tringali,
  Tripathee, Troiano, Trovato, Trozzo, Trudeau, Tsai, Tsai, Tsang, Tsang, Tsao,
  Tse, Tso, Tsubono, Tsuchida, Tsukada, Tsuna, Tsutsui, Tsuzuki, Turbang,
  Turconi, Tuyenbayev, Ubhi, Uchikata, Uchiyama, Udall, Ueda, Uehara, Ueno,
  Ueshima, Unnikrishnan, Uraguchi, Urban, Ushiba, Utina, Vahlbruch, Vajente,
  Vajpeyi, Valdes, Valentini, Valsan, van Bakel, van Beuzekom, van~den Brand,
  Van Den~Broeck, Vander-Hyde, van~der Schaaf, van Heijningen, Vanosky, van
  Putten, van Remortel, Vardaro, Vargas, Varma, Vas\'uth, Vecchio, Vedovato,
  Veitch, Veitch, Venneberg, Venugopalan, Verkindt, Verma, Verma, Veske,
  Vetrano, Vicer\'e, Vidyant, Viets, Vijaykumar, Villa-Ortega, Vinet, Virtuoso,
  Vitale, Vo, Vocca, von Reis, von Wrangel, Vorvick, Vyatchanin, Wade, Wade,
  Wagner, Walet, Walker, Wallace, Wallace, Walsh, Wang, Wang, Wang, Ward,
  Warner, Was, Washimi, Washington, Watchi, Weaver, Webster, Weinert,
  Weinstein, Weiss, Weller, Wellmann, Wen, We\ss{}els, Wette, Whelan, White,
  Whiting, Whittle, Wilken, Williams, Williams, Williamson, Willis, Willke,
  Wilson, Winkler, Wipf, Wlodarczyk, Woan, Woehler, Wofford, Wong, Wu, Wu, Wu,
  Wu, Wysocki, Xiao, Xu, Yamada, Yamamoto, Yamamoto, Yamamoto, Yamamoto,
  Yamashita, Yamazaki, Yang, Yang, Yang, Yang, Yang, Yap, Yeeles, Yelikar,
  Ying, Yokogawa, Yokoyama, Yokozawa, Yoo, Yoshioka, Yu, Yu, Yuzurihara,
  Zadro\ifmmode~\dot{z}\else \.{z}\fi{}ny, Zanolin, Zeidler, Zelenova, Zendri,
  Zevin, Zhan, Zhang, Zhang, Zhang, Zhang, Zhang, Zhao, Zhao, Zhao, Zhao,
  Zheng, Zhou, Zhou, Zhu, Zhu, Zimmerman, Zlochower, Zucker, \&
  Zweizig}]{Ligo_Virgo_2021}
Abbott, R., Abbott, T.~D., Acernese, F., {et~al.} 2023, Phys. Rev. X, 13,
  011048, \dodoi{10.1103/PhysRevX.13.011048}

\bibitem[{Agrawal {et~al.}(2023)Agrawal, Hurley, Stevenson, Rodriguez, Szécsi,
  \& Kemp}]{Agrawal:2023}
Agrawal, P., Hurley, J., Stevenson, S., {et~al.} 2023, Monthly Notices of the
  Royal Astronomical Society, 525, 933, \dodoi{10.1093/mnras/stad2334}

\bibitem[{{Arca Sedda} {et~al.}(2023){Arca Sedda}, {Naoz}, \&
  {Kocsis}}]{ArcaSedda:2023agnReview}
{Arca Sedda}, M., {Naoz}, S., \& {Kocsis}, B. 2023, Universe, 9, 138,
  \dodoi{10.3390/universe9030138}

\bibitem[{{Bavera} {et~al.}(2022{\natexlab{a}}){Bavera}, {Fishbach}, {Zevin},
  {Zapartas}, \& {Fragos}}]{Bavera:2022spinredshift}
{Bavera}, S.~S., {Fishbach}, M., {Zevin}, M., {Zapartas}, E., \& {Fragos}, T.
  2022{\natexlab{a}}, \aap, 665, A59, \dodoi{10.1051/0004-6361/202243724}

\bibitem[{{Bavera} {et~al.}(2022{\natexlab{b}}){Bavera}, {Franciolini},
  {Cusin}, {Riotto}, {Zevin}, \& {Fragos}}]{Bavera:2022}
{Bavera}, S.~S., {Franciolini}, G., {Cusin}, G., {et~al.} 2022{\natexlab{b}},
  \aap, 660, A26, \dodoi{10.1051/0004-6361/202142208}

\bibitem[{Bavera {et~al.}(2021)Bavera, Fragos, Zevin, Berry, Marchant, Andrews,
  Coughlin, Dotter, Kovlakas, Misra, Serra-Perez, Qin, Rocha, Rom{\'{a}
  }n-Garza, Tran, \& Zapartas}]{Bavera_2021}
Bavera, S.~S., Fragos, T., Zevin, M., {et~al.} 2021, Astronomy {\&}
  Astrophysics, 647, A153, \dodoi{10.1051/0004-6361/202039804}

\bibitem[{Belczynski {et~al.}(2022)Belczynski, Doctor, Zevin, Olejak, Banerje,
  \& Chattopadhyay}]{Belczynski:2022massredshift}
Belczynski, K., Doctor, Z., Zevin, M., {et~al.} 2022, The Astrophysical
  Journal, 935, 126, \dodoi{10.3847/1538-4357/ac8167}

\bibitem[{{Belczynski} {et~al.}(2016{\natexlab{a}}){Belczynski}, {Holz},
  {Bulik}, \& {O'Shaughnessy}}]{Belczynski:2016}
{Belczynski}, K., {Holz}, D.~E., {Bulik}, T., \& {O'Shaughnessy}, R.
  2016{\natexlab{a}}, \nat, 534, 512, \dodoi{10.1038/nature18322}

\bibitem[{{Belczynski} {et~al.}(2016{\natexlab{b}}){Belczynski}, {Holz},
  {Bulik}, \& {O'Shaughnessy}}]{Belczynski_2016}
---. 2016{\natexlab{b}}, \nat, 534, 512, \dodoi{10.1038/nature18322}

\bibitem[{Belczynski {et~al.}(2017)Belczynski, Ryu, Perna, Berti, Tanaka, \&
  Bulik}]{Belczynski:2017}
Belczynski, K., Ryu, T., Perna, R., {et~al.} 2017, Monthly Notices of the Royal
  Astronomical Society, 471, 4702, \dodoi{10.1093/mnras/stx1759}

\bibitem[{{Belczynski} {et~al.}(2018){Belczynski}, {Bulik}, {Olejak},
  {Chruslinska}, {Singh}, {Pol}, {Zdunik}, {O'Shaughnessy}, {McLaughlin},
  {Lorimer}, {Korobkin}, {van den Heuvel}, {Davies}, \&
  {Holz}}]{Belczynski:2018}
{Belczynski}, K., {Bulik}, T., {Olejak}, A., {et~al.} 2018, arXiv e-prints,
  arXiv:1812.10065, \dodoi{10.48550/arXiv.1812.10065}

\bibitem[{{Belczynski} {et~al.}(2022){Belczynski}, {Romagnolo}, {Olejak},
  {Klencki}, {Chattopadhyay}, {Stevenson}, {Coleman Miller}, {Lasota}, \&
  {Crowther}}]{Belczynski:2022}
{Belczynski}, K., {Romagnolo}, A., {Olejak}, A., {et~al.} 2022, \apj, 925, 69,
  \dodoi{10.3847/1538-4357/ac375a}

\bibitem[{{Belgacem} {et~al.}(2019){Belgacem}, {Dirian}, {Foffa}, {Howell},
  {Maggiore}, \& {Regimbau}}]{Belgacem:2019}
{Belgacem}, E., {Dirian}, Y., {Foffa}, S., {et~al.} 2019, \jcap, 2019, 015,
  \dodoi{10.1088/1475-7516/2019/08/015}

\bibitem[{{Berger}(2014)}]{Berger:2014}
{Berger}, E. 2014, \araa, 52, 43, \dodoi{10.1146/annurev-astro-081913-035926}

\bibitem[{Biscoveanu {et~al.}(2022)Biscoveanu, Callister, Haster, Ng, Vitale,
  \& Farr}]{Biscoveanu:2022}
Biscoveanu, S., Callister, T.~A., Haster, C.-J., {et~al.} 2022, The
  Astrophysical Journal Letters, 932, L19, \dodoi{10.3847/2041-8213/ac71a8}

\bibitem[{Boesky {et~al.}(2024)Boesky, Broekgaarden, \& Berger}]{Boesky_2023}
Boesky, A., Broekgaarden, F.~S., \& Berger, E. 2024, \dodoi{TBD}

\bibitem[{{Borhanian} \& {Sathyaprakash}(2022)}]{Borhanian:2022arXiv220211048B}
{Borhanian}, S., \& {Sathyaprakash}, B.~S. 2022, arXiv e-prints,
  arXiv:2202.11048, \dodoi{10.48550/arXiv.2202.11048}

\bibitem[{{Broekgaarden} {et~al.}(2019){Broekgaarden}, {Justham}, {de Mink},
  {Gair}, {Mandel}, {Stevenson}, {Barrett}, {Vigna-G{\'o}mez}, \&
  {Neijssel}}]{Broekgaarden_2019}
{Broekgaarden}, F.~S., {Justham}, S., {de Mink}, S.~E., {et~al.} 2019, \mnras,
  490, 5228, \dodoi{10.1093/mnras/stz2558}

\bibitem[{Broekgaarden {et~al.}(2021)Broekgaarden, Berger, Neijssel,
  Vigna-G{\'{o} }mez, Chattopadhyay, Stevenson, Chruslinska, Justham, de~Mink,
  \& Mandel}]{Broekgaarden_2021}
Broekgaarden, F.~S., Berger, E., Neijssel, C.~J., {et~al.} 2021, Monthly
  Notices of the Royal Astronomical Society, 508, 5028,
  \dodoi{10.1093/mnras/stab2716}

\bibitem[{Broekgaarden {et~al.}(2022)Broekgaarden, Berger, Stevenson, Justham,
  Mandel, Chru{\'{s} }li{\'{n}}ska, van Son, Wagg, Vigna-G{\'{o}}mez, de~Mink,
  Chattopadhyay, \& Neijssel}]{Broekgaarden_2022}
Broekgaarden, F.~S., Berger, E., Stevenson, S., {et~al.} 2022, Monthly Notices
  of the Royal Astronomical Society, \dodoi{10.1093/mnras/stac1677}

\bibitem[{{Callister} \& {Farr}(2023)}]{Callister:2023}
{Callister}, T.~A., \& {Farr}, W.~M. 2023, arXiv e-prints, arXiv:2302.07289,
  \dodoi{10.48550/arXiv.2302.07289}

\bibitem[{Cheng {et~al.}(2023)Cheng, Zevin, \& Vitale}]{QiuCheng:2023}
Cheng, A.~Q., Zevin, M., \& Vitale, S. 2023, The Astrophysical Journal, 955,
  127, \dodoi{10.3847/1538-4357/aced98}

\bibitem[{Choksi {et~al.}(2019)Choksi, Volonteri, Colpi, Gnedin, \&
  Li}]{Choksi:2019}
Choksi, N., Volonteri, M., Colpi, M., Gnedin, O.~Y., \& Li, H. 2019, The
  Astrophysical Journal, 873, 100, \dodoi{10.3847/1538-4357/aaffde}

\bibitem[{{Chru{\'s}li{\'n}ska} \& {Nelemans}(2019)}]{Chruslinska:2019obsSFRD}
{Chru{\'s}li{\'n}ska}, M., \& {Nelemans}, G. 2019, \mnras, 488, 5300,
  \dodoi{10.1093/mnras/stz2057}

\bibitem[{Chruslinska {et~al.}(2018)Chruslinska, Nelemans, \&
  Belczynski}]{Chruslinska_2018}
Chruslinska, M., Nelemans, G., \& Belczynski, K. 2018, Monthly Notices of the
  Royal Astronomical Society, 482, 5012, \dodoi{10.1093/mnras/sty3087}

\bibitem[{Chruślińska(2022)}]{chruslinska_2022}
Chruślińska, M. 2022, Chemical evolution of the Universe and its consequences
  for gravitational-wave astrophysics,  arXiv,
  \dodoi{10.48550/ARXIV.2206.10622}

\bibitem[{{Chu} {et~al.}(2022){Chu}, {Yu}, \& {Lu}}]{Chu:2022}
{Chu}, Q., {Yu}, S., \& {Lu}, Y. 2022, \mnras, 509, 1557,
  \dodoi{10.1093/mnras/stab2882}

\bibitem[{Collaboration {et~al.}(2018)Collaboration, Price-Whelan, Sipőcz,
  Günther, Lim, Crawford, Conseil, Shupe, Craig, Dencheva, Ginsburg,
  VanderPlas, Bradley, Pérez-Suárez, de~Val-Borro, Contributors), Aldcroft,
  Cruz, Robitaille, Tollerud, Committee), Ardelean, Babej, Bach, Bachetti,
  Bakanov, Bamford, Barentsen, Barmby, Baumbach, Berry, Biscani, Boquien,
  Bostroem, Bouma, Brammer, Bray, Breytenbach, Buddelmeijer, Burke, Calderone,
  Rodríguez, Cara, Cardoso, Cheedella, Copin, Corrales, Crichton, D’Avella,
  Deil, Depagne, Dietrich, Donath, Droettboom, Earl, Erben, Fabbro, Ferreira,
  Finethy, Fox, Garrison, Gibbons, Goldstein, Gommers, Greco, Greenfield,
  Groener, Grollier, Hagen, Hirst, Homeier, Horton, Hosseinzadeh, Hu, Hunkeler,
  Ivezić, Jain, Jenness, Kanarek, Kendrew, Kern, Kerzendorf, Khvalko, King,
  Kirkby, Kulkarni, Kumar, Lee, Lenz, Littlefair, Ma, Macleod, Mastropietro,
  McCully, Montagnac, Morris, Mueller, Mumford, Muna, Murphy, Nelson, Nguyen,
  Ninan, Nöthe, Ogaz, Oh, Parejko, Parley, Pascual, Patil, Patil, Plunkett,
  Prochaska, Rastogi, Janga, Sabater, Sakurikar, Seifert, Sherbert,
  Sherwood-Taylor, Shih, Sick, Silbiger, Singanamalla, Singer, Sladen, Sooley,
  Sornarajah, Streicher, Teuben, Thomas, Tremblay, Turner, Terrón, van
  Kerkwijk, de~la Vega, Watkins, Weaver, Whitmore, Woillez, Zabalza, \&
  Contributors)}]{2018AJ....156..123A}
Collaboration, T.~A., Price-Whelan, A.~M., Sipőcz, B.~M., {et~al.} 2018, The
  Astronomical Journal, 156, 123, \dodoi{10.3847/1538-3881/aabc4f}

\bibitem[{Collette(2013)}]{collette_python_hdf5_2014}
Collette, A. 2013, Python and HDF5 (O'Reilly)

\bibitem[{Colombo {et~al.}(2022)Colombo, Salafia, Gabrielli, Ghirlanda,
  Giacomazzo, Perego, \& Colpi}]{Colombo:2022}
Colombo, A., Salafia, O.~S., Gabrielli, F., {et~al.} 2022, The Astrophysical
  Journal, 937, 79, \dodoi{10.3847/1538-4357/ac8d00}

\bibitem[{{de Kool}(1990)}]{deKool_1990}
{de Kool}, M. 1990, \apj, 358, 189, \dodoi{10.1086/168974}

\bibitem[{{De Luca} {et~al.}(2020){De Luca}, {Desjacques}, {Franciolini}, \&
  {Riotto}}]{DeLuca:2020}
{De Luca}, V., {Desjacques}, V., {Franciolini}, G., \& {Riotto}, A. 2020,
  \jcap, 2020, 028, \dodoi{10.1088/1475-7516/2020/11/028}

\bibitem[{{De Luca} {et~al.}(2021){De Luca}, {Franciolini}, {Pani}, \&
  {Riotto}}]{DeLuca:2021}
{De Luca}, V., {Franciolini}, G., {Pani}, P., \& {Riotto}, A. 2021, \jcap,
  2021, 039, \dodoi{10.1088/1475-7516/2021/11/039}

\bibitem[{de~Mink \& Mandel(2016)}]{deMinkMandel:2016}
de~Mink, S.~E., \& Mandel, I. 2016, Monthly Notices of the Royal Astronomical
  Society, 460, 3545, \dodoi{10.1093/mnras/stw1219}

\bibitem[{{Dominik} {et~al.}(2013){Dominik}, {Belczynski}, {Fryer}, {Holz},
  {Berti}, {Bulik}, {Mandel}, \& {O'Shaughnessy}}]{Dominik:2013}
{Dominik}, M., {Belczynski}, K., {Fryer}, C., {et~al.} 2013, \apj, 779, 72,
  \dodoi{10.1088/0004-637X/779/1/72}

\bibitem[{{Dorozsmai} \& {Toonen}(2022)}]{Dorozsmai:2022}
{Dorozsmai}, A., \& {Toonen}, S. 2022, arXiv e-prints, arXiv:2207.08837,
  \dodoi{10.48550/arXiv.2207.08837}

\bibitem[{{Dorozsmai} \& {Toonen}(2024)}]{Dorozsmai:2024}
---. 2024, \mnras, \dodoi{10.1093/mnras/stae152}

\bibitem[{du Buisson {et~al.}(2020)du Buisson, Marchant, Podsiadlowski,
  Kobayashi, Abdalla, Taylor, Mandel, de Mink, Moriya, \&
  Langer}]{duBuisson:2020}
du Buisson, L., Marchant, P., Podsiadlowski, P., {et~al.} 2020, Monthly
  Notices of the Royal Astronomical Society, 499, 5941,
  \dodoi{10.1093/mnras/staa3225}

\bibitem[{{Edelman} {et~al.}(2023){Edelman}, {Farr}, \&
  {Doctor}}]{edelman:2023}
{Edelman}, B., {Farr}, B., \& {Doctor}, Z. 2023, \apj, 946, 16,
  \dodoi{10.3847/1538-4357/acb5ed}

\bibitem[{{Eggleton} {et~al.}(1989){Eggleton}, {Fitchett}, \&
  {Tout}}]{Eggleton_1989}
{Eggleton}, P.~P., {Fitchett}, M.~J., \& {Tout}, C.~A. 1989, \apj, 347, 998,
  \dodoi{10.1086/168190}

\bibitem[{{Eldridge} {et~al.}(2019){Eldridge}, {Stanway}, \&
  {Tang}}]{Eldridge_2018}
{Eldridge}, J.~J., {Stanway}, E.~R., \& {Tang}, P.~N. 2019, \mnras, 482, 870,
  \dodoi{10.1093/mnras/sty2714}

\bibitem[{{Evans} {et~al.}(2021){Evans}, {Adhikari}, {Afle}, {Ballmer},
  {Biscoveanu}, {Borhanian}, {Brown}, {Chen}, {Eisenstein}, {Gruson}, {Gupta},
  {Hall}, {Huxford}, {Kamai}, {Kashyap}, {Kissel}, {Kuns}, {Landry}, {Lenon},
  {Lovelace}, {McCuller}, {Ng}, {Nitz}, {Read}, {Sathyaprakash}, {Shoemaker},
  {Slagmolen}, {Smith}, {Srivastava}, {Sun}, {Vitale}, \&
  {Weiss}}]{CosmicExplorer:2021evans}
{Evans}, M., {Adhikari}, R.~X., {Afle}, C., {et~al.} 2021, arXiv e-prints,
  arXiv:2109.09882, \dodoi{10.48550/arXiv.2109.09882}

\bibitem[{{Evans} {et~al.}(2023){Evans}, {Corsi}, {Afle}, {Ananyeva}, {Arun},
  {Ballmer}, {Bandopadhyay}, {Barsotti}, {Baryakhtar}, {Berger}, {Berti},
  {Biscoveanu}, {Borhanian}, {Broekgaarden}, {Brown}, {Cahillane}, {Campbell},
  {Chen}, {Daniel}, {Dhani}, {Driggers}, {Effler}, {Eisenstein}, {Fairhurst},
  {Feicht}, {Fritschel}, {Fulda}, {Gupta}, {Hall}, {Hammond}, {Hannuksela},
  {Hansen}, {Haster}, {Kacanja}, {Kamai}, {Kashyap}, {Shapiro Key},
  {Khadkikar}, {Kontos}, {Kuns}, {Landry}, {Landry}, {Lantz}, {Li}, {Lovelace},
  {Mandic}, {Mansell}, {Martynov}, {McCuller}, {Miller}, {Nitz}, {Owen},
  {Palomba}, {Read}, {Phurailatpam}, {Reddy}, {Richardson}, {Rollins},
  {Romano}, {Sathyaprakash}, {Schofield}, {Shoemaker}, {Sigg}, {Singh},
  {Slagmolen}, {Sledge}, {Smith}, {Soares-Santos}, {Strunk}, {Sun}, {Tanner},
  {van Son}, {Vitale}, {Willke}, {Yamamoto}, \& {Zucker}}]{CosmicExplorer:2023}
{Evans}, M., {Corsi}, A., {Afle}, C., {et~al.} 2023, arXiv e-prints,
  arXiv:2306.13745, \dodoi{10.48550/arXiv.2306.13745}

\bibitem[{{Fishbach} {et~al.}(2018){Fishbach}, {Holz}, \&
  {Farr}}]{Fishbach:2018}
{Fishbach}, M., {Holz}, D.~E., \& {Farr}, W.~M. 2018, \apjl, 863, L41,
  \dodoi{10.3847/2041-8213/aad800}

\bibitem[{Fishbach {et~al.}(2022)Fishbach, Kimball, \&
  Kalogera}]{Fishbach:2022}
Fishbach, M., Kimball, C., \& Kalogera, V. 2022, The Astrophysical Journal
  Letters, 935, L26, \dodoi{10.3847/2041-8213/ac86c4}

\bibitem[{Fishbach {et~al.}(2021)Fishbach, Doctor, Callister, Edelman, Ye,
  Essick, Farr, Farr, \& Holz}]{Fishbach:2021}
Fishbach, M., Doctor, Z., Callister, T., {et~al.} 2021, The Astrophysical
  Journal, 912, 98, \dodoi{10.3847/1538-4357/abee11}

\bibitem[{{Fong} {et~al.}(2015){Fong}, {Berger}, {Margutti}, \&
  {Zauderer}}]{Fong:2015}
{Fong}, W., {Berger}, E., {Margutti}, R., \& {Zauderer}, B.~A. 2015, \apj, 815,
  102, \dodoi{10.1088/0004-637X/815/2/102}

\bibitem[{Fragione \& Kocsis(2018)}]{FragioneKocsis:2018}
Fragione, G., \& Kocsis, B. 2018, Phys. Rev. Lett., 121, 161103,
  \dodoi{10.1103/PhysRevLett.121.161103}

\bibitem[{Franciolini {et~al.}(2022)Franciolini, Cotesta, Loutrel, Berti, Pani,
  \& Riotto}]{Franciolini:2022}
Franciolini, G., Cotesta, R., Loutrel, N., {et~al.} 2022, Phys. Rev. D, 105,
  063510, \dodoi{10.1103/PhysRevD.105.063510}

\bibitem[{Giacobbo \& Mapelli(2018)}]{Giacobbo_2018}
Giacobbo, N., \& Mapelli, M. 2018, Monthly Notices of the Royal Astronomical
  Society, 480, 2011, \dodoi{10.1093/mnras/sty1999}

\bibitem[{{Giacobbo} {et~al.}(2018){Giacobbo}, {Mapelli}, \&
  {Spera}}]{Giacobbo_2018_2}
{Giacobbo}, N., {Mapelli}, M., \& {Spera}, M. 2018, \mnras, 474, 2959,
  \dodoi{10.1093/mnras/stx2933}

\bibitem[{{Godfrey} {et~al.}(2023){Godfrey}, {Edelman}, \&
  {Farr}}]{Godfrey:2023}
{Godfrey}, J., {Edelman}, B., \& {Farr}, B. 2023, arXiv e-prints,
  arXiv:2304.01288, \dodoi{10.48550/arXiv.2304.01288}

\bibitem[{{Gupta} {et~al.}(2023){Gupta}, {Afle}, {Arun}, {Bandopadhyay},
  {Baryakhtar}, {Biscoveanu}, {Borhanian}, {Broekgaarden}, {Corsi}, {Dhani},
  {Evans}, {Hall}, {Hannuksela}, {Kacanja}, {Kashyap}, {Khadkikar}, {Kuns},
  {Li}, {Miller}, {Nitz}, {Owen}, {Palomba}, {Pearce}, {Phurailatpam},
  {Rajbhandari}, {Read}, {Romano}, {Sathyaprakash}, {Shoemaker}, {Singh},
  {Vitale}, {Barsotti}, {Berti}, {Cahillane}, {Chen}, {Fritschel}, {Haster},
  {Landry}, {Lovelace}, {McClelland}, {Slagmolen}, {Smith}, {Soares-Santos},
  {Sun}, {Tanner}, {Yamamoto}, \& {Zucker}}]{Gupta:2023}
{Gupta}, I., {Afle}, C., {Arun}, K.~G., {et~al.} 2023, arXiv e-prints,
  arXiv:2307.10421, \dodoi{10.48550/arXiv.2307.10421}

\bibitem[{Harris {et~al.}(2020)Harris, Millman, van~der Walt, Gommers,
  Virtanen, Cournapeau, Wieser, Taylor, Berg, Smith, Kern, Picus, Hoyer, van
  Kerkwijk, Brett, Haldane, Fernández~del Río, Wiebe, Peterson,
  Gérard-Marchant, Sheppard, Reddy, Weckesser, Abbasi, Gohlke, \&
  Oliphant}]{2020NumPy-Array}
Harris, C.~R., Millman, K.~J., van~der Walt, S.~J., {et~al.} 2020, Nature, 585,
  357–362, \dodoi{10.1038/s41586-020-2649-2}

\bibitem[{Hijikawa {et~al.}(2021)Hijikawa, Tanikawa, Kinugawa, Yoshida, \&
  Umeda}]{Hijikawa:2021}
Hijikawa, K., Tanikawa, A., Kinugawa, T., Yoshida, T., \& Umeda, H. 2021,
  Monthly Notices of the Royal Astronomical Society: Letters, 505, L69,
  \dodoi{10.1093/mnrasl/slab052}

\bibitem[{{Hunter}(2007)}]{2007CSE.....9...90H}
{Hunter}, J.~D. 2007, Computing in Science and Engineering, 9, 90,
  \dodoi{10.1109/MCSE.2007.55}

\bibitem[{{Hurley} {et~al.}(2000){Hurley}, {Pols}, \& {Tout}}]{Hurley_2000}
{Hurley}, J.~R., {Pols}, O.~R., \& {Tout}, C.~A. 2000, \mnras, 315, 543,
  \dodoi{10.1046/j.1365-8711.2000.03426.x}

\bibitem[{{Hurley} {et~al.}(2002){Hurley}, {Tout}, \& {Pols}}]{Hurley_2002}
{Hurley}, J.~R., {Tout}, C.~A., \& {Pols}, O.~R. 2002, \mnras, 329, 897,
  \dodoi{10.1046/j.1365-8711.2002.05038.x}

\bibitem[{{Iacovelli} {et~al.}(2022){Iacovelli}, {Mancarella}, {Foffa}, \&
  {Maggiore}}]{Iacovelli:2022}
{Iacovelli}, F., {Mancarella}, M., {Foffa}, S., \& {Maggiore}, M. 2022, \apj,
  941, 208, \dodoi{10.3847/1538-4357/ac9cd4}

\bibitem[{{Ivanova} {et~al.}(2020){Ivanova}, {Justham}, \&
  {Ricker}}]{ivanova:2020book}
{Ivanova}, N., {Justham}, S., \& {Ricker}, P. 2020, {Common Envelope
  Evolution}, \dodoi{10.1088/2514-3433/abb6f0}

\bibitem[{{Kinugawa} {et~al.}(2014){Kinugawa}, {Inayoshi}, {Hotokezaka},
  {Nakauchi}, \& {Nakamura}}]{Kinugawa:2014}
{Kinugawa}, T., {Inayoshi}, K., {Hotokezaka}, K., {Nakauchi}, D., \&
  {Nakamura}, T. 2014, \mnras, 442, 2963, \dodoi{10.1093/mnras/stu1022}

\bibitem[{Kinugawa {et~al.}(2020)Kinugawa, Nakamura, \& Nakano}]{Kinugawa:2020}
Kinugawa, T., Nakamura, T., \& Nakano, H. 2020, Monthly Notices of the Royal
  Astronomical Society, 498, 3946, \dodoi{10.1093/mnras/staa2511}

\bibitem[{{Kippenhahn} \& {Weigert}(1990)}]{KippenhahnWeigert:1990}
{Kippenhahn}, R., \& {Weigert}, A. 1990, {Stellar Structure and Evolution}

\bibitem[{Klencki {et~al.}(2018)Klencki, Moe, Gladysz, Chruslinska, Holz, \&
  Belczynski}]{Klencki_2018}
Klencki, J., Moe, M., Gladysz, W., {et~al.} 2018, Astronomy {\&} Astrophysics,
  619, A77, \dodoi{10.1051/0004-6361/201833025}

\bibitem[{Kluyver {et~al.}(2016)Kluyver, Ragan-Kelley, P{\'e}rez, Granger,
  Bussonnier, Frederic, Kelley, Hamrick, Grout, Corlay,
  {et~al.}}]{kluyver2016jupyter}
Kluyver, T., Ragan-Kelley, B., P{\'e}rez, F., {et~al.} 2016, in ELPUB, 87--90

\bibitem[{{Kritos} {et~al.}(2022){Kritos}, {Strokov}, {Baibhav}, \&
  {Berti}}]{Kritos:2022}
{Kritos}, K., {Strokov}, V., {Baibhav}, V., \& {Berti}, E. 2022, arXiv
  e-prints, arXiv:2210.10055, \dodoi{10.48550/arXiv.2210.10055}

\bibitem[{{Kruckow} {et~al.}(2018){Kruckow}, {Tauris}, {Langer}, {Kramer}, \&
  {Izzard}}]{Kruckow:2018}
{Kruckow}, M.~U., {Tauris}, T.~M., {Langer}, N., {Kramer}, M., \& {Izzard},
  R.~G. 2018, \mnras, 481, 1908, \dodoi{10.1093/mnras/sty2190}

\bibitem[{{Langer}(2012)}]{Langer:2012}
{Langer}, N. 2012, \araa, 50, 107, \dodoi{10.1146/annurev-astro-081811-125534}

\bibitem[{{Laplace} {et~al.}(2020){Laplace}, {G{\"o}tberg}, {de Mink},
  {Justham}, \& {Farmer}}]{Laplace:2020}
{Laplace}, E., {G{\"o}tberg}, Y., {de Mink}, S.~E., {Justham}, S., \& {Farmer},
  R. 2020, \aap, 637, A6, \dodoi{10.1051/0004-6361/201937300}

\bibitem[{{Lehoucq} {et~al.}(2023){Lehoucq}, {Dvorkin}, {Srinivasan},
  {Pellouin}, \& {Lamberts}}]{Lehoucq:2023}
{Lehoucq}, L., {Dvorkin}, I., {Srinivasan}, R., {Pellouin}, C., \& {Lamberts},
  A. 2023, \mnras, 526, 4378, \dodoi{10.1093/mnras/stad2917}

\bibitem[{Lipunov {et~al.}(2017)Lipunov, Kornilov, Gorbovskoy, Tiurina,
  Balanutsa, \& Kuznetsov}]{Lipunov_2017}
Lipunov, V., Kornilov, V., Gorbovskoy, E., {et~al.} 2017, New Astronomy, 51,
  122, \dodoi{https://doi.org/10.1016/j.newast.2016.08.017}

\bibitem[{Liu \& Bromm(2021)}]{Liu:2021}
Liu, B., \& Bromm, V. 2021, Monthly Notices of the Royal Astronomical Society,
  506, 5451, \dodoi{10.1093/mnras/stab2028}

\bibitem[{{Ma} {et~al.}(2016){Ma}, {Hopkins}, {Faucher-Gigu{\`e}re}, {Zolman},
  {Muratov}, {Kere{\v{s}}}, \& {Quataert}}]{Ma_2016}
{Ma}, X., {Hopkins}, P.~F., {Faucher-Gigu{\`e}re}, C.-A., {et~al.} 2016,
  \mnras, 456, 2140, \dodoi{10.1093/mnras/stv2659}

\bibitem[{Madau \& Dickinson(2014)}]{Madau_Dickinson_2014}
Madau, P., \& Dickinson, M. 2014, Annual Review of Astronomy and Astrophysics,
  52, 415, \dodoi{10.1146/annurev-astro-081811-125615}

\bibitem[{{Madau} \& {Fragos}(2017)}]{Madau_Fragos_2017}
{Madau}, P., \& {Fragos}, T. 2017, \apj, 840, 39,
  \dodoi{10.3847/1538-4357/aa6af9}

\bibitem[{{Maggiore} {et~al.}(2020){Maggiore}, {Van Den Broeck}, {Bartolo},
  {Belgacem}, {Bertacca}, {Bizouard}, {Branchesi}, {Clesse}, {Foffa},
  {Garc{\'\i}a-Bellido}, {Grimm}, {Harms}, {Hinderer}, {Matarrese}, {Palomba},
  {Peloso}, {Ricciardone}, \& {Sakellariadou}}]{EinsteinTelescope:2020Maggiore}
{Maggiore}, M., {Van Den Broeck}, C., {Bartolo}, N., {et~al.} 2020, \jcap,
  2020, 050, \dodoi{10.1088/1475-7516/2020/03/050}

\bibitem[{{Mandel} \& {Broekgaarden}(2022)}]{MandelBroekgaarden:2022}
{Mandel}, I., \& {Broekgaarden}, F.~S. 2022, Living Reviews in Relativity, 25,
  1, \dodoi{10.1007/s41114-021-00034-3}

\bibitem[{Mandel \& de Mink(2016)}]{MandeldeMink:2016}
Mandel, I., \& de Mink, S.~E. 2016, Monthly Notices of the Royal Astronomical
  Society, 458, 2634, \dodoi{10.1093/mnras/stw379}

\bibitem[{{Mapelli}(2021)}]{Mapelli:2021review}
{Mapelli}, M. 2021, in Handbook of Gravitational Wave Astronomy, 16,
  \dodoi{10.1007/978-981-15-4702-7_16-1}

\bibitem[{Mapelli {et~al.}(2022)Mapelli, Bouffanais, Santoliquido, Arca Sedda,
  \& Artale}]{Mapelli:2022}
Mapelli, M., Bouffanais, Y., Santoliquido, F., Arca Sedda, M., \& Artale,
  M.~C. 2022, Monthly Notices of the Royal Astronomical Society, 511, 5797,
  \dodoi{10.1093/mnras/stac422}

\bibitem[{{Martinez} {et~al.}(2020){Martinez}, {Fragione}, {Kremer},
  {Chatterjee}, {Rodriguez}, {Samsing}, {Ye}, {Weatherford}, {Zevin}, {Naoz},
  \& {Rasio}}]{Martinez:2020}
{Martinez}, M. A.~S., {Fragione}, G., {Kremer}, K., {et~al.} 2020, \apj, 903,
  67, \dodoi{10.3847/1538-4357/abba25}

\bibitem[{{McKernan} {et~al.}(2022){McKernan}, {Ford}, {Callister}, {Farr},
  {O'Shaughnessy}, {Smith}, {Thrane}, \& {Vajpeyi}}]{McKernan:2022}
{McKernan}, B., {Ford}, K.~E.~S., {Callister}, T., {et~al.} 2022, \mnras, 514,
  3886, \dodoi{10.1093/mnras/stac1570}

\bibitem[{Mehta {et~al.}(2023)Mehta, Olsen, Wadekar, Roulet, Venumadhav,
  Mushkin, Zackay, \& Zaldarriaga}]{mehta2023new}
Mehta, A.~K., Olsen, S., Wadekar, D., {et~al.} 2023, New binary black hole
  mergers in the LIGO-Virgo O3b data.
\newblock \doarXiv{2311.06061}

\bibitem[{{Mennekens} \& {Vanbeveren}(2016)}]{Mennekens_2016}
{Mennekens}, N., \& {Vanbeveren}, D. 2016, \aap, 589, A64,
  \dodoi{10.1051/0004-6361/201628193}

\bibitem[{{Naidu} {et~al.}(2022){Naidu}, {Ji}, {Conroy}, {Bonaca}, {Ting},
  {Zaritsky}, {van Son}, {Broekgaarden}, {Tacchella}, {Chandra}, {Caldwell},
  {Cargile}, \& {Speagle}}]{Naidu:2022}
{Naidu}, R.~P., {Ji}, A.~P., {Conroy}, C., {et~al.} 2022, \apjl, 926, L36,
  \dodoi{10.3847/2041-8213/ac5589}

\bibitem[{{Neijssel} {et~al.}(2019){Neijssel}, {Vigna-G{\'o}mez}, {Stevenson},
  {Barrett}, {Gaebel}, {Broekgaarden}, {de Mink}, {Sz{\'e}csi}, {Vinciguerra},
  \& {Mandel}}]{Neijssel_2019}
{Neijssel}, C.~J., {Vigna-G{\'o}mez}, A., {Stevenson}, S., {et~al.} 2019,
  \mnras, 490, 3740, \dodoi{10.1093/mnras/stz2840}

\bibitem[{Ng {et~al.}(2021)Ng, Vitale, Farr, \& Rodriguez}]{Ng_2021}
Ng, K. K.~Y., Vitale, S., Farr, W.~M., \& Rodriguez, C.~L. 2021, The
  Astrophysical Journal Letters, 913, L5, \dodoi{10.3847/2041-8213/abf8be}

\bibitem[{{Nitz} {et~al.}(2023){Nitz}, {Kumar}, {Wang}, {Kastha}, {Wu},
  {Sch{\"a}fer}, {Dhurkunde}, \& {Capano}}]{Nitz:2023-4-OGC}
{Nitz}, A.~H., {Kumar}, S., {Wang}, Y.-F., {et~al.} 2023, \apj, 946, 59,
  \dodoi{10.3847/1538-4357/aca591}

\bibitem[{{Nugent} {et~al.}(2022){Nugent}, {Fong}, {Dong}, {Leja}, {Berger},
  {Zevin}, {Chornock}, {Cobb}, {Kelley}, {Kilpatrick}, {Levan}, {Margutti},
  {Paterson}, {Perley}, {Escorial}, {Smith}, \& {Tanvir}}]{Nugent:2022}
{Nugent}, A.~E., {Fong}, W.-F., {Dong}, Y., {et~al.} 2022, \apj, 940, 57,
  \dodoi{10.3847/1538-4357/ac91d1}

\bibitem[{{Olejak} {et~al.}(2022){Olejak}, {Fryer}, {Belczynski}, \&
  {Baibhav}}]{Olejak:2022supernova}
{Olejak}, A., {Fryer}, C.~L., {Belczynski}, K., \& {Baibhav}, V. 2022, \mnras,
  516, 2252, \dodoi{10.1093/mnras/stac2359}

\bibitem[{{Olsen} {et~al.}(2022){Olsen}, {Venumadhav}, {Mushkin}, {Roulet},
  {Zackay}, \& {Zaldarriaga}}]{Olsen:2022}
{Olsen}, S., {Venumadhav}, T., {Mushkin}, J., {et~al.} 2022, \prd, 106, 043009,
  \dodoi{10.1103/PhysRevD.106.043009}

\bibitem[{{Panter} {et~al.}(2004){Panter}, {Heavens}, \&
  {Jimenez}}]{Panter_2004}
{Panter}, B., {Heavens}, A.~F., \& {Jimenez}, R. 2004, \mnras, 355, 764,
  \dodoi{10.1111/j.1365-2966.2004.08355.x}

\bibitem[{{Payne} \& {Thrane}(2023)}]{Payne:2023}
{Payne}, E., \& {Thrane}, E. 2023, Physical Review Research, 5, 023013,
  \dodoi{10.1103/PhysRevResearch.5.023013}

\bibitem[{{Perez} \& {Granger}(2007)}]{2007CSE.....9c..21P}
{Perez}, F., \& {Granger}, B.~E. 2007, Computing in Science and Engineering, 9,
  21, \dodoi{10.1109/MCSE.2007.53}

\bibitem[{Peters(1964)}]{Peters_1964}
Peters, P.~C. 1964, Phys. Rev., 136, B1224, \dodoi{10.1103/PhysRev.136.B1224}

\bibitem[{Pols {et~al.}(1998)Pols, Schröder, Hurley, Tout, \&
  Eggleton}]{Pols_1998}
Pols, O.~R., Schröder, K.-P., Hurley, J.~R., Tout, C.~A., \& Eggleton, P.~P.
  1998, Monthly Notices of the Royal Astronomical Society, 298, 525,
  \dodoi{10.1046/j.1365-8711.1998.01658.x}

\bibitem[{{Punturo} {et~al.}(2010){Punturo}, {Abernathy}, {Acernese}, {Allen},
  {Andersson}, {Arun}, {Barone}, {Barr}, {Barsuglia}, {Beker}, {Beveridge},
  {Birindelli}, {Bose}, {Bosi}, {Braccini}, {Bradaschia}, {Bulik}, {Calloni},
  {Cella}, {Chassande Mottin}, {Chelkowski}, {Chincarini}, {Clark}, {Coccia},
  {Colacino}, {Colas}, {Cumming}, {Cunningham}, {Cuoco}, {Danilishin},
  {Danzmann}, {De Luca}, {De Salvo}, {Dent}, {De Rosa}, {Di Fiore}, {Di
  Virgilio}, {Doets}, {Fafone}, {Falferi}, {Flaminio}, {Franc}, {Frasconi},
  {Freise}, {Fulda}, {Gair}, {Gemme}, {Gennai}, {Giazotto}, {Glampedakis},
  {Granata}, {Grote}, {Guidi}, {Hammond}, {Hannam}, {Harms}, {Heinert},
  {Hendry}, {Heng}, {Hennes}, {Hild}, {Hough}, {Husa}, {Huttner}, {Jones},
  {Khalili}, {Kokeyama}, {Kokkotas}, {Krishnan}, {Lorenzini}, {L{\"u}ck},
  {Majorana}, {Mandel}, {Mandic}, {Martin}, {Michel}, {Minenkov}, {Morgado},
  {Mosca}, {Mours}, {M{\"u}ller{\textendash}Ebhardt}, {Murray}, {Nawrodt},
  {Nelson}, {Oshaughnessy}, {Ott}, {Palomba}, {Paoli}, {Parguez},
  {Pasqualetti}, {Passaquieti}, {Passuello}, {Pinard}, {Poggiani}, {Popolizio},
  {Prato}, {Puppo}, {Rabeling}, {Rapagnani}, {Read}, {Regimbau}, {Rehbein},
  {Reid}, {Rezzolla}, {Ricci}, {Richard}, {Rocchi}, {Rowan}, {R{\"u}diger},
  {Sassolas}, {Sathyaprakash}, {Schnabel}, {Schwarz}, {Seidel}, {Sintes},
  {Somiya}, {Speirits}, {Strain}, {Strigin}, {Sutton}, {Tarabrin},
  {Th{\"u}ring}, {van den Brand}, {van Leewen}, {van Veggel}, {van den Broeck},
  {Vecchio}, {Veitch}, {Vetrano}, {Vicere}, {Vyatchanin}, {Willke}, {Woan},
  {Wolfango}, \& {Yamamoto}}]{EinsteinTelescope:2010Punturo}
{Punturo}, M., {Abernathy}, M., {Acernese}, F., {et~al.} 2010, Classical and
  Quantum Gravity, 27, 194002, \dodoi{10.1088/0264-9381/27/19/194002}

\bibitem[{{Qin} {et~al.}(2018){Qin}, {Fragos}, {Meynet}, {Andrews},
  {S{\o}rensen}, \& {Song}}]{Qin:2018}
{Qin}, Y., {Fragos}, T., {Meynet}, G., {et~al.} 2018, \aap, 616, A28,
  \dodoi{10.1051/0004-6361/201832839}

\bibitem[{{Raidal} {et~al.}(2019){Raidal}, {Spethmann}, {Vaskonen}, \&
  {Veerm{\"a}e}}]{Raidal:2019}
{Raidal}, M., {Spethmann}, C., {Vaskonen}, V., \& {Veerm{\"a}e}, H. 2019,
  \jcap, 2019, 018, \dodoi{10.1088/1475-7516/2019/02/018}

\bibitem[{{Ray} {et~al.}(2023){Ray}, {Hernandez}, {Mohite}, {Creighton}, \&
  {Kapadia}}]{Ray:2023}
{Ray}, A., {Hernandez}, I.~M., {Mohite}, S., {Creighton}, J., \& {Kapadia}, S.
  2023, \apj, 957, 37, \dodoi{10.3847/1538-4357/acf452}

\bibitem[{{Regimbau} {et~al.}(2012){Regimbau}, {Dent}, {Del Pozzo},
  {Giampanis}, {Li}, {Robinson}, {Van Den Broeck}, {Meacher}, {Rodriguez},
  {Sathyaprakash}, \& {W{\'o}jcik}}]{Regimbau:2012}
{Regimbau}, T., {Dent}, T., {Del Pozzo}, W., {et~al.} 2012, \prd, 86, 122001,
  \dodoi{10.1103/PhysRevD.86.122001}

\bibitem[{{Reitze} {et~al.}(2019){Reitze}, {Adhikari}, {Ballmer}, {Barish},
  {Barsotti}, {Billingsley}, {Brown}, {Chen}, {Coyne}, {Eisenstein}, {Evans},
  {Fritschel}, {Hall}, {Lazzarini}, {Lovelace}, {Read}, {Sathyaprakash},
  {Shoemaker}, {Smith}, {Torrie}, {Vitale}, {Weiss}, {Wipf}, \&
  {Zucker}}]{CosmicExplorer:2019reitze}
{Reitze}, D., {Adhikari}, R.~X., {Ballmer}, S., {et~al.} 2019, in Bulletin of
  the American Astronomical Society, Vol.~51, 35,
  \dodoi{10.48550/arXiv.1907.04833}

\bibitem[{Riley {et~al.}(2021)Riley, Mandel, Marchant, Butler, Nathaniel,
  Neijssel, Shortt, \& Vigna-Gómez}]{Riley:2021}
Riley, J., Mandel, I., Marchant, P., {et~al.} 2021, Monthly Notices of the
  Royal Astronomical Society, 505, 663, \dodoi{10.1093/mnras/stab1291}

\bibitem[{Rodriguez {et~al.}(2016)Rodriguez, Chatterjee, \&
  Rasio}]{Rodriguez:2016}
Rodriguez, C.~L., Chatterjee, S., \& Rasio, F.~A. 2016, Phys. Rev. D, 93,
  084029, \dodoi{10.1103/PhysRevD.93.084029}

\bibitem[{Rodriguez \& Loeb(2018)}]{RodriguezLoeb:2018}
Rodriguez, C.~L., \& Loeb, A. 2018, The Astrophysical Journal Letters, 866, L5,
  \dodoi{10.3847/2041-8213/aae377}

\bibitem[{{Romagnolo} {et~al.}(2023){Romagnolo}, {Belczynski}, {Klencki},
  {Agrawal}, {Shenar}, \& {Sz{\'e}csi}}]{Romagnolo:2023}
{Romagnolo}, A., {Belczynski}, K., {Klencki}, J., {et~al.} 2023, \mnras, 525,
  706, \dodoi{10.1093/mnras/stad2366}

\bibitem[{{Rom{\'a}n-Garza} {et~al.}(2021){Rom{\'a}n-Garza}, {Bavera},
  {Fragos}, {Zapartas}, {Misra}, {Andrews}, {Coughlin}, {Dotter}, {Kovlakas},
  {Serra}, {Qin}, {Rocha}, \& {Tran}}]{RomanGarza:2020}
{Rom{\'a}n-Garza}, J., {Bavera}, S.~S., {Fragos}, T., {et~al.} 2021, \apjl,
  912, L23, \dodoi{10.3847/2041-8213/abf42c}

\bibitem[{Santoliquido {et~al.}(2021)Santoliquido, Mapelli, Giacobbo,
  Bouffanais, \& Artale}]{Santoliquido_2021}
Santoliquido, F., Mapelli, M., Giacobbo, N., Bouffanais, Y., \& Artale, M.~C.
  2021, Monthly Notices of the Royal Astronomical Society, 502, 4877,
  \dodoi{10.1093/mnras/stab280}

\bibitem[{Santoliquido {et~al.}(2023)Santoliquido, Mapelli, Iorio, Costa,
  Glover, Hartwig, Klessen, \& Merli}]{Santoliquido:2023}
Santoliquido, F., Mapelli, M., Iorio, G., {et~al.} 2023, Monthly Notices of the
  Royal Astronomical Society, 524, 307, \dodoi{10.1093/mnras/stad1860}

\bibitem[{{Sathyaprakash} {et~al.}(2012){Sathyaprakash}, {Abernathy},
  {Acernese}, {Ajith}, {Allen}, {Amaro-Seoane}, {Andersson}, {Aoudia}, {Arun},
  {Astone}, {Krishnan}, {Barack}, {Barone}, {Barr}, {Barsuglia}, {Bassan},
  {Bassiri}, {Beker}, {Beveridge}, {Bizouard}, {Bond}, {Bose}, {Bosi},
  {Braccini}, {Bradaschia}, {Britzger}, {Brueckner}, {Bulik}, {Bulten},
  {Burmeister}, {Calloni}, {Campsie}, {Carbone}, {Cella}, {Chalkley},
  {Chassande-Mottin}, {Chelkowski}, {Chincarini}, {Di Cintio}, {Clark},
  {Coccia}, {Colacino}, {Colas}, {Colla}, {Corsi}, {Cumming}, {Cunningham},
  {Cuoco}, {Danilishin}, {Danzmann}, {Daw}, {De Salvo}, {Del Pozzo}, {Dent},
  {De Rosa}, {Di Fiore}, {Emilio}, {Di Virgilio}, {Dietz}, {Doets}, {Dueck},
  {Edwards}, {Fafone}, {Fairhurst}, {Falferi}, {Favata}, {Ferrari}, {Ferrini},
  {Fidecaro}, {Flaminio}, {Franc}, {Frasconi}, {Freise}, {Friedrich}, {Fulda},
  {Gair}, {Galimberti}, {Gemme}, {Genin}, {Gennai}, {Giazotto}, {Glampedakis},
  {Gossan}, {Gouaty}, {Graef}, {Graham}, {Granata}, {Grote}, {Guidi}, {Hallam},
  {Hammond}, {Hannam}, {Harms}, {Haughian}, {Hawke}, {Heinert}, {Hendry},
  {Heng}, {Hennes}, {Hild}, {Hough}, {Huet}, {Husa}, {Huttner}, {Iyer},
  {Jones}, {Jones}, {Kamaretsos}, {Kant Mishra}, {Kawazoe}, {Khalili}, {Kley},
  {Kokeyama}, {Kokkotas}, {Kroker}, {Kumar}, {Kuroda}, {Lagrange}, {Lastzka},
  {Li}, {Lorenzini}, {Losurdo}, {L{\"u}ck}, {Majorana}, {Malvezzi}, {Mandel},
  {Mandic}, {Marka}, {Marin}, {Marion}, {Marque}, {Martin}, {McLeod},
  {Mckechan}, {Mehmet}, {Michel}, {Minenkov}, {Morgado}, {Morgia}, {Mosca},
  {Moscatelli}, {Mours}, {M{\"u}ller-Ebhardt}, {Murray}, {Naticchioni},
  {Nawrodt}, {Nelson}, {O' Shaughnessy}, {Ott}, {Palomba}, {Paoli}, {Parguez},
  {Pasqualetti}, {Passaquieti}, {Passuello}, {Perciballi}, {Piergiovanni},
  {Pinard}, {Pitkin}, {Plastino}, {Plissi}, {Poggiani}, {Popolizio}, {Porter},
  {Prato}, {Prodi}, {Punturo}, {Puppo}, {Rabeling}, {Racz}, {Rapagnani}, {Re},
  {Read}, {Regimbau}, {Rehbein}, {Reid}, {Ricci}, {Richard}, {Robinson},
  {Rocchi}, {Romano}, {Rowan}, {R{\"u}diger}, {Samblowski}, {Santamar{\'\i}a},
  {Sassolas}, {Schilling}, {Schmidt}, {Schnabel}, {Schutz}, {Schwarz}, {Scott},
  {Seidel}, {Sintes}, {Somiya}, {Sopuerta}, {Sorazu}, {Speirits}, {Storchi},
  {Strain}, {Strigin}, {Sutton}, {Tarabrin}, {Taylor}, {Th{\"u}rin},
  {Tokmakov}, {Tonelli}, {Tournefier}, {Vaccarone}, {Vahlbruch}, {van den
  Brand}, {Van Den Broeck}, {van der Putten}, {van Veggel}, {Vecchio},
  {Veitch}, {Vetrano}, {Vicere}, {Vyatchanin}, {We{\ss}els}, {Willke},
  {Winkler}, {Woan}, {Woodcraft}, \&
  {Yamamoto}}]{EinsteinTelescope:2012Sathyaprakash}
{Sathyaprakash}, B., {Abernathy}, M., {Acernese}, F., {et~al.} 2012, Classical
  and Quantum Gravity, 29, 124013, \dodoi{10.1088/0264-9381/29/12/124013}

\bibitem[{Singh {et~al.}(2022)Singh, Bulik, Belczynski, \& Askar}]{Singh_2022}
Singh, N., Bulik, T., Belczynski, K., \& Askar, A. 2022, Astronomy {\&}
  Astrophysics, 667, A2, \dodoi{10.1051/0004-6361/202142856}

\bibitem[{{Spera} {et~al.}(2022){Spera}, {Trani}, \& {Mencagli}}]{Spera:2022}
{Spera}, M., {Trani}, A.~A., \& {Mencagli}, M. 2022, Galaxies, 10, 76,
  \dodoi{10.3390/galaxies10040076}

\bibitem[{{Stevenson} {et~al.}(2017){Stevenson}, {Vigna-G{\'o}mez}, {Mandel},
  {Barrett}, {Neijssel}, {Perkins}, \& {de Mink}}]{Stevenson_2017}
{Stevenson}, S., {Vigna-G{\'o}mez}, A., {Mandel}, I., {et~al.} 2017, Nature
  Communications, 8, 14906, \dodoi{10.1038/ncomms14906}

\bibitem[{{Team COMPAS: Riley} {et~al.}(2022){Team COMPAS: Riley}, {Agrawal},
  {Barrett}, {Boyett}, {Broekgaarden}, {Chattopadhyay}, {Gaebel}, {Gittins},
  {Hirai}, {Howitt}, {Justham}, {Khandelwal}, {Kummer}, {Lau}, {Mandel}, {de
  Mink}, {Neijssel}, {Riley}, {van Son}, {Stevenson}, {Vigna-Gomez},
  {Vinciguerra}, {Wagg}, \& {Willcox}}]{COMPAS_2022}
{Team COMPAS: Riley}, J., {Agrawal}, P., {Barrett}, J.~W., {et~al.} 2022,
  \apjs, 258, 34, \dodoi{10.3847/1538-4365/ac416c}

\bibitem[{{Tong} {et~al.}(2022){Tong}, {Galaudage}, \& {Thrane}}]{Tong:2022}
{Tong}, H., {Galaudage}, S., \& {Thrane}, E. 2022, \prd, 106, 103019,
  \dodoi{10.1103/PhysRevD.106.103019}

\bibitem[{Tout {et~al.}(1996)Tout, Pols, Eggleton, \& Han}]{Tout_1996}
Tout, C.~A., Pols, O.~R., Eggleton, P.~P., \& Han, Z. 1996, Monthly Notices of
  the Royal Astronomical Society, 281, 257, \dodoi{10.1093/mnras/281.1.257}

\bibitem[{van Son {et~al.}(2022)van Son, de~Mink, Callister, Justham, Renzo,
  Wagg, Broekgaarden, Kummer, Pakmor, \& Mandel}]{van_Son_2022}
van Son, L. A.~C., de~Mink, S.~E., Callister, T., {et~al.} 2022, The
  Astrophysical Journal, 931, 17, \dodoi{10.3847/1538-4357/ac64a3}

\bibitem[{{van Son} {et~al.}(2022){van Son}, {de Mink}, {Renzo}, {Justham},
  {Zapartas}, {Breivik}, {Callister}, {Farr}, \& {Conroy}}]{vanSon:2022}
{van Son}, L.~A.~C., {de Mink}, S.~E., {Renzo}, M., {et~al.} 2022, \apj, 940,
  184, \dodoi{10.3847/1538-4357/ac9b0a}

\bibitem[{{\noopsort{Van Rossum}}{van Rossum}(1995)}]{CS-R9526}
{\noopsort{Van Rossum}}{van Rossum}, G. 1995, Python tutorial, Tech. Rep.
  CS-R9526, Centrum voor Wiskunde en Informatica (CWI), Amsterdam

\bibitem[{{Venumadhav} {et~al.}(2019){Venumadhav}, {Zackay}, {Roulet}, {Dai},
  \& {Zaldarriaga}}]{Venumadhav:2019}
{Venumadhav}, T., {Zackay}, B., {Roulet}, J., {Dai}, L., \& {Zaldarriaga}, M.
  2019, \prd, 100, 023011, \dodoi{10.1103/PhysRevD.100.023011}

\bibitem[{{Venumadhav} {et~al.}(2020){Venumadhav}, {Zackay}, {Roulet}, {Dai},
  \& {Zaldarriaga}}]{Venumadhav:2020}
---. 2020, \prd, 101, 083030, \dodoi{10.1103/PhysRevD.101.083030}

\bibitem[{{Vigna-G{\'o}mez} {et~al.}(2018){Vigna-G{\'o}mez}, {Neijssel},
  {Stevenson}, {Barrett}, {Belczynski}, {Justham}, {de Mink}, {M{\"u}ller},
  {Podsiadlowski}, {Renzo}, {Sz{\'e}csi}, \& {Mandel}}]{VignaGomez_2018}
{Vigna-G{\'o}mez}, A., {Neijssel}, C.~J., {Stevenson}, S., {et~al.} 2018,
  \mnras, 481, 4009, \dodoi{10.1093/mnras/sty2463}

\bibitem[{{Vijaykumar} {et~al.}(2023){Vijaykumar}, {Fishbach}, {Adhikari}, \&
  {Holz}}]{Vijaykumar:2023}
{Vijaykumar}, A., {Fishbach}, M., {Adhikari}, S., \& {Holz}, D.~E. 2023, arXiv
  e-prints, arXiv:2312.03316, \dodoi{10.48550/arXiv.2312.03316}

\bibitem[{{Vink} {et~al.}(2000){Vink}, {de Koter}, \& {Lamers}}]{Vink_2000}
{Vink}, J.~S., {de Koter}, A., \& {Lamers}, H.~J.~G.~L.~M. 2000, \aap, 362,
  295.
\newblock \doarXiv{astro-ph/0008183}

\bibitem[{{Vink} {et~al.}(2001){Vink}, {de Koter}, \& {Lamers}}]{Vink_2001}
{Vink}, J.~S., {de Koter}, A., \& {Lamers}, H. J. G. L.~M. 2001, A\&A, 369,
  574, \dodoi{10.1051/0004-6361:20010127}

\bibitem[{Virtanen {et~al.}(2020)Virtanen, Gommers, Oliphant, Haberland, Reddy,
  Cournapeau, Burovski, Peterson, Weckesser, Bright, {van der Walt}, Brett,
  Wilson, Millman, Mayorov, Nelson, Jones, Kern, Larson, Carey, Polat, Feng,
  Moore, {VanderPlas}, Laxalde, Perktold, Cimrman, Henriksen, Quintero, Harris,
  Archibald, Ribeiro, Pedregosa, {van Mulbregt}, \& {SciPy 1.0
  Contributors}}]{2020SciPy-NMeth}
Virtanen, P., Gommers, R., Oliphant, T.~E., {et~al.} 2020, Nature Methods, 17,
  261, \dodoi{10.1038/s41592-019-0686-2}

\bibitem[{Vitale {et~al.}(2019)Vitale, Farr, Ng, \& Rodriguez}]{Vitale:2019}
Vitale, S., Farr, W.~M., Ng, K. K.~Y., \& Rodriguez, C.~L. 2019, The
  Astrophysical Journal Letters, 886, L1, \dodoi{10.3847/2041-8213/ab50c0}

\bibitem[{{Wadekar} {et~al.}(2023){Wadekar}, {Roulet}, {Venumadhav}, {Mehta},
  {Zackay}, {Mushkin}, {Olsen}, \& {Zaldarriaga}}]{Wadekar:2023}
{Wadekar}, D., {Roulet}, J., {Venumadhav}, T., {et~al.} 2023, arXiv e-prints,
  arXiv:2312.06631, \dodoi{10.48550/arXiv.2312.06631}

\bibitem[{{Webbink}(1984)}]{Webbink_1984}
{Webbink}, R.~F. 1984, \apj, 277, 355, \dodoi{10.1086/161701}

\bibitem[{{Ye} \& {Fishbach}(2024)}]{Ye_2024}
{Ye}, C.~S., \& {Fishbach}, M. 2024, arXiv e-prints, arXiv:2402.12444.
\newblock \doarXiv{2402.12444}

\bibitem[{{Zackay} {et~al.}(2019){Zackay}, {Venumadhav}, {Dai}, {Roulet}, \&
  {Zaldarriaga}}]{2019PhRvD.100b3007Z}
{Zackay}, B., {Venumadhav}, T., {Dai}, L., {Roulet}, J., \& {Zaldarriaga}, M.
  2019, \prd, 100, 023007, \dodoi{10.1103/PhysRevD.100.023007}

\bibitem[{{Zevin} {et~al.}(2022){Zevin}, {Nugent}, {Adhikari}, {Fong}, {Holz},
  \& {Kelley}}]{Zevin_2022}
{Zevin}, M., {Nugent}, A.~E., {Adhikari}, S., {et~al.} 2022, \apjl, 940, L18,
  \dodoi{10.3847/2041-8213/ac91cd}

\bibitem[{{Zevin} {et~al.}(2021){Zevin}, {Bavera}, {Berry}, {Kalogera},
  {Fragos}, {Marchant}, {Rodriguez}, {Antonini}, {Holz}, \&
  {Pankow}}]{Zevin:2021}
{Zevin}, M., {Bavera}, S.~S., {Berry}, C. P.~L., {et~al.} 2021, \apj, 910, 152,
  \dodoi{10.3847/1538-4357/abe40e}

\end{thebibliography}
\bibliographystyle{aasjournal}

\appendix

\section{The impact of metallicity on stellar evolution} \label{sec:impact_of_metallicity}

Metallicity plays a major role in both single and binary star evolution \citep{Neijssel_2019, Broekgaarden_2022, chruslinska_2022}, predominantly as a direct result of its impact on stellar winds \citep{Vink_2001}.  Winds from massive stars are driven by spectral lines from metals. Wind strips mass from stars, so stars with high initial metallicity lose significantly more mass in their lifetimes than metal-poor stars. This leads stars born with greater proportions of metals to form \acp{NS} instead of \acp{BH}.
The effects of metallicity on \textit{binary} stellar evolution are numerous and even more complex than for single stars. Overall, stronger stellar winds at higher metallicity lead to a decrease in mergers of all types in our simulations due to a range of indirect effects including mass loss associated with stellar winds leading to loss of angular momentum in a binary. This causes greater widening of binaries at high metallicities, increasing inspiral times \citep{Peters_1964} and the number of systems that disrupt, ultimately leading to a decrease in mergers from metal-rich progenitors. More mass loss through stellar winds also leads to systems with smaller stellar envelopes when the binary enters the \ac{CE} phase. This also leads to less shrinking and fewer systems that can merge in a Hubble time \citep[see][and references therein]{Neijssel_2019, Broekgaarden_2022, chruslinska_2022}. Another important effect is that metallicity plays a considerable role in determining stellar radial expansion during the star's evolution with higher metallicities, typically leading to more radial expansion in our simulations \citep{Hurley_2000,Hurley_2002, Romagnolo:2023}, increasing the probability of unstable mass transfer thus encouraging stellar mergers \citep[cf.][]{Neijssel_2019}.

\section{Metallicity-dependent star formation history} \label{sec:appendix-metallicity-dependent-star-formation-history}

\begin{figure*}[h!]
  \centering
  \includegraphics[width=0.5\linewidth]{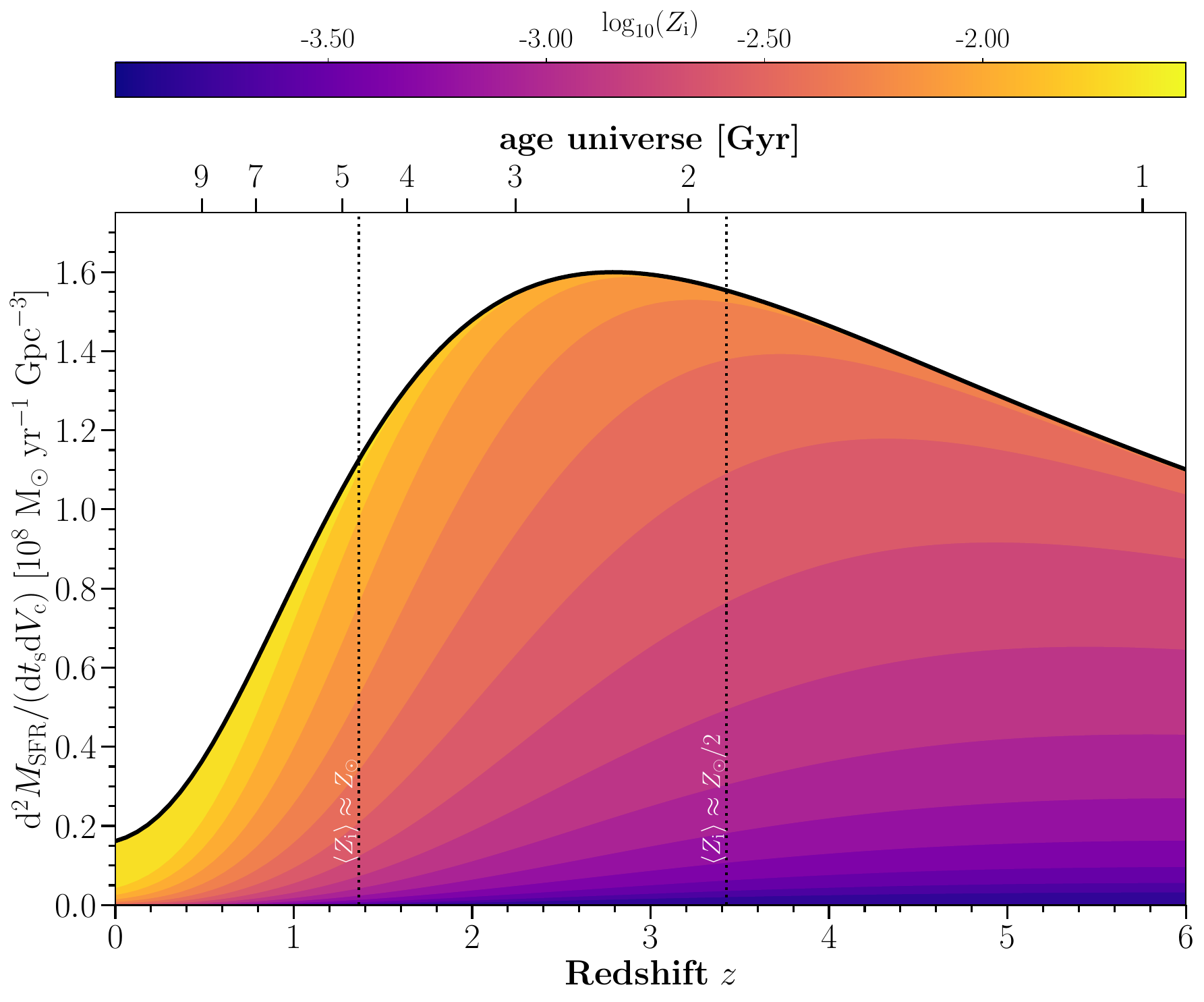}
  \caption{The assumed SFRD colored by stars' initial metallicity, $Z_i$. The dotted lines show the location of the median star formation rate for a few metallicities (i.e. half of the stars form with higher metallicity and half with lower metallicity than the labeled value). }
  \label{fig:SFRD_by_Z}
\end{figure*}

Figure \ref{fig:SFRD_by_Z} shows the assumed metallicity-dependent SFRD as a function of redshift, $\mathcal{S}(Z_i,z)$. At high redshifts, most stars form with low birth metallicities $Z_i \lesssim \Zsun/5$. At higher redshift, around $z\gtrsim 3$, most stars instead form from higher metallicity $Z_i \gtrsim \Zsun/2$. We showed in Figure~\ref{fig:formation_rate_A} that the formation yield of BBH mergers in our simulations drastically drops for stars formed with $Z_i\gtrsim 5$. This figure implies that metallicity effects will therefore lead to a drop in the formation of merging BBHs at redshifts $z \lesssim 3$.  

\section{The shape of the merger rate due to different delay time magnitudes}
\label{sec:appendix_delay_time_shift}

\begin{figure*}[h!]
    \centering
    \includegraphics[width=\linewidth]{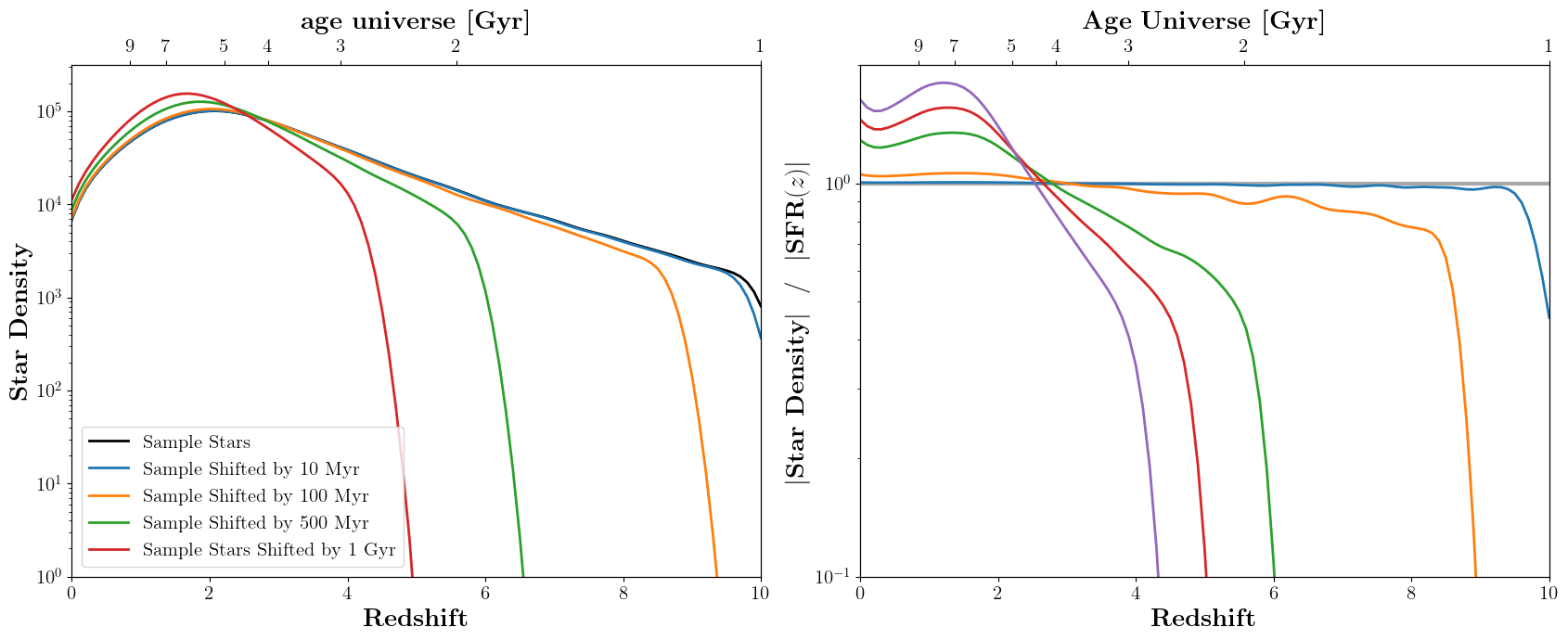}
    \caption{The effect of delay time on the distribution of stars throughout cosmic history. We use rejection sampling of one million stars to acquire a sample of stars reflecting the assumed \ac{SFRD} \citep{Madau_Fragos_2017}. Then, we add $10\, \textrm{Myr}$, $100\, \textrm{Myr}$, and $1\, \textrm{Gyr}$ to their birth ages. The left panel shows the original distribution and the distributions of the stars shifted by the three different example delay times. In the right panel, we normalize each distribution and then divide by the normalized \ac{SFRD}. In doing so, we can see how delay time effects cause the population of stars at low redshift to be greater than the \ac{SFRD}.}
    \label{fig:sample_stars_shifted}
\end{figure*}

In Figure \ref{fig:sample_stars_shifted}, we show how delay times affect a synthetic distribution of stars throughout cosmic history. We rejection sample the \ac{SFRD} from one million stars out to redshift $10$, which we plot in black. We then age the sample stars by $10$ Myr, $100$ Myr, and $1$ Gyr and plot their distributions. In the left panel, we see that the stars travel to lower redshift throughout their lifetimes; in other words, the distributions of the aged stars shift increasingly from the birth distribution toward lower redshift. Furthermore, this effect only causes a notable deviation from the star formation rate when the delay time is of order $500 \textrm{ Myr}$. A non-trivial aspect of this shift is the fact that the relationship between time and redshift is non-linear. As we can see from the difference between the x-axes on the bottom and top of the panels, time passes faster at higher redshift. This implies that a star born at a high redshift should travel more redshift during its lifetime before merging than if it were born at low redshift.

In the right panel of Figure \ref{fig:sample_stars_shifted}, we take the distributions from the left panel, normalize them, and then divide by the normalized assumed \ac{SFRD} (similarly to the right panel of Figure \ref{fig:merger_rate_A}). The horizontal grey line is what we would observe if the distribution of stars exactly followed the \ac{SFRD}. Again, we observe that the longer stars are allowed to evolve, the lower redshift they travel to. The stars that were aged by $1$ Gyr in particular show the extent to which delay times can cause distributions to shift to lower redshift, as the distribution is far above \ac{SFRD} (gray line) for reshifts lower than $z \sim 3$, and higher for redshifts above. 

\section{BBH formation channels as a function of redshift}\label{sec:appendix:BBH-formation-channels-as-a-function-of-redshift}

\begin{figure*}[h!]
    \centering
    \includegraphics[width=0.92\linewidth]{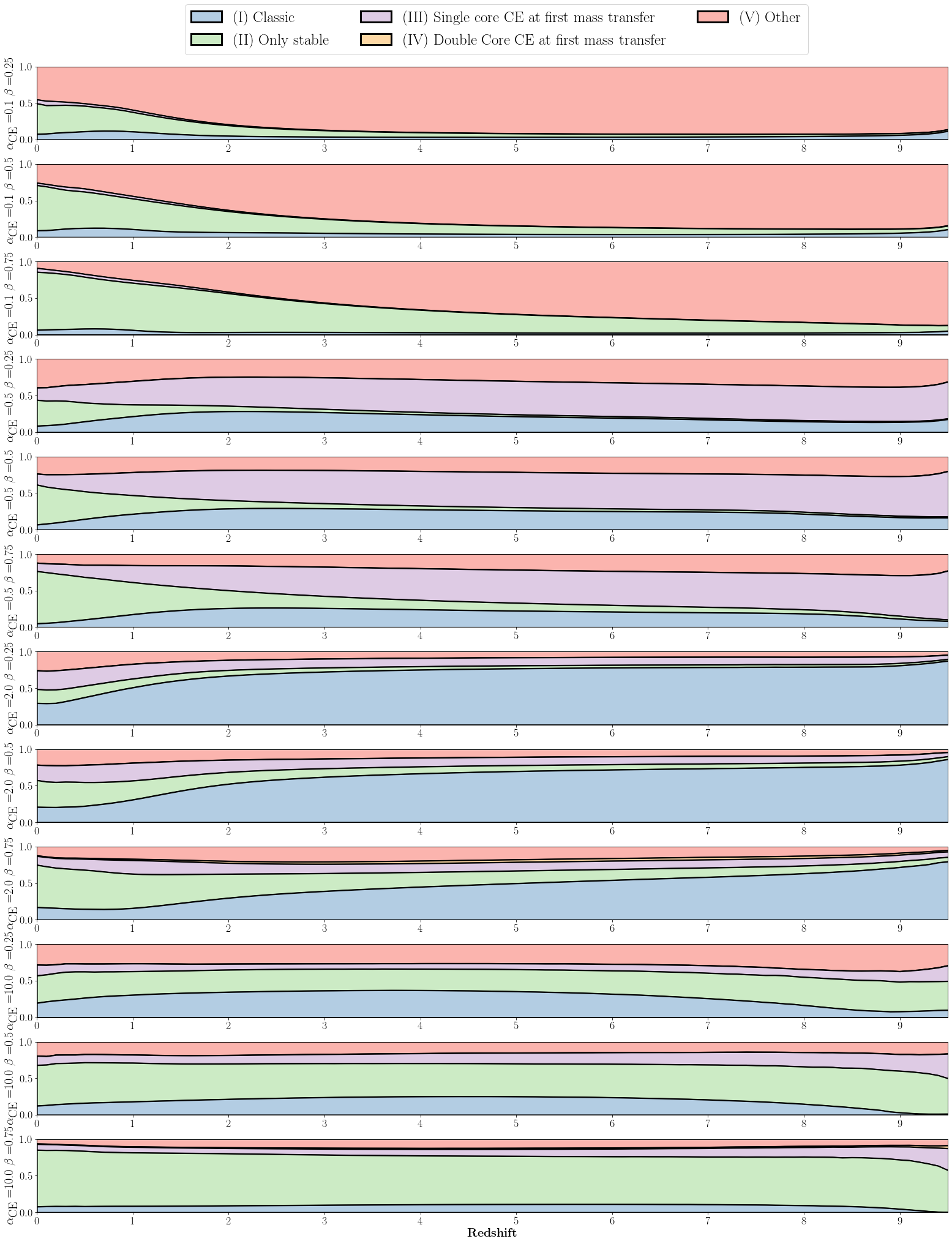}
    \caption{The percent contributions of each formation channel (as described in \citealt{Broekgaarden_2021}) for the merger rates as a function of redshift of all models in grid A.}
    \label{fig:all_merger_rates_by_formation_channel_A}
\end{figure*}

As was discussed in the main body of this paper, the delay time distribution is strongly correlated with the formation channels that lead to mergers. In Figure \ref{fig:all_merger_rates_by_formation_channel_A}, we show the percent contribution of the main formation channels as described in \citep{Broekgaarden_2021} to the merger rate as a function of redshift for all models in grid A. Clearly, the formation pathways that \ac{BBH} merger progenitors follow are highly dependent on physical assumptions---the contributions of each formation channel to the merger rate are vastly different from model to model.
Furthermore, the proportion of mergers created by each channel changes considerably throughout cosmic history, again likely as a result of the difference in the initial metallicity of  \acp{BBH} progenitors at different redshifts and their interplay with the model prescriptions. Noticeably, trends in channel contribution also appears to be similar for each group of models with the same assumed $\alphaCE$. This observation is the cause of the delay time distributions being grouped by $\alphaCE$---the shape of the median delay times throughout cosmic history are similar because the proportion of mergers created by each channel is the same. Furthermore, the models with $\alphaCE = 2.0$, which have the most distinct median delay time shape, have a drastic increase in the proportion of mergers formed through the classic channel (with a common-envelope phase) which corresponds to a drastic decrease in median delay time from $z=0$ to $z\approx 3$ as at higher redshift the BBH systems form through the classic-CE channel that leads to shorter delay times compared to the only stable mass transfer channel. This is in agreement with studies such as \citep[][]{Olejak:2022supernova, van_Son_2022} which found that channels that include \ac{CE}s tend to have lower delay times.
Ultimately, the most important conclusion from this figure is that the delay time distribution is strongly dependent on the channels that the \ac{BBH} mergers follow, and changes in channel contributions throughout cosmic history are highly non-trivial and dependent on the model assumptions.

\section{Deviations between simulations and toy models}
\begin{table*}
\caption{The minimum, maximum, and median of the normalized BBH merger rates for each model divided by normalized delayed variations of the SFRD for redshift values $0$ - $9$. The first three columns use a toy model rate by rescaling an SFRD with no delay, the middle three colums use as toy model a SFRD with a constant delay of $20\ \textrm{Myr}$, and the last three columns use for the toy model a scaled SFRD with a delay time distribution using: $t^{-1}$ with min($t_\textrm{delay}) = 20\ \textrm{Myr}$.}

\resizebox{1.2\textwidth}{!}{%
\hspace{-2.6cm}
\begin{tabular}{c|ccc|ccc|ccc}
    \hline \hline
    & \multicolumn{3}{c|}{No delay}   &   \multicolumn{3}{c|}{$t_{\rm{delay}} \propto \delta(t - 20 \textrm{\ Myr}) $}   &  \multicolumn{3}{c}{$t_{\rm{delay}} \propto 1/t $} \\ \hline  \hline
    Redshift        &   \hspace{0.2cm}   Minimum  \hspace{0.2cm}     &    \hspace{0.2cm} Median   \hspace{0.2cm}   &  \hspace{0.2cm}   Maximum  \hspace{0.2cm}   &  \hspace{0.2cm}   Minimum  \hspace{0.2cm}     &    \hspace{0.2cm} Median   \hspace{0.2cm}   &  \hspace{0.2cm}   Maximum  \hspace{0.2cm}  &   \hspace{0.2cm} Minimum    \hspace{0.2cm}   &   \hspace{0.2cm}  Median   \hspace{0.2cm}   &   \hspace{0.2cm}  Maximum    \hspace{0.2cm}   \\ \hline 
    $z=0$   &   1.27   &    2.43   &   3.48   &   1.74   &   3.33   &   4.76   &   0.21   &   0.41   &   0.59    \\   \hline
    $z=1$   &   0.94   &    1.26   &   1.62   &   0.90   &   1.22   &   1.56   &   0.59   &   0.80   &   1.02    \\   \hline
    $z=2$   &   0.90   &    0.95   &   1.00   &   0.89   &   0.94   &   0.98   &   1.07   &   1.13   &   1.18    \\   \hline
    $z=3$   &   0.79   &    0.90   &   1.02   &   0.78   &   0.89   &   1.00   &   0.97   &   1.11   &   1.26    \\   \hline
    $z=4$   &   0.73   &    0.97   &   1.10   &   0.72   &   0.95   &   1.09   &   0.89   &   1.18   &   1.35    \\   \hline
    $z=5$   &   0.67   &    0.99   &   1.12   &   0.68   &   1.01   &   1.15   &   0.94   &   1.39   &   1.58    \\   \hline
    $z=6$   &   0.59   &    0.96   &   1.10   &   0.63   &   1.02   &   1.16   &   1.03   &   1.67   &   1.90    \\   \hline
    $z=7$   &   0.49   &    0.87   &   1.01   &   0.56   &   0.99   &   1.16   &   1.10   &   1.94   &   2.26    \\   \hline
    $z=8$   &   0.37   &    0.70   &   0.86   &   0.46   &   0.88   &   1.08   &   1.12   &   2.16   &   2.63    \\   \hline
    $z=9$   &   0.20   &    0.41   &   0.59   &   0.30   &   0.61   &   0.89   &   1.03   &   2.07   &   3.00    \\   \hline
    \hline \hline
    \end{tabular}%
\hspace{5cm}
}
\label{tab:merger_rate_SFR_deviations}
\end{table*}

Table~\ref{tab:merger_rate_SFR_deviations} shows the deviations at different redshifts between our simulated merger rates and the toy model merger rates.

\end{document}